\documentclass[usenatbib]{amsart}

\usepackage[foot]{amsaddr}                  
\usepackage{amsmath}                        
\usepackage{amssymb}                        
\usepackage{dsfont} 	                    
\usepackage{bm}                             
\usepackage{mathtools}                      
\usepackage[hypertexnames=false]{hyperref}  
\usepackage[scientific-notation=true]{siunitx}
\usepackage{blkarray}
\usepackage{enumerate}
\usepackage{centernot}
\usepackage{paralist}
\usepackage{stackrel}
\usepackage[table]{xcolor}
\usepackage[colorinlistoftodos,prependcaption,textsize=tiny]{todonotes} 
\usepackage{pagenote}
\usepackage{setspace}
\usepackage[english]{babel}
\usepackage{subfig}
\usepackage{dirtytalk}
\usepackage[authoryear]{natbib}

\usepackage{algorithm}
\usepackage{algpseudocode}

\usepackage{listings}
\usepackage{color}

\usepackage{amsthm}
\theoremstyle{plain}
\theoremstyle{definition}

\theoremstyle{remark}

\DeclareMathOperator{\tr}{tr}               
\newcommand{\LA}{\mathbf{\Lambda}}
\newcommand{\LAt}{\mathbf{\Lambda}^{\top}}
\newcommand{\PS}{\mathbf{\Psi}}

\newcommand{\PH}{\mathbf{\Phi}}

\newcommand{\covmat}{\boldsymbol{\sigma}}
\newcommand{\covmath}{\hat{\boldsymbol{\sigma}}}

\newcommand{\Covmat}{\boldsymbol{\Sigma}}
\newcommand{\Covmath}{\hat{\boldsymbol{\Sigma}}}
\newcommand{\Cormat}{\mathbf{R}}
\newcommand{\Cormath}{\hat{\mathbf{R}}}
\newcommand{\Y}{\mathbf{Y}}
\newcommand{\Yh}{\hat{\mathbf{Y}}}
\newcommand{\y}{\mathbf{y}}

\newcommand{\G}{\mathbf{G}}
\newcommand{\Gh}{\hat{\mathbf{G}}}
\newcommand{\g}{\mathbf{g}}
\newcommand{\gh}{\hat{\mathbf{g}}}
\newcommand{\E}{\mathbf{E}}

\newcommand{\e}{\boldsymbol{\epsilon}}

\newcommand{\err}{\boldsymbol{\varepsilon}}

\newcommand{\X}{\mathbf{X}}

\newcommand{\K}{\mathbf{K}}
\newcommand{\Z}{\mathbf{Z}}

\newcommand{\Vh}{\hat{\mathbf{V}}}
\newcommand{\I}{\mathbf{I}}
\newcommand{\params}{\boldsymbol{\Theta}}

\makeatletter
\def\squarecorner#1{
	\pgf@x=\the\wd\pgfnodeparttextbox%
	\pgfmathsetlength\pgf@xc{\pgfkeysvalueof{/pgf/inner xsep}}%
	\advance\pgf@x by 2\pgf@xc%
	\pgfmathsetlength\pgf@xb{\pgfkeysvalueof{/pgf/minimum width}}%
	\ifdim\pgf@x<\pgf@xb%
	\pgf@x=\pgf@xb%
	\fi%
	\pgf@y=\ht\pgfnodeparttextbox%
	\advance\pgf@y by\dp\pgfnodeparttextbox%
	\pgfmathsetlength\pgf@yc{\pgfkeysvalueof{/pgf/inner ysep}}%
	\advance\pgf@y by 2\pgf@yc%
	\pgfmathsetlength\pgf@yb{\pgfkeysvalueof{/pgf/minimum height}}%
	\ifdim\pgf@y<\pgf@yb%
	\pgf@y=\pgf@yb%
	\fi%
	\ifdim\pgf@x<\pgf@y%
	\pgf@x=\pgf@y%
	\else
	\pgf@y=\pgf@x%
	\fi
	\pgf@x=#1.5\pgf@x%
	\advance\pgf@x by.5\wd\pgfnodeparttextbox%
	\pgfmathsetlength\pgf@xa{\pgfkeysvalueof{/pgf/outer xsep}}%
	\advance\pgf@x by#1\pgf@xa%
	\pgf@y=#1.5\pgf@y%
	\advance\pgf@y by-.5\dp\pgfnodeparttextbox%
	\advance\pgf@y by.5\ht\pgfnodeparttextbox%
	\pgfmathsetlength\pgf@ya{\pgfkeysvalueof{/pgf/outer ysep}}%
	\advance\pgf@y by#1\pgf@ya%
}
\makeatother

\pgfdeclareshape{square}{
	\savedanchor\northeast{\squarecorner{}}
	\savedanchor\southwest{\squarecorner{-}}
	
	\foreach \x in {east,west} \foreach \y in {north,mid,base,south} {
		\inheritanchor[from=rectangle]{\y\space\x}
	}
	\foreach \x in {east,west,north,mid,base,south,center,text} {
		\inheritanchor[from=rectangle]{\x}
	}
	\inheritanchorborder[from=rectangle]
	\inheritbackgroundpath[from=rectangle]
}

\def\ImportFromMnSymbol#1{%
	\DeclareFontFamily{U} {MnSymbol#1}{}
	\DeclareFontShape{U}{MnSymbol#1}{m}{n}{
		<-6> MnSymbol#15
		<6-7> MnSymbol#16
		<7-8> MnSymbol#17
		<8-9> MnSymbol#18
		<9-10> MnSymbol#19
		<10-12> MnSymbol#110
		<12-> MnSymbol#112}{}
	\DeclareFontShape{U}{MnSymbol#1}{b}{n}{
		<-6> MnSymbol#1-Bold5
		<6-7> MnSymbol#1-Bold6
		<7-8> MnSymbol#1-Bold7
		<8-9> MnSymbol#1-Bold8
		<9-10> MnSymbol#1-Bold9
		<10-12> MnSymbol#1-Bold10
		<12-> MnSymbol#1-Bold12}{}
	\DeclareSymbolFont{MnSy#1} {U} {MnSymbol#1}{m}{n}
}
\newcommand\DeclareMnSymbol[4]{\DeclareMathSymbol{#1}{#2}{MnSy#3}{#4}}
\ImportFromMnSymbol{A}
\DeclareMnSymbol{\ConIndepNat}{\mathrel}{A}{225}

\def\ConIndepNatt{\ConIndepNat}

\makeatletter
\newcommand{\addresseshere}{%
  \enddoc@text\let\enddoc@text\relax
}

\definecolor{lightgrey}{rgb}{0.9,0.9,0.9}
\definecolor{darkgreen}{rgb}{0,0.6,0}
\calclayout

\DeclareMathOperator{\vect}{vec}
\usepackage{enumitem}
\usepackage{tikz}
\tikzstyle{node}=[very thick, circle, draw=black, minimum size=22, inner sep=0.8, outer sep=0.6]

\usetikzlibrary{calc}
\usetikzlibrary{arrows.meta}
\usetikzlibrary{chains}
\usetikzlibrary{arrows, arrows.meta}

\DeclareRobustCommand{\rvdots}{%
	\vbox{
		\baselineskip4\p@\lineskiplimit\z@
		\kern-\p@
		\hbox{.}\hbox{.}\hbox{.}
}}

\begin{document}
\sloppy

\title[Genomic Prediction: Genetic Latent Factor Approach]
{Improving Genomic Prediction using High-dimensional Secondary Phenotypes: the Genetic Latent Factor Approach}

\author{Killian A.C.\ Melsen$^{1*}$}
\author{Jonathan F.\ Kunst$^1$}
\author{José Crossa$^2$}
\author{Margaret R.\ Krause$^3$}
\author{Fred A.\ van Eeuwijk$^1$}
\author{Willem Kruijer$^1$}
\author{Carel F.W.\ Peeters$^1$}

\address{$^1$Mathematical \& Statistical Methods Group (Biometris), Wageningen University \& Research, PO Box 16, 6700 AA, Wageningen, The Netherlands}
\address{$^2$Global Wheat Program, International Maize and Wheat Improvement Centre (CIMMYT), Texcoco, Mexico}
\address{$^3$College of Agricultural Sciences, Oregon State University, Corvallis, OR, USA}

\email{$^*$killian.melsen@wur.nl}

\begin{abstract}
\label{abstract}
Decreasing costs and new technologies have led to an increase in the amount of data available to plant breeding programs.
High-throughput phenotyping (HTP) platforms routinely generate high-dimensional datasets of secondary features that may be used to improve genomic prediction accuracy.
However, integration of these data comes with challenges such as multicollinearity, parameter estimation in $p > n$ settings, and the computational complexity of many standard approaches.
Several methods have emerged to analyze such data, but interpretation of model parameters often remains challenging.
We propose genetic latent factor best linear unbiased prediction (glfBLUP), a prediction pipeline that reduces the dimensionality of the original secondary HTP data using generative factor analysis.
In short, glfBLUP uses redundancy filtered and regularized genetic and residual correlation matrices to fit a maximum likelihood factor model and estimate genetic latent factor scores.
These latent factors are subsequently used in multi-trait genomic prediction.
Our approach performs better than alternatives in extensive simulations and a real-world application, while producing easily interpretable and biologically relevant parameters.
We discuss several possible extensions and highlight glfBLUP as the basis for a flexible and modular multi-trait genomic prediction framework.

\end{abstract}

\maketitle


\section{Introduction}
\label{s:intro}
Genomic prediction has revolutionized plant breeding programs over the past two decades.
Its use allows maintaining and evaluating breeding populations larger than ever before.
As a result, breeding programs have seen increases in efficiency leading to more rapid development of varieties that meet today's challenges \citep{Crossaetal2013}.
Genomic prediction is a powerful tool in ensuring food security in the face of a growing population and more extreme climate conditions \citep{Harfoucheetal2019}.

The practice of genomic prediction involves training a statistical model using evaluated genotypes, followed by making trait predictions for untrialed test genotypes \citep{Meuwissenetal_2001}.
The model is mainly used in a univariate form (gBLUP, genomic best linear unbiased prediction) where the predictions are sums of genomic marker effects \citep{Endelman_2011}.
These predictions allow breeders to make selections without the need for extensive field trials.

Recent years have seen an increase in the use of high-throughput phenotyping (HTP) technologies resulting in the characterization of individuals in more aspects than just those of the genomic and trait phenotypic data.
Examples include the collection of hyperspectral reflectivity measurements on the crop canopy in large field trials, metabolic profiling of plant leaves, and gene expression profiling using DNA microarrays \citep[e.g.,][]{Krauseetal_2019}.
These secondary phenotypes can be collected for either the training set alone (termed CV1, cross-validation scenario 1), or both the training and test individuals (CV2, cross-validation scenario 2) \citep{Runcie&cheng_2019}.
If of sufficient quality, the secondary data can contain valuable information that readily improves genomic predictions for the trait of interest (focal trait).
There are several such examples where integration of genomic and phenomic data leads to higher accuracies \citep{Adaketal_2023}

Univariate genomic prediction uses only genomic information and focal trait information to obtain predictions.
To integrate phenomics and genomics, the univariate model can be extended to a multivariate form relatively easily by modeling the focal trait as well as the secondary phenotypes as functions of the SNP markers \citep{Arouisseetal_2021}.
Direct integration of HTP data to improve predictions is challenging for several reasons, however.
First of all, secondary HTP data are often high-dimensional complicating parameter estimation.
Second, subsets of secondary features are often highly correlated.
This multicollinearity can greatly complicate the interpretation of model parameters \citep{Fotheringham&Oshan_2016}.
Finally, even if the number of observations were greater than the number of features, and features were not highly correlated, direct integration would still be computationally demanding as the calculation of the BLUPs requires inverting a symmetric matrix whose dimension equals the number of genotypes times the number of features.

\subsection{Current methodology}
A number of methods have been developed to deal with the high dimensionality, multicollinearity, and computational complexity.
\citet{Lopez-Cruzetal_2020} proposed linearly combining the secondary features into a single regularized selection index (siBLUP), thereby reducing the dimensionality to one.
The lsBLUP method introduced by \citet{Arouisseetal_2021} is similar in the sense that it reduces the dimensionality to one, but uses a LASSO prediction of the focal trait using the HTP features instead of a regularized selection index.
Another alternative is MegaLMM as proposed by \citet{Runcieetal_2021}.
MegaLMM constructs an arbitrary number of independent factors and unique trait contributions that are linear combinations of the original features.
Neural networks and deep learning have also been investigated as options for the integration of secondary HTP and genomic data \citep{Togninallietal_2023}.
Due to their flexible nature they present an attractive option to model complex interactions between many features.

\subsection{The proposed model \& overview}
We introduce glfBLUP (genetic latent factor BLUP) to deal with the challenges of secondary HTP data.
It is based on the idea that HTP features typically form a set of many noisy measurements of a much lower dimensional set of latent features \citep{Münchetal_2022}.
In short, we use factor analysis to reduce the dimensionality of the HTP data to a data-driven number of uncorrelated latent factors which are subsequently used in multivariate genomic prediction.
We avoid any manual hyperparameter tuning by using a data-driven approach to determine the dimensionality of the factor model.
Our approach is also flexible in the sense that it can accommodate latent structures of any dimension lower than the Ledermann bound \citep{Ledermann1937}.
This flexibility means that complex structures underlying high-dimensional data can be accurately modeled.
Furthermore, the factor loadings and genetic correlations between factors and the focal trait are interpretable and reveal the relevant biological patterns in the otherwise noisy HTP data.

The remainder of this paper is organized as follows.
We elaborate the used notation and the proposed factor analytic multivariate genomic prediction approach in Section \ref{s:model}.
Section \ref{s:sim} provides details and results on a comprehensive simulation study comparing the performances of different methods as well as the importance of various genetic parameters.
In Section \ref{s:cimmyt} we apply glfBLUP and alternatives to a wheat yield and hyperspectral reflectivities dataset to evaluate their real-world performance and show how glfBLUP produces biologically relevant and interpretable parameters.
Finally, we discuss several possible extensions to the proposed approach in Section \ref{s:discuss}.

\section{Model}\label{s:model}
\subsection{Notation}\label{s:not}
We denote matrices and vectors by boldface upper- and lower-case symbols, and scalars by lower-case symbols.
Furthermore, we use $\Covmat$ and $\Cormat$ to denote covariance and correlation matrices.
We additionally use the superscripts $p$, $g$, and $\epsilon$ to denote phenotypic, genetic, and residual covariance and correlation matrices.
For example, the symbol $\Covmat^\epsilon$ denotes the residual covariance matrix.

Let $\mathbf{A} \in \mathbb{R}^{n \times (p+1)}$ be some data matrix containing $p$ secondary features and a single focal trait for $n$ individuals.
While we thus use $p$ for the number of secondary features as well as phenotypic covariance matrices, its meaning should be clear from the context.
We assume that genotypes are replicated $r$ times following a balanced design so that the total number of individuals $n=n_gr$, where $n_g$ and $r$ represent the number of genotypes and number of replicates per genotype.
Note that we assume a balanced design for notational simplicity, but that it is not a requirement.
We use two indices to specify the rows of data matrices so that $\mathbf{a}_{(i|j)}$ represents the $1 \times (p + 1)$ data vector for replicate $i$ given genotype $j$.
We use the subscripts $s$ and $f$ to denote a matrix containing only secondary features or a vector containing only the focal trait.
As an example, $\mathbf{A}_s$ is a data matrix containing secondary features, and $\mathbf{a}_{s(i|j)}$ is the row vector of secondary features for genotype $j$, replicate $i$.
The kinship matrix is denoted by $\K$.

We use $\circ$ and $\otimes$ to denote the Hadamard and Kronecker products.
We use $diag(s_{11},\dots, s_{pp})$ to refer to a $p$-dimensional diagonal matrix with elements $s_{11},\dots, s_{pp}$.
We denote the column-wise vectorization of a matrix by $\vect(\mathbf{A})$ and use $\vect^{-1}(\mathbf{A})$ for the reverse operation.
The cardinality of a set $\mathcal{A}$ is denoted by $\text{card}(\mathcal{A})$.
Some further information and examples on the partitioning of data and covariance matrices, as well as some general background information on genomic prediction, can be found in SM Section 1.

\subsection{Modeling of secondary features}\label{s:sec}
The pipeline we propose uses a fully unsupervised dimensionality reduction before including factor scores as additional traits in multi-variate genomic prediction.
This section contains details on the steps involved in the dimensionality reduction.
Section \ref{s:foc} then provides some details on the multivariate genomic prediction.
\subsubsection{Model and assumptions}\label{sec:secmodel_assumptions}
Consider the secondary feature phenotypic data matrix $\Y_{s} \in \mathbb{R}^{n \times p}$.
We assume a balanced design for simplicity, but note that this is not required.
In a balanced design $n=n_gr$, i.e., the number of observations is equal to the number of genotypes multiplied by the number of replicates per genotype.
Assuming no covariates, all secondary phenotypes are composed of genetic and residual components:
\begin{equation*}
	\Y_{s} = \G_{s} + \E_{s} \text{.}
\end{equation*}
We assume that the genetic components are independent of the residual components so that
\begin{equation*}
	\vect({\Y_{s}}) = \vect({\G_{s}}) + \vect({\E_{s}}) \sim \mathcal{N}_{np}(\mathbf{0}, \Covmat^g_{ss} \otimes \Z\K\Z^\top + \Covmat^\epsilon_{ss} \otimes \I_{n}) \text{,}
\end{equation*}
where $\Z\K\Z^\top$ and $\I_n$ are the row covariances of $\G_{s}$ and $\E_{s}$, respectively.
The $n \times n_g$ incidence matrix $\Z$ links individuals to genotypes.
Similarly, $\Covmat^g_{ss}$ and $\Covmat^\epsilon_{ss}$ are the column covariances of $\G_{s}$ and $\E_{s}$.
We use the subscript ${ss}$ to denote that $\Covmat^g_{ss}$ and $\Covmat^\epsilon_{ss}$ contain genetic and residual covariances among secondary features.

While we require plot-level data to estimate covariance matrices, two-stage genomic prediction typically uses best linear unbiased estimates (BLUEs).
For a completely randomized design we can obtain these BLUEs for a single genotype $j$ by simply taking the genotypic mean of all replicates $i=1,\dots,r$ \citep{Kruijeretal2020}:
\begin{equation}\label{eq:BLUE}
	\bar{\y}_{s(j)}=\g_{s(j)}+\bar{\e}_{s(j)}=\g_{s(j)}+\dfrac{1}{r}\sum_{i=1}^{r}\e_{s(i|j)} \text{.}
\end{equation}
Note that as the residual vectors $\e_{s(i|j)}$ of the replicates of genotype $j$ are \textit{iid} random vectors, basic covariance algebra shows that the covariance of $\bar{\e}_{s(j)}$ is simply $r^{-1}\Covmat_{ss}^\epsilon$ \citep{Kruijeretal2020} (see also Supplementary Material (SM) Section 2).
As a result, $\bar{\y}_{s(j)} \sim \mathcal{N}_p(\mathbf{0}, \Covmat_{ss}^p)$, where $\Covmat_{ss}^p=\Covmat_{ss}^g + r^{-1}\Covmat_{ss}^\epsilon$, assuming for a moment that $\K \circ \I_{n_g} = \I_{n_g}$.
Also note that the assumption of \textit{iid} errors is valid only if data is truly from a completely randomized design.
If some pre-processing is applied in a two-stage modeling approach, e.g., spatial correction, the assumption is an approximation only.

In glfBLUP, we use a factor analytic structure for $\Covmat_{ss}^g$.
This follows from the assumption that the genetic components $\g_{s(j)}$ can be written as a weighted sum of $m$ latent factors and $p$ factor model errors:
\begin{equation}\label{eq:G_as_factors}
	\g_{s(j)}^\top=\LA\boldsymbol{\xi}_{j}+\err_{j} \text{,}
\end{equation}
where $\LA$ is a $p \times m$ matrix containing factor loadings, $\boldsymbol{\xi}_{j}$ is an $m \times 1$ vector containing the latent factor scores for genotype $j$, and $\err_j$ contains the factor model errors.
Substituting (\ref{eq:G_as_factors}) into (\ref{eq:BLUE}) gives us the full model for the secondary phenotypic data:
\begin{equation}\label{eq:full_model}
	\bar{\y}_{s(j)}^\top=\LA\boldsymbol{\xi}_{j}+\err_{j}+\bar{\e}_{s(j)}^\top \text{.}
\end{equation}
We now make the following assumptions:
(i) $\text{rank}(\LA) = m < p$,
(ii) $\err_{j}\sim \mathcal{N}_{p}(\mathbf{0}, \PS)$, where $\PS = \mathbb{E}(\err_{j}\err_{j}^\top)$ and $\PS \equiv diag(\psi_{11},\dots,\psi_{pp})$ with $\psi_{qq}>0,\forall qq$,
(iii) $\boldsymbol{\xi}_{j}\sim \mathcal{N}_{m}(\mathbf{0}, \I_{m})$,
(iv) $\boldsymbol{\xi}_{j} \ConIndepNatt \err_{j'}, \forall j,j'$,
(v) $\boldsymbol{\xi}_{j} \ConIndepNatt \bar{\e}_{s(j')}, \forall j,j'$, and
(vi) $\err_{j} \ConIndepNatt \bar{\e}_{s(j')}, \forall j,j'$.

These assumptions imply that the genetic covariance matrix $\Covmat_{ss}^{g}$ equals the model-implied genetic covariance matrix $\Covmat_{ss}^{g}(\params)=\LA\LAt+\PS=\mathbb{E}\left[(\LA\boldsymbol{\xi}_{j}+\err_{j})(\LA\boldsymbol{\xi}_{j}+\err_{j})^\top\right]$ where $\params=\{m, \LA, \PS\}$.
Therefore, $\bar{\y}_{s(j)}\sim \mathcal{N}_{p}(\mathbf{0}, \LA\LAt+\PS+r^{-1}\Covmat_{ss}^{\epsilon})$.
The covariance of the genetic components $\g_{s(j)}$ can thus be decomposed into common ($\LA\LAt$) and unique ($\PS$) elements.

We will use the model above to project the observed secondary features onto a lower-dimensional latent space in six steps.
We first use the plot-level data to obtain estimates for the genetic and residual correlation matrices of the observed secondary features (I).
We then redundancy filter the secondary data (II) to ensure there are no pairs of features with near perfect genetic correlations, followed by regularization of the filtered genetic and residual correlation matrices (III).
We then estimate the latent dimension (IV) and fit the factor model (V).
Next, we obtain estimates for the latent factor scores (VI), completing the dimension reduction.
The remainder of this section is devoted to the details of these six steps.

\subsubsection{Covariance matrix estimation (step I)}\label{sec:covmat_estimation}
The factor analytic approach in glfBLUP largely revolves around the genetic and residual covariance matrices of the observed secondary data.
We can obtain estimates of these matrices by using REML (restricted maximum likelihood) and the kinship matrix $\K$ for a small number of features.
When the number of secondary features $p$ exceeds $5 \text{-} 10$, however, the computational complexity quickly becomes prohibitive and REML estimates can no longer be obtained \citep{Zhou&Stephens2014}.
As this is usually the case for high-dimensional HTP data, we use data containing genotypic replicates and correct for experimental design factors and spatial trends to obtain pseudo-CRD (pseudo completely randomized design) data as a pre-processing step.
Note again that in this case, assuming balanced data, the total number of records is simply the number of genotypes times the number of replicates, i.e., $n=n_gr$.
We then assume that $\Z\K\Z^\top = \Z\Z^\top$, i.e., $\K=\I_{n_g}$, and obtain estimates for $\Covmat_{ss}^{g}$ and $\Covmat_{ss}^{\epsilon}$ using sums of squares as measures for between and within-genotype variation \citep{Ott&Longnecker2015}:
\begin{equation}\label{eq:SS}
	\begin{split}
		\mathbf{MS}_{ss}^g&=(\bar{\Y}_s^g-\bar{\Y}_s)^\top(\bar{\Y}_s^g-\bar{\Y}_s)(n_{g}-1)^{-1} \\
		\mathbf{MS}_{ss}^\epsilon&=(\bar{\Y}_s^g-\Y_s)^\top(\bar{\Y}_s^g-\Y_s)(n-n_{g})^{-1} \\
		\Covmath_{ss}^g&=(\mathbf{MS}_{ss}^g-\mathbf{MS}_{ss}^\epsilon) r^{-1} \\
		\Covmath_{ss}^\epsilon&=\mathbf{MS}_{ss}^\epsilon \text{,}
	\end{split}
\end{equation}
where $\Y_s$ is the $n \times p$ training set data matrix containing the secondary features. Overall means and genotypic means are located in $\bar{\Y}_s$ and $\bar{\Y}_s^g$, respectively.
A well-known issue with the estimation of such covariance matrices using multivariate sums of squares is the fact that sufficient within-genotype variation can result in $\Covmath_{ss}^{g}$ having negative eigenvalues \citep{Meyer&Kirkpatrick2010} as it is not a moment matrix but rather the difference of two moment matrices (as opposed to $\Covmath_{ss}^{\epsilon}$ which \emph{is} a moment matrix and therefore at worst positive-semidefinite if $p > n$).
We deal with this issue by simply finding the nearest positive-definite matrix using Higham's algorithm \citep{Higham2002}.
From here on, $\Covmath_{ss}^{g}$ refers to this nearest positive definite matrix.

We scale the covariance matrices as our approach requires them to be on the correlation scale for regularization:
\begin{equation*}\label{eq:covtocor}
	\begin{split}
		\Cormath_{ss}^{g}&=(\Covmath_{ss}^{g}\circ \I_p)^{-1/2} \Covmath_{ss}^g (\Covmath_{ss}^{g}\circ \I_p)^{-1/2} \\
		\Cormath_{ss}^{\epsilon}&=(\Covmath_{ss}^{\epsilon}\circ \I_p)^{-1/2} \Covmath_{ss}^{\epsilon} (\Covmath_{ss}^{\epsilon}\circ \I_p)^{-1/2} \text{.}
	\end{split}
\end{equation*}
We emphasize that the assumption of independent genetic effects is only made for the genetic and residual covariance estimators in equation (\ref{eq:SS}).
We use a marker-based kinship matrix in the final genomic prediction step.
The assumption of diagonal kinship ($\K=\I_{n_g}$) in (\ref{eq:SS}) simply means that the subsequent factor analysis is based on the covariance between the total (rather than only the additive) genetic effects.

\subsubsection{Redundancy filtering (step II)}
Given the fact that multicollinearity is common in HTP data, some of the $p$ secondary features will likely show near perfect genetic correlation and are therefore redundant.
We filter the genetic correlation matrix using the algorithm described by \citet{Peetersetal2019} with an absolute genetic correlation threshold $\tau$ (i.e., after filtering, no two features will have an absolute genetic correlation $|\rho^g|$ equal to, or exceeding $\tau$).
This filtering step produces $\Cormath_{ss}^{g*}\in \mathbb{R}^{p^*\times p^*}$ containing only the $p^*$ secondary features that pass filtering ($p^*\leq p$). We subset the residual correlation matrix and data matrix so they too only contain the remaining $p^*$ features.

\subsubsection{Regularization (step III)}\label{sec:regularization}
The filtered estimates $\Cormath_{ss}^{g*}$ and $\Cormath_{ss}^{\epsilon*}$ may still be ill-conditioned or singular which is why regularization is required.
To achieve this, we use penalized maximum likelihood forms in the remainder of the glfBLUP method \citep{Peetersetal2019, vanWieringen&Peeters2016}:
\begin{equation}\label{eq:identity_reg}
	\begin{split}
		\Cormath_{ss}^{g*}(\vartheta_g)&=(1-\vartheta_g)\Cormath_{ss}^{g*}+\vartheta_g \I_{p^*} \\
		\Cormath_{ss}^{\epsilon*}(\vartheta_\epsilon)&=(1-\vartheta_\epsilon)\Cormath_{ss}^{\epsilon*}+\vartheta_\epsilon \I_{p^*} \text{,}
	\end{split}
\end{equation}
where $\vartheta_g, \vartheta_\epsilon\in(0,1]$.
We find the optimal values for the penalty parameters $\vartheta_g^\ddagger$ and $\vartheta_\epsilon^\ddagger$ through $K$-fold cross-validation where we split the training data by dividing the $n_g$ genotypes over $K$ disjoint folds and minimize the following expression using Brent's algorithm \citep{Brent1973}:
\begin{equation}\label{eq:reg_minimization}
	\varphi^K(\vartheta):=\dfrac{1}{K}\sum_{k=1}^{K}w_k \left\{\text{ln}\Big|\Bigl(\Cormat_{ss}^{*}(\vartheta)\Bigr)_{\neg k}\Big|+\tr \left[\Bigl(\Cormat_{ss}^{*}\Bigr)_k\Bigl(\Cormat_{ss}^{*}(\vartheta)\Bigr)_{\neg k}^{-1}\right]\right\} \text{,}
\end{equation}
where $\left(\Cormat_{ss}^*(\vartheta)\right)_{\neg k}$ is the penalized genetic or residual correlation matrix given the corresponding penalty $\vartheta$ estimated using the $K-1$ folds.
The non-penalized correlation matrix $(\Cormat_{ss}^{*})_k$ is estimated using the left-out fold.
Each left-out fold is weighed by $w_k$, which is equal to the number of genotypes or individuals in that fold for the genetic and residual correlation matrix, respectively.
See \citet{Peetersetal2019} for further details.
We use an efficient implementation to minimize (\ref{eq:reg_minimization}) which results in iterations of complexity $O(p^*)$ as opposed to $O(p^{*3})$ due to the repeated inversion of $\Cormat_{ss}^{*}(\vartheta)_{\neg k}$ required in the naive minimization.
This leads to faster optimization, especially if optimal penalty values are close to the bounds.
Details regarding our implementation can be found in SM Section 3.
Note that both $\Cormath_{ss}^{g*}(\vartheta_g^\ddagger)$ and $\Cormath_{ss}^{\epsilon*}(\vartheta_\epsilon^\ddagger)$ balance the unbiased but high-variance estimated correlation matrices with the biased but stable identity matrix, optimizing the bias-variance trade-off.
While our approach uses the identity matrix, we may use another target matrix $\mathbf{T}$ in (\ref{eq:identity_reg}) instead if prior information is available in the form of, for example, covariance estimates from earlier trials.
The only requirement is that $\mathbf{T}$ is a positive-definite correlation matrix.

\subsubsection{Latent dimension (step IV)}
In principle, the model specified in (\ref{eq:full_model}) is valid for any latent dimension $m \leq \frac{1}{2} \left[2p + 1 - (8p + 1)^{1/2}\right]$ \citep{Ledermann1937}.
We must specify an estimate for $m$ prior to fitting the factor model and obtaining an estimate for the other parameters in $\params$, however.
To do so we use a lower eigenvalue bound based on the Marchenko-Pastur distribution \citep{Marcenko&pastur1967}:
\begin{equation*}\label{eq:MP}
	\omega_+=\left(1+\sqrt{\dfrac{p^*}{n_{g}}}\right)^2 \text{,}
\end{equation*}
where $n_g$ is the number of genotypes used to estimate the genetic covariance matrix as described in Section \ref{sec:covmat_estimation}.
We then determine $\tilde{m}$:
\begin{equation*}
	\tilde{m} := \text{card}\left\{d\left[\Cormath_{ss}^{g*}(\vartheta_g^\ddagger)\right] > \omega_+\right\} \text{,}
\end{equation*}
where $d[\Cormath_{ss}^{g*}(\vartheta_g^\ddagger)]$ are the eigenvalues of $\Cormath_{ss}^{g*}(\vartheta_g^\ddagger)$.
Consequently, the latent dimension we use is simply the number of eigenvalues of the regularized correlation-scale estimate for $\Covmat_{ss}^{g*}$, $\Cormath_{ss}^{g*}(\vartheta_g^\ddagger)$, that exceed $\omega_+$.

\subsubsection{Factor analytic modeling (step V)}\label{sec:FAP}
With regularized estimates for the correlation matrices and latent dimension available, we now proceed with the factor analytic dimension reduction of the redundancy filtered secondary data.
In the context of plant breeding, we are mainly interested in genetic correlations and will therefore fit the factor model using a reduced form of the secondary data covariance matrix.
Recall that the full covariance of the BLUEs given in Section \ref{sec:secmodel_assumptions} was $\Covmat_{ss}^{p*}=\Covmat_{ss}^{g*} + r^{-1}\Covmat_{ss}^{\epsilon*}$, with stars added to denote the redundancy filtering.
The reduced covariance that we use to fit the factor model is thus $\Covmat_{ss}^{p*}-r^{-1}\Covmat_{ss}^{\epsilon*}=\Covmat_{ss}^{g*}=\LA\LAt+\PS$.
Given the fact that maximum-likelihood estimation of $\params$ is scale-free \citep{Peeters2012}, we opt to estimate the loadings and uniquenesses using correlation rather than covariance matrices.
We denote these correlation-scale parameters by $\params_R=\{m, \LA_R, \PS_R\}$.
Furthermore, we recognize that the sample genetic correlation matrix $\Cormath_{ss}^{g*}$ is a sufficient statistic for the population correlation matrix $\Cormat_{ss}^{g*}$, meaning the maximum likelihood estimate $\hat{\params}_R$, given a latent dimension $\tilde{m}$, can be found by minimizing the following discrepancy function \citep{Jöreskog1967}:
\begin{equation}\label{eq:discrepancy}
	F\left[\Cormat_{ss}^{g*}(\params_R);\Cormath_{ss}^{g*};\tilde{m}\right]=\ln\big|\Cormat_{ss}^{g*}(\params_R)\big| + \tr\left[\Cormath_{ss}^{g*}\Cormat_{ss}^{g*}(\params_R)^{-1}\right] - \ln\big|\Cormath_{ss}^{g*}\big| - p^* \text{,}
\end{equation}
where we constrain $\LA_R^\top\PS_R^{-1}\LA_R$ to be a diagonal matrix with ordered diagonal elements for identification.
We replace $\Cormath_{ss}^{g*}$ by the regularized estimate $\Cormath_{ss}^{g*}(\vartheta_g^\ddagger)$ as minimizing (\ref{eq:discrepancy}) requires a positive-definite and ideally well-conditioned sample genetic correlation matrix \citep{Peetersetal2019}.
We employ a Varimax rotation once we have obtained the estimates for $\LA_R$ and $\PS_R$ to facilitate interpretability \citep{Kaiser1958}.
In essence, the varimax rotation attempts to bring the loadings matrix close to having simple structure, i.e., a structure in which variables load highly on a single factor and have near zero loadings on the other factors.
To avoid notational clutter we simply use $\hat{\LA}_R$ to refer to either the rotated or non-rotated estimated loadings.
The described factor analytic model is implemented using an efficient matrix-free estimation method \citep{Daietal2020}.

\subsubsection{Estimation of factor scores (step VI)}
\sloppypar{We would now like to obtain estimates for the factor scores $\boldsymbol{\xi}_{j}$.
	These factor scores are essentially estimates of the locations of genotypes in latent space, based on the many noisy but observable secondary feature measurements.
	They can thus be viewed as the \emph{unobserved} phenotypes of a small number of latent traits that generate the \emph{observed} phenotypes for the many secondary features.
	We will use the joint distribution of the secondary feature BLUEs and latent factors (\ref{eq:joint_distro}) in a slightly modified Thomson regression approach \citep{Thomson1939} to obtain $\hat{\boldsymbol{\xi}}_j$.
	Recalling that we estimated $\params_R$ on the correlation scale, we first convert $\hat{\LA}_R$ and $\hat{\PS}_R$ back to the covariance scale:
	\begin{equation*}
		\hat{\LA}=\mathbf{D}\hat{\LA}_R \text{ and }
		\hat{\PS}=\mathbf{D}\hat{\PS}_R\mathbf{D} \text{, where }
		\mathbf{D}=(\Covmath_{ss}^{g*} \circ \I_{p^*})^{1/2} \text{.}
	\end{equation*}
	We also scale back the regularized residual correlation matrix obtained using (\ref{eq:SS}) and (\ref{eq:identity_reg}):
	\begin{equation}\label{eq:Ve_scaling}
		\Covmath_{ss}^{\epsilon}(\vartheta_\epsilon^\ddagger)=(\Covmath_{ss}^{\epsilon*}\circ\I_{p^*})^{1/2}\Cormath_{ss}^{\epsilon*}(\vartheta_\epsilon^\ddagger)(\Covmath_{ss}^{\epsilon*}\circ\I_{p^*})^{1/2} \text{.}
	\end{equation}
	We can now obtain the matrix $\mathbf{P}$ mapping the redundancy filtered secondary features to their projections on the lower-dimensional latent space through $\mathbb{E}(\boldsymbol{\xi}_j|\bar{\y}_{s(j)}^*)$.
	To obtain this expectation, we first use standard covariance algebra to obtain the joint distribution of the secondary feature BLUEs and latent factors as
	\begin{equation}\label{eq:joint_distro}
		\begin{bmatrix}
			\bar{\y}_{s(j)}^{*\top} \\
			\boldsymbol{\xi}_j
		\end{bmatrix} \sim \mathcal{N}_{p+m}\left[\mathbf{0}, \Covmat(\params)_\xi^{y}\right] \text{,}
	\end{equation}
	where
	\begin{equation*}
		\Covmat(\params)_\xi^{y}=
		\begin{bmatrix}
			\Covmat(\params)_{yy} 	& \Covmat(\params)_{y\xi} \\
			\Covmat(\params)_{\xi y} & \Covmat(\params)_{\xi\xi}
		\end{bmatrix}=
		\begin{bmatrix}
			\LA\LAt + \PS + r^{-1}\Covmat_{ss}^{\epsilon} 	& \LA \\
			\LAt											& \I_{m}
		\end{bmatrix}=
		\begin{bmatrix}
			\Covmat_{ss}^{g} + r^{-1}\Covmat_{ss}^{\epsilon} 	& \LA \\
			\LAt												& \I_{m}
		\end{bmatrix} \text{.}
	\end{equation*}
	Given that the joint distribution (\ref{eq:joint_distro}) is normal, we can obtain the conditional expectation as $\Covmat(\params)_{\xi y}\left[\Covmat(\params)_{yy}\right]^{-1}\bar{\y}_{s(j)}^*$ \citep{Anderson2003}.
	Using the Woodbury matrix identity \citep{Woodbury1950}, we rewrite this last expression to produce:
	\begin{equation*}
		\begin{split}
			\tilde{\boldsymbol{\xi}}_j&=\Big[\I_{\tilde{m}}+\hat{\LA}^\top(\hat{\PS} + r^{-1}\Covmath_{ss}^{\epsilon*})^{-1}\hat{\LA}\Big]^{-1}\hat{\LA}^\top(\hat{\PS} + r^{-1}\Covmath_{ss}^{\epsilon*})^{-1}\bar{\y}_{s(j)}^{*\top} \\
			&=\mathbf{P}^\top\bar{\y}_{s(j)}^{*\top} \text{.}
		\end{split}
	\end{equation*}
	Replacing $\Covmath_{ss}^{\epsilon*}$ by the regularized estimate obtained using (\ref{eq:SS}), (\ref{eq:identity_reg}), and (\ref{eq:Ve_scaling}) gives us the final expression for the estimation of the factor scores for all genotypes:
	\begin{equation}\label{eq:factorscores1}
		\begin{split}
			\tilde{\boldsymbol{\Xi}}&=\bar{\Y}_s^*\left[\hat{\PS} + r^{-1}\Covmath_{ss}^{\epsilon*}(\vartheta_\epsilon^\ddagger)\right]^{-1}\hat{\LA}\left\{\I_{\tilde{m}}+\hat{\LA}^\top\left[\hat{\PS} + r^{-1}\Covmath_{ss}^{\epsilon*}(\vartheta_\epsilon^\ddagger)\right]^{-1}\hat{\LA}\right\}^{-1} \\
			&=\bar{\Y}_s^*\mathbf{P} \text{.}
		\end{split}
	\end{equation}
	Through this projection we thus obtain factor scores that are essentially posterior estimates of latent phenotypes given the observed data.
	Instead of a full Bayesian approach we use empirical Bayes by using the maximum likelihood estimates of the parameters associated with the factor model in (\ref{eq:factorscores1}).
	Furthermore, note that $\text{var}(\bar{\y}_{s(j)}^*) \rightarrow \Covmat_{ss}^{g*}$ as $r \rightarrow \infty$, and as a result the projection $\mathbf{P}$ approaches the matrix that would be obtained using the ordinary Thomson regression approach.
	We thus account for the extra uncertainty due to a limited number of genotypic replicates.}

In (\ref{eq:factorscores1}) we apply the projection to the BLUEs (that is, the genotypic means).
In practice we project the plot-level data, however, to obtain plot level factor scores $\tilde{\boldsymbol{\Xi}} \in \mathbb{R}^{n \times \tilde{m}}$.
We later use these plot-level factor scores and sums of squares to obtain estimates for the genetic and residual covariances of the latent factors and focal trait.

One of the characteristics of Thomson regression that we maintain in our modified approach is the orthogonality of the factor scores.
We have thus reduced the possibly high-dimensional and collinear secondary data to a much smaller set of orthogonal factors, without losing interpretability.
These orthogonal factors can easily and effectively be used in any multivariate genomic prediction approach.
We use the next section to describe the multivariate genomic prediction step using the factor scores, target trait, and marker-based kinship matrix, thus finally integrating genomic and secondary feature data.

\subsection{Modeling of latent factors and focal trait}\label{s:foc}
Having reduced the dimensionality of the secondary data we now proceed with the multivariate genomic prediction.
The remainder of this section contains a description of the model and assumptions, estimation of covariance matrices, selection of relevant factors, and calculations resulting in the final focal trait predictions.
\subsubsection{Model and assumptions}\label{sec:focmodel_assumptions}
Consider the data matrix $\Y_{\tilde{\Xi}f} \in \mathbb{R}^{n \times (\tilde{m} + 1)}$ containing the latent factor scores and focal trait.
The matrix $\Y_{\tilde{\Xi}f}$ can be partitioned:
\begin{equation}\label{eq:datamatrix}
	\Y_{\tilde{\Xi}f}=\begin{bmatrix}
		\Y_{\tilde{\Xi}} & \y_{f}
	\end{bmatrix} \text{,}
\end{equation}
where the matrix $\Y_{\tilde{\Xi}}$ is the $n \times \tilde{m}$ matrix containing the plot-level factor scores and $\y_f$ is the $n \times 1$ column vector containing the focal trait.
Like for the secondary features, these factor scores and focal trait are composed of genetic and residual components:
\begin{equation*}
	\Y_{\tilde{\Xi}f} = \G_{\tilde{\Xi}f} + \E_{\tilde{\Xi}f} \text{.}
\end{equation*}
We again assume that the genetic components are independent of the residual components so that
\begin{equation*}
	\vect({\Y_{\tilde{\Xi}f}}) = \vect({\G_{\tilde{\Xi}f}}) + \vect({\E_{\tilde{\Xi}f}}) \sim \mathcal{N}_{np}(\mathbf{0}, \Covmat^g \otimes \Z\K\Z^\top + \Covmat^\epsilon \otimes \I_{n}) \text{,}
\end{equation*}
where $\Z\K\Z^\top$ and $\I_n$ are the row covariances of $\G_{\tilde{\Xi}f}$ and $\E_{\tilde{\Xi}f}$, respectively.
The $n \times n_g$ incidence matrix $\Z$ links individuals to genotypes.
Similarly, $\Covmat^g$ and $\Covmat^\epsilon$ are the column covariances of $\G_{\tilde{\Xi}f}$ and $\E_{\tilde{\Xi}f}$.
These covariance matrices can be block-partitioned:
\begin{equation*}\label{eq:covmat_partitioning}
	\begin{split}
		\Covmat^g &=
		\begin{pmatrix}
			\Covmat_{\tilde{\Xi}\tilde{\Xi}}^g & \covmat_{\tilde{\Xi}f}^g \\
			\covmat_{f\tilde{\Xi}}^g & \sigma_{ff}^g \\
		\end{pmatrix} = 
		\begin{pmatrix}
			\Covmat_{\tilde{\Xi}.}^g \\
			\covmat_{f.}^g
		\end{pmatrix} \\
		\Covmat^\epsilon &=
		\begin{pmatrix}
			\Covmat_{\tilde{\Xi}\tilde{\Xi}}^\epsilon & \covmat_{\tilde{\Xi}f}^\epsilon \\
			\covmat_{f\tilde{\Xi}}^\epsilon & \sigma_{ff}^\epsilon \\
		\end{pmatrix} =
		\begin{pmatrix}
			\Covmat_{\tilde{\Xi}.}^\epsilon \\
			\covmat_{f.}^\epsilon
		\end{pmatrix} \text{,}
	\end{split}
\end{equation*}
where $\Covmat_{\tilde{\Xi}\tilde{\Xi}}^g$ and $\Covmat_{\tilde{\Xi}\tilde{\Xi}}^\epsilon$ contain the genetic and residual covariances among the latent factors present in the sample, $\sigma_{ff}^g$ and $\sigma_{ff}^\epsilon$ are the scalar focal trait genetic and residual variances, and $\covmat_{\tilde{\Xi}f}^g=(\covmat_{f\textbf{s}}^g)^\top$ and $\covmat_{\textbf{s}f}^\epsilon = (\covmat_{f\textbf{s}}^\epsilon)^\top$ contain the genetic and residual covariances between latent factors and the focal trait.
We estimate the above covariance matrices using the plot-level data as described in Section \ref{sec:covmat_estimation}.
\subsubsection{Factor selection and multivariate genomic prediction}
We now convert the plot-level data to BLUEs $\bar{\Y}_{\tilde{\Xi} f}$ by taking the genotypic means.
Note that we have not used the focal trait data $\y_f$ in the modeling of the secondary features.
As a result, not all of the $\tilde{m}$ factors may be relevant to the multivariate genomic prediction problem.
As a basic form of subset selection we perform an exhaustive search for the best subset of $\tilde{m}^\star \leq \tilde{m}$ factors for predicting the focal trait BLUEs $\bar{\y}_{f}$ using the factor score BLUEs $\bar{\Y}_{\tilde{\Xi}}$ in multiple linear regression, using adjusted $R^2$ as the accuracy measure \citep{Lumley2020}.
Following this search, we subset the BLUEs to obtain $\bar{\Y}_{\tilde{\Xi}^\star f} \in \mathbb{R}^{n \times (\tilde{m}^\star + 1)}$ and subset the estimated covariance matrices of factor scores and focal trait accordingly:
\begin{equation*}
	\begin{split}
		\hat{\Covmat}^{g\star} &=\begin{pmatrix}
			\hat{\Covmat}_{\tilde{\Xi}^\star \tilde{\Xi}^\star}^g & \hat{\covmat}_{\tilde{\Xi}^\star f}^g \\
			\hat{\covmat}_{f \tilde{\Xi}^\star}^g & \hat{\sigma}_{ff}^g \\
		\end{pmatrix} \\
		\hat{\Covmat}^{\epsilon\star} &=\begin{pmatrix}
			\hat{\Covmat}_{\tilde{\Xi}^\star \tilde{\Xi}^\star}^\epsilon & \hat{\covmat}_{\tilde{\Xi}^\star f}^\epsilon \\
			\hat{\covmat}_{f \tilde{\Xi}^\star}^\epsilon & \hat{\sigma}_{ff}^\epsilon \\
		\end{pmatrix} \text{.}
	\end{split}
\end{equation*}
We finally divide $\hat{\Covmat}^{\epsilon\star}$ by the number of replicates and combine secondary and genomic data using the standard multivariate BLUP equations (see SM Section 1 and, e.g., \citep{Runcie&cheng_2019, Arouisseetal_2021}).

The integration of secondary and genomic data thus follows from the use of the marker-based kinship matrix and the inclusion of the factor scores as additional traits in the multivariate genomic prediction.
If secondary data are only available for the training set (CV1), the extra features refine estimates of genetic (i.e., marker) effects for training genotypes.
These improved training set estimates are then translated to test set estimates.
In CV2, where secondary data are available for the test set too, the extra features provide direct information on the genetic effects for the test set, improving estimates of the focal trait directly.
To evaluate the BLUP equations described in SM Section 1 we use a fast algorithm that decreases the complexity of these calculations from $O(n_g^3(\tilde{m}+1)^3)$ to $O(n_g(\tilde{m}+1)^2+(\tilde{m}+1)^3)$ \citep{Dahletal2013}.

On a final note, consider that as the dimensionality reduction described in Section \ref{s:sec} is entirely unsupervised, glfBLUP is flexible with regards to what secondary feature data are used.
If secondary features are only available for the training set (CV1), estimation of the covariance matrices, redundancy filtering, regularization, estimation of $\params$, etc.\ is based only on training data and factor scores are only obtained for the training set.
In CV2, where secondary features are also available for the test set, these data can be included effortlessly as well using the multivariate BLUP equations found in SM Section 1.

\section{Simulations}
\label{s:sim}
We compared glfBLUP to a number of alternative approaches in terms of performance in different high-dimensional data settings.
We compared glfBLUP to siBLUP, lsBLUP, MegaLMM, deep learning (multiMLP: multivariate multilayer perceptron), and the univariate gBLUP model, see SM Section 4 for details on these methods.
The following two subsections provide details on the simulation process as well as the results.
\subsection{Setup}\label{sec:simsetup}
We simulated a range of different high-dimensional datasets to evaluate model performance under varying circumstances (Figure \ref{fig:SEMS1}).
We used $n_g=500$ randomly selected genotypes (1500 SNP markers) from the RegMap \emph{Arabidopsis thaliana} panel \citep{Hortonetal2012} to construct a kinship matrix and simulated $r=2$ replicates for a total number of $n=n_gr=1000$ individuals.
Consider the model for the observed secondary features and focal trait:
\begin{equation}
	\Y=\Z\left[\left(\boldsymbol{\Xi}\LA^\top + \boldsymbol{\Gamma}\right)\mathbf{S}^g\right] + \E\mathbf{S}^\epsilon \text{,}
\end{equation}
where $\Y \in \mathbb{R}^{n \times (p+1)}$ is the data matrix, $\boldsymbol{\Xi}$ is the $n_g \times m$ matrix of factor scores, $\LA$ is the $(p+1) \times m$ matrix of loadings, $\boldsymbol{\Gamma} \in \mathbb{R}^{n_g \times (p+1)}$ is the matrix containing factor model errors, and $\E$ is the $n \times (p+1)$ matrix containing correlation scale residuals.
The $n \times n_g$ design matrix $\Z$ links records to genotypes.
The matrices $\mathbf{S}^g$ and $\mathbf{S}^\epsilon$ are diagonal matrices containing the genetic and residual standard deviations of the secondary features and focal trait.
The correlation scale genetic components of the phenotypes are thus defined as $\G = \boldsymbol{\Xi}\LA^\top + \boldsymbol{\Gamma}$.

To simulate the data we first randomly sampled $m=8$ factor scores for each genotype from a multivariate normal distribution taking relationships between genotypes as defined in the kinship into account:
\begin{equation}
	\vect(\boldsymbol{\Xi}) \sim \mathcal{N}_{n_gm}\left(\mathbf{0}, \PH \otimes \K \right) \text{,}
\end{equation}
where $\PH = \I_m$, thus assuming that factors are independent.
Next we sampled the factor model errors:
\begin{equation}
	\vect(\boldsymbol{\Gamma}) \sim \mathcal{N}_{n_gp}\left(\mathbf{0}, \PS \otimes \K \right) \text{,}
\end{equation}
where $\PS \equiv diag(\psi_{1},\dots,\psi_{p+1})$ is a diagonal matrix with unique variances.
The first $p=800$ unique variances corresponding to the secondary features are sampled from a uniform distribution: $\psi_{q} \sim \mathcal{U}\left[1-(0.3)^2,\,1-(0.8)^2\right]$, ensuring that the loadings for these features are between $0.3$ and $0.8$ given that features load on a single factor as described next.
The last unique variance corresponding to the focal trait is one of the parameters we varied so that $\psi_{p+1} \in \{0.8, 0.5, 0.2\}$.

We constructed the loadings matrix $\LA \in \mathbb{R}^{(p+1) \times m}$ so that eight blocks of $100$ secondary features ($s_1,\dots,s_p$) load on $m=8$ genetic latent factors with a single feature loading on a single factor.
Part of the focal trait ($y$) phenotype is generated by $m_s=4$ latent signal factors ($LSF$) through loadings.
The focal trait does not load on the remaining $m_n=4$ factors (latent noise factors, $LNF$), meaning that the $400$ secondary features generated by the $LNFs$ are not genetically correlated to the focal trait.
The loading for each of the secondary features $s_q$ was obtained as $\sqrt{1 - \psi_q}$ and given a random sign ensuring that the genetic component of each secondary feature is simulated on the correlation scale.
The loadings of the focal trait on the $m_s=4$ latent signal factors were obtained as $\sqrt{(1 - \psi_{p+1})/m_s}$.
The focal trait thus loads equally on all four latent signal factors.

The genetic components and corresponding correlation matrix can now be obtained as $\G = \boldsymbol{\Xi}\LA^\top + \boldsymbol{\Gamma}$ and $\Cormat^g = \LA\PH\LA^\top + \PS$.
The non-genetic or residual components are sampled from a multivariate normal distribution:
\begin{equation}\label{sim_res_cors1}
	\vect(\E) \sim \mathcal{N}_{np}\left(\mathbf{0}, \Cormat^\epsilon \otimes \I_n \right) \text{,}
\end{equation}
where $\Cormat^\epsilon$ is a correlation matrix containing random residual correlations (mean of $0$ and standard deviation of approximately $0.05$) among the $800$ observed features and between those features and the focal trait.
We then finally obtain the phenotypic data:
\begin{equation}
	\Y = \Z\left(\G\mathbf{S}^g\right) + \E\mathbf{S}^\epsilon \text{,}
\end{equation}
where the genetic and residual standard deviations in $\mathbf{S}^g$ and $\mathbf{S}^\epsilon$ are obtained through the heritabilities of the secondary features and focal trait.
Using the secondary feature heritabilities $h^2(s) \in \{0.5, 0.7, 0.9\}$ we obtain the genetic and residual standard deviations as $\sigma_g(s) = \sqrt{h^2(s)}$ and $\sigma_\epsilon(s) = \sqrt{1 - h^2(s)}$, given that we restrict $\sigma^2_g(s) + \sigma^2_\epsilon(s)$ to be $1$.
We use the focal trait heritabilities $h^2(y) \in \{0.1, 0.3, 0.5, 0.7, 0.9\}$ to similarly obtain the standard deviations for the focal trait.
These heritabilities together with the focal trait unique variance are the three parameters we vary.
Figure \ref{fig:SEMS1} provides a graphical overview of the simulation setting.

In addition to using random residual correlations as given in (\ref{sim_res_cors1}), we also simulated datasets using a low-rank structure for the residual correlations:
\begin{equation}\label{sim_res_cors2}
	\vect(\E) \sim \mathcal{N}_{np}\left(\mathbf{0}, \LA_\epsilon\LA_\epsilon^\top + \PS_{\epsilon} \otimes \I_n \right) \text{,}
\end{equation}
where $\LA_\epsilon \in \mathbb{R}^{(p+1)\times 1}$ contains residual loadings of the secondary features and focal trait on a single factor $\xi_\epsilon$ (Figure \ref{fig:SEMS2}).
For each simulated dataset, a random subset of $80$ out of $800$ secondary features load on $\xi_\epsilon$.
Residual loadings and residual unique variances for these features are simulated in the same way as the loadings and unique variances underlying the genetic correlations.
The unique residual variance for the focal trait was equal to the unique genetic variance, i.e., $0.8$, $0.5$, or $0.2$.
The $720$ secondary features that have no residual loadings on $\xi_\epsilon$ had unique residual variances of 1.
A graphical overview of this simulation setting can be found in Figure \ref{fig:SEMS2}.

\begin{figure}[h]
	\centering
	\begin{tikzpicture}
		\def\Sx{0};
		\def\LFx{5.8};
		\def\Yx{10};
		
		\node[rectangle, draw=black, very thick, minimum width=70, minimum height=15] (S1) at (\Sx, 0) {$s_1,\dots , s_{100}$};
		\node[rectangle, draw=black, very thick, minimum width=70, minimum height=15] (S2) at (\Sx, -0.8) {$s_{101},\dots , s_{200}$};
		\node[rectangle, draw=black, very thick, minimum width=70, minimum height=15] (S3) at (\Sx, -1.6) {$s_{201},\dots , s_{300}$};
		\node[rectangle, draw=black, very thick, minimum width=70, minimum height=15] (S4) at (\Sx, -2.4) {$s_{301},\dots , s_{400}$};
		
		\node[rectangle, draw=black, very thick, minimum width=70, minimum height=15] (S5) at (\Sx, -4.0) {$s_{401},\dots , s_{500}$};
		\node[rectangle, draw=black, very thick, minimum width=70, minimum height=15] (S6) at (\Sx, -4.8) {$s_{501},\dots , s_{600}$};
		\node[rectangle, draw=black, very thick, minimum width=70, minimum height=15] (S7) at (\Sx, -5.6) {$s_{601},\dots , s_{700}$};
		\node[rectangle, draw=black, very thick, minimum width=70, minimum height=15] (S8) at (\Sx, -6.4) {$s_{701},\dots , s_{800}$};
		
		\node[node, draw=black, minimum size=35] (LsF1) at (\LFx-0.6, 0.0) {$LSF_1$};
		\node[node, draw=black, minimum size=35] (LsF2) at (\LFx+0.6,-0.8) {$LSF_2$};
		\node[node, draw=black, minimum size=35] (LsF3) at (\LFx-0.6,-1.6) {$LSF_3$};
		\node[node, draw=black, minimum size=35] (LsF4) at (\LFx+0.6,-2.4) {$LSF_4$};
		
		\node[node, draw=black, minimum size=35] (LnF1) at (\LFx-0.6,-4.0) {$LNF_1$};
		\node[node, draw=black, minimum size=35] (LnF2) at (\LFx+0.6,-4.8) {$LNF_2$};
		\node[node, draw=black, minimum size=35] (LnF3) at (\LFx-0.6,-5.6) {$LNF_3$};
		\node[node, draw=black, minimum size=35] (LnF4) at (\LFx+0.6,-6.4) {$LNF_4$};
		
		\node[square, draw=black, very thick, minimum size=26] (Y) at (\Yx,-1.2) {$y$};
		
		\draw[>= latex, ->, thick] (LsF1) to [out=180,in=0] (S1);
		\draw[>= latex, ->, thick] (LsF2) to [out=180,in=0] (S2);
		\draw[>= latex, ->, thick] (LsF3) to [out=180,in=0] (S3);
		\draw[>= latex, ->, thick] (LsF4) to [out=180,in=0] (S4);
		
		\draw[>= latex, ->, thick] (LnF1) to [out=180,in=0] (S5);
		\draw[>= latex, ->, thick] (LnF2) to [out=180,in=0] (S6);
		\draw[>= latex, ->, thick] (LnF3) to [out=180,in=0] (S7);
		\draw[>= latex, ->, thick] (LnF4) to [out=180,in=0] (S8);
		
		\draw[>= latex, ->, thick] (LsF1) to [out=0,in=160] (Y);
		\draw[>= latex, ->, thick] (LsF2) to [out=0,in=172] (Y);
		\draw[>= latex, ->, thick] (LsF3) to [out=0,in=-172] (Y);
		\draw[>= latex, ->, thick] (LsF4) to [out=0,in=-160] (Y);
		
		\node[above=0.5, align=center] at (\Sx, -0.2) {$h^2(s)\in\{0.5,\,0.7,\,0.9\}$\\$\sigma_g^2(s)+\sigma_\epsilon^2(s)=1$};
		\node[above=0.5, align=center] at (\Yx, -1.1) {$h^2(y)\in\{0.1,\,0.3,\,0.5,\,0.7,\,0.9\}$\\$\sigma_g^2(y)+\sigma_\epsilon^2(y)=1$};
		\node[align=center] at (\Yx, -2.5) {$\psi_{p+1}\in\{0.8,\,0.5,\,0.2\}$};
		\node[align=center] at (\Sx+0.8, -3.2) {$\psi_{q}\sim \pm \mathcal{U}\left[1-(0.3)^2,\,1-(0.8)^2\right]$};
	\end{tikzpicture}
	\caption{The basic model underlying the simulated datasets.
		Groups of observed secondary features are shown as rectangles while genetic latent factors are shown as circles.
		Secondary feature unique variances $\psi_q$ were sampled from a uniform distribution $\psi_{q} \sim \mathcal{U}\left[1-(0.3)^2,\,1-(0.8)^2\right]$.
		The focal trait unique variance $\psi_{p+1}$ was $0.8$, $0.5$, or $0.2$.
		Note that the latent noise factors ($LNF$) and thus $S_{401},\dots,S_{800}$ are not genetically correlated to the focal trait.
		Random residual correlations among secondary features and between the secondary features and focal trait are not shown.}
	\label{fig:SEMS1}
\end{figure}

We generated $100$ datasets for each of the $90$ combinations obtained by varying the secondary feature heritabilities $h^2(s) \in \{0.5,\,0.7,\,0.9\}$, focal trait heritabilities $h^2(y) \in \{0.1,\,0.3,\,0.5,\,0.7,\,0.9\}$, focal trait unique variances $\psi_{p+1} \in \{0.8,\,0.5,\,0.2\}$, and residual correlation structure (random residual correlations or low-rank structure).
These combinations represent a realistic spectrum of data settings ranging from settings where the quality of secondary features is high while focal trait measurements are noisy, to settings where the opposite is true.
For each of the $9{,}000$ datasets, $300$ genotypes were randomly selected as the training set.
Prediction accuracy was calculated as the Pearson correlation between predictions and the simulated focal trait genetic component for the test set.
We did not redundancy filter the simulated data for any of the methods.
In addition to the models, we obtained benchmark predictions by using the true simulated latent signal factors as secondary features in the BLUP equations.
We only evaluated MegaLMM and the neural networks for the first $20$ datasets of each combination due to time and computational constraints.
\begin{figure}[h]
	\centering
	\includegraphics[width=\textwidth]{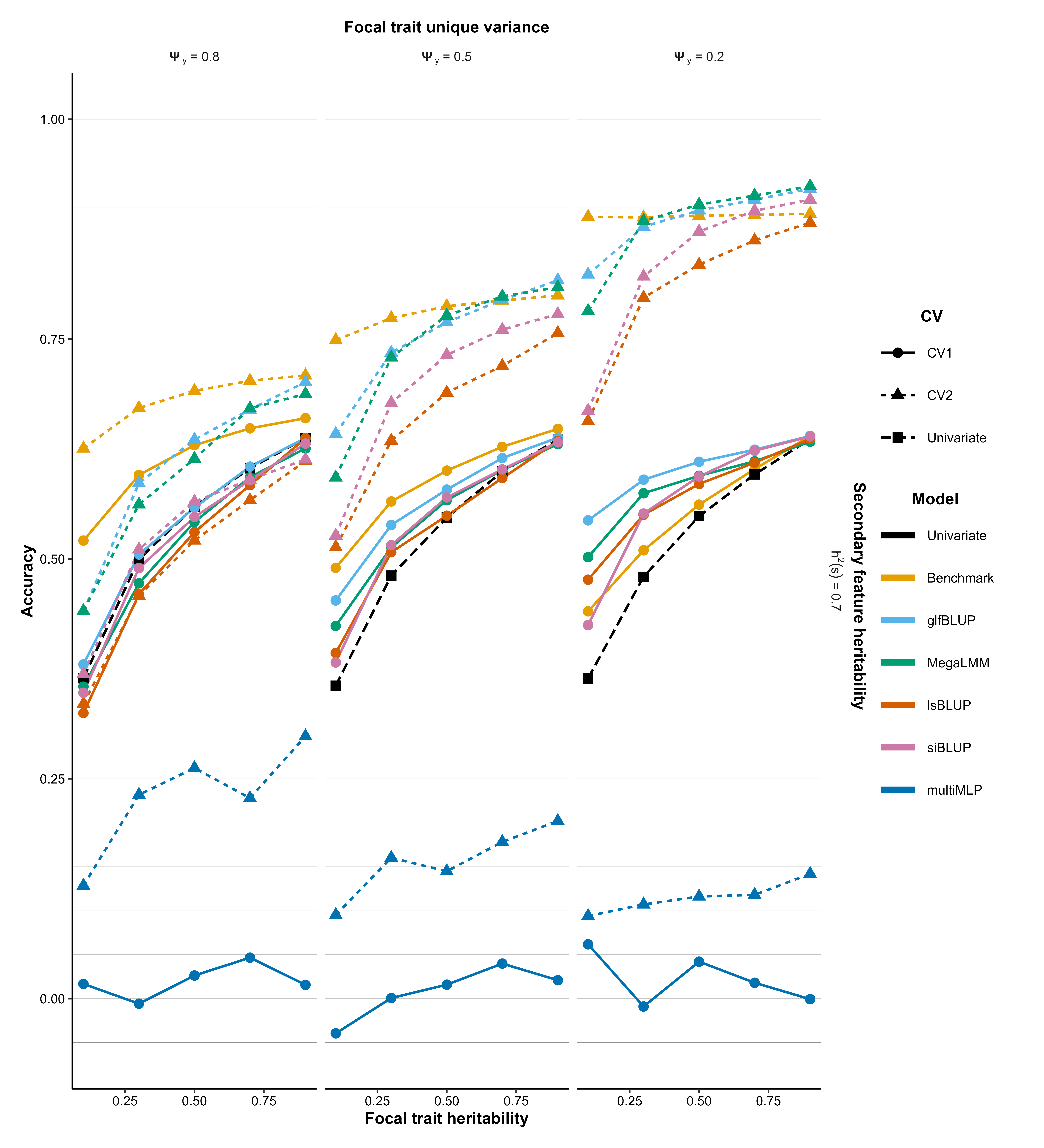}
	\caption{Prediction accuracies for several methods in the CV1 and CV2 scenarios as we vary the focal trait communality and heritability at a secondary feature heritability of $0.7$ (random residual correlations). The benchmark predictions use the true simulated factor scores and covariance matrices for multivariate genomic prediction. Further details regarding these simulations can be found in Section \ref{sec:simsetup}. Results for more secondary feature heritabilities can be found in the Appendix}
	\label{fig:SGP_results}
\end{figure}

\subsection{Results}
The impact of genetic parameters like the focal trait heritability and unique variance is clear (Figure \ref{fig:SGP_results}).
With the exception of the high focal trait unique variance setting, the CV1 accuracy of most multivariate approaches is, at worst, on par with that of the univariate model across the range of focal trait heritabilities.
Furthermore, CV1 accuracies of all methods, except the neural networks, converge to more or less the same point ($\rho\approx0.64$) as $h^2(y)$ increases, highlighting that secondary features bring the biggest benefit when the focal trait is hard to measure accurately and thus has a low heritability.

Most methods significantly outperform the univariate model in CV2, provided the quality of the secondary features is high enough.
Looking at the setting with the highest quality secondary features, i.e., $\boldsymbol{\psi}_y=0.2$, improvements with respect to the univariate approach reach approximately $40\%$ to $125\%$ depending on the focal trait heritability and model.
Overall, glfBLUP performs best, closely followed by MegaLMM.
The difference between MegaLMM and glfBLUP is mainly visible at the lowest focal trait heritability.
This might be due to the supervised nature of MegaLMM, where noisy focal trait data is detrimental when included in the modelling.
The lsBLUP and siBLUP approaches show similar performance with siBLUP being slightly better.
While the CV2 accuracy for the neural networks is higher than its CV1 accuracy, both are significantly lower than any of the other models, including the univariate approach.
We provide results for two more secondary feature heritabilities in the Appendix.

We also provide results for varying focal trait heritabilities, focal trait unique variances, and secondary feature heritabilities combined with a low-rank structure for the residual correlations in the Appendix (Figure \ref{fig:SGP_extra_results_V2}).
Simulating data using a low-rank structure for the residual correlations instead of random noise impacts relative model performance in several ways.
MegaLMM performance increases slightly, while the impact on glfBLUP's performance is negligible.
This results in an average accuracy for MegaLMM that is now slightly higher than for glfBLUP.
Accuracies for lsBLUP decrease significantly while siBLUP performance increases, especially at lower focal trait unique variances.
Deep learning performance is unaffected, but remains low.

These results are interesting for several reasons.
We have introduced additional complexity by simulating data using a low-rank structure for the residual covariances while keeping the structure underlying the genetics the same.
This low rank structure for the residual covariances may appear to be confounded with the low-rank structure for the genetics from a modeling perspective.
Intuitively, this additional complexity would result in lower prediction accuracies with models that fail to separate genetic and non-genetic sources of variation seeing the largest reduction in performance.
While we observed a drop in accuracy for lsBLUP, both siBLUP and MegaLMM show an increase in prediction accuracy.
The siBLUP and MegaLMM models both apply dimension reduction to the phenotypic data, rather than the genetic components.
MegaLMM decomposes the phenotypic data on all traits into factors and residuals that are then modeled independently, while siBLUP uses phenotypic covariances between secondary features and the focal trait to construct the selection index.
MegaLMM and siBLUP thus seem to use the low-rank residual structure that is present in the phenotypic data as signal, possibly explaining the higher prediction accuracies.
The glfBLUP approach is practically unaffected, showing prediction accuracies that are very similar regardless of whether a low-rank or random residual structure is simulated.
This indicates that glfBLUP appropriately separates genetic and non-genetic sources of variation before performing the dimensionality reduction as the change in residual structure does not impact the genomic predictions.

\section{Analysis of CIMMYT wheat data}
\label{s:cimmyt}
In addition to the extensive simulations provided in the previous section we also evaluated the different approaches in terms of accuracy and interpretability using wheat data from Centro Internacional de Mejoramiento de Maíz y Trigo (CIMMYT) \citep{Krauseetal_2019}.
The following two sections provide details on the dataset, pre-processing, and results.
\subsection{Data description and pre-processing}
The CIMMYT wheat dataset contains measurements on grain yield and crop canopy hyperspectral reflectivity at $62$ wavelengths between $398$nm and $847$nm.
The data were collected for four breeding cycles and five managed treatments within each cycle.
We limited ourselves to two subsets of the data to compare glfBLUP and the other methods.
The subsets we analyzed (breeding cycle 2014--2015, treatment B5IR and HEAT) contain $39$ trials, $1094$ unique genotypes, and reflectance measurements at ten and four timepoints for a total of $620$ and $248$ timepoint-wavelength combinations (referred to as the secondary features), respectively.
The B5IR treatment represents optimal growing conditions with five flood irrigations during the growth cycle with sowing in Late November/early December.
The HEAT treatment consists of five furrow irrigations and a sowing date in late February, resulting in higher temperatures throughout the growth cycle.
Note that \say{trial} refers to a factor that is part of the experimental design and not a trial in the multi-environment trial (MET) context.
The timepoints span the vegetative, heading (flowering), and grain-filling growth phases.
Each trial had $28$ trial-specific genotypes and two check genotypes that occurred in every trial, all replicated three times.
Trials were structured according to a resolvable incomplete block design.
Marker data were available for $1033$ of the $1094$ genotypes.
One genotype had missing yield for the B5IR treatment, leaving us with $1032$ genotypes for that subset.
Yield was missing for $214$ of $1033$ genotypes for the HEAT treatment, leaving us with $819$ for further analyses.

We first resolved the experimental design for each secondary feature and yield using the following SpATS (Spatial Analysis of Field Trials with Splines) model implemented in LMMsolver \citep{Rodríguez-Álvarezetal_2018, Pérez-Valenciaetal_2022, Boer_2023}:
\begin{equation}\label{eq:hypermodel}
	y_{(i|j)klmt}= \mu + g_{jt} + t_{kt} + r_{lt} + c_{mt} + f_t\left(r,c\right) +\epsilon_{(i|j)klmt} \text{,}
\end{equation}
where $y_{(i|j)klmt}$ is yield or any of the secondary features for replicate $i$ given genotype $j$ in trial $k$ with row number $l$ and column number $m$, measured at timepoint $t$.
The overall mean is given by $\mu$ and $g_{jt}$ is the fixed genotypic effect.
The random trial effect $t_{kt}$ is assumed to be independently, identically, and normally distributed with mean zero and variance $\sigma_t^2$: $t_k\sim \mathcal{N}(0, \sigma_t^2)$.
Similarly, $r_{lt}\sim \mathcal{N}(0, \sigma_r^2)$, $c_{mt}\sim \mathcal{N}(0, \sigma_c^2)$, and $\epsilon_{(i|j)klmt}\sim \mathcal{N}(0, \sigma_\epsilon^2)$ for the random row and column effects, as well as the residual effect.
The two-dimensional smooth surface over row and column positions is defined by $f_t\left(r,c\right)$.
We then created the pseudo-CRD data for each timepoint by simply adding the estimated genetic effects and residuals from model (\ref{eq:hypermodel}):
\begin{equation*}
	y_{(q|j)t}=\mu+\hat{g}_{jt}+\hat{\epsilon}_{(i|j)klmt} \text{.}
\end{equation*}
Note that due to the elimination of the design factors we can replace the indices $i$, $k$, $l$, and $m$ by a single index $q$ that denotes replicate $q$ of genotype $j$.
For the secondary features we then fit three-level nested hierarchical growth models through the ten (B5IR) and four (HEAT) timepoints and took fitted values as our final corrected data \citep{Pérez-Valenciaetal_2022}.
We finally created $250$ datasets for each managed treatment by randomly sampling $12$ trials as the test set and the remaining trials as the training set.
We note that by creating training and test sets after pre-processing the data we are violating the independence between training and test set to some extent.
Fully separated pre-processing of the training and test data is infeasible, however, as we need the complete data to perform the spatial corrections.
Furthermore, full separation would require fitting an infeasibly large number of linear mixed models for pre-processing.

We scaled and centered yield as well as the hyperspectral features separately for the training and test set prior to analyzing the data.
We calculated the CV1 prediction accuracy as the correlation between predictions and the BLUEs ($\hat{g}_j$) obtained from model (\ref{eq:hypermodel}).
For the CV2 prediction accuracies we used the parametric correction described by \citet{Runcie&cheng_2019}.
In addition to including test set hyperspectral measurements from all timepoints (CV2), we also evaluated the impact of including only measurements from the vegetative phase (CV2VEG), again using the parametric correction.
We used a redundancy filtering threshold $\tau = 0.95$ for all multivariate approaches as this provides a good trade-off between excluding redundant features, reducing computational complexity, and maintaining enough features to identify latent factors.
Details on the methods other than glfBLUP can be found in SM Section 4.

\subsection{Results}\label{hyper:results}
\begin{figure}[h]
	\centering
	\includegraphics[width=\textwidth]{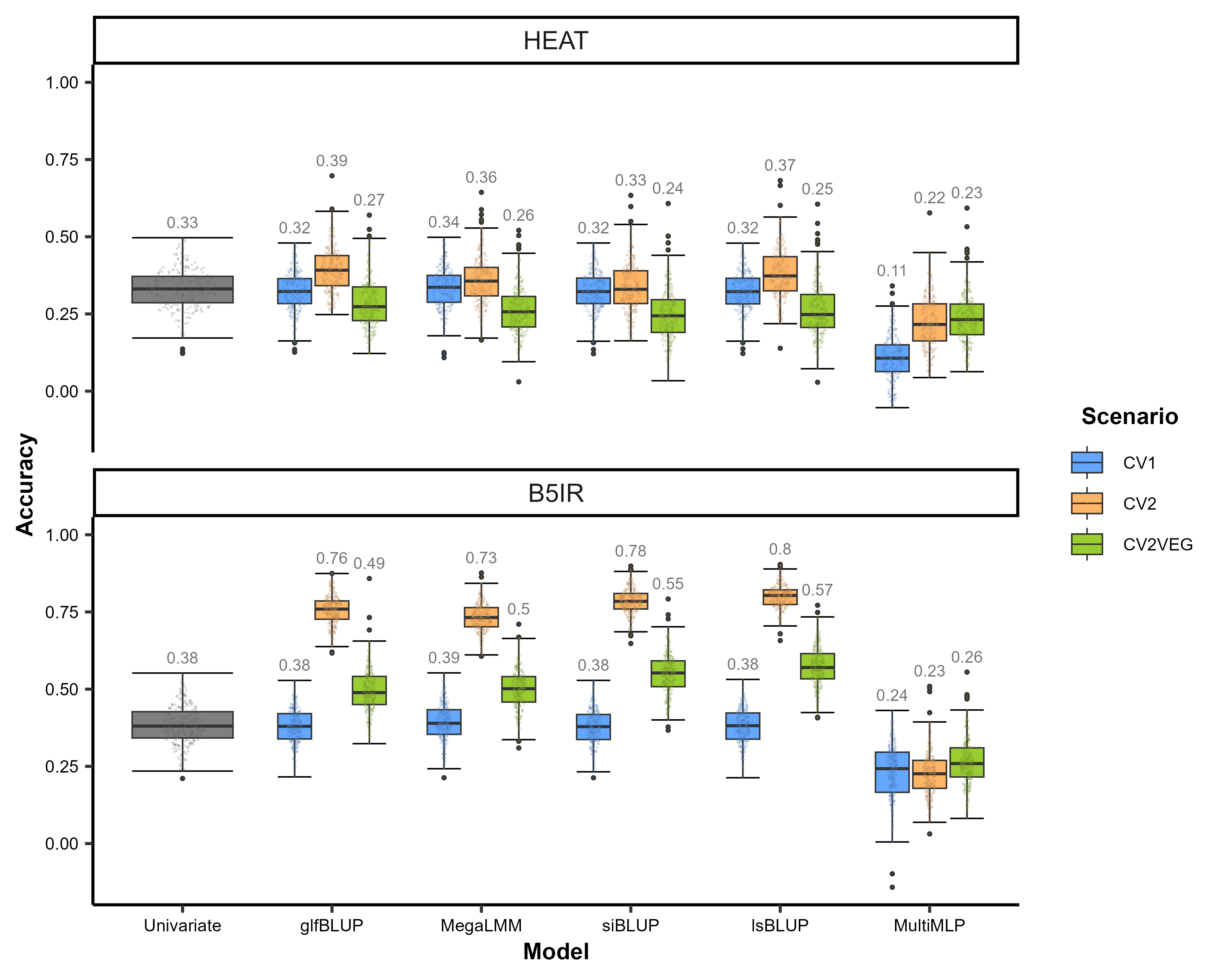}
	\caption{Prediction accuracies for several methods in the CV1, CV2, and CV2VEG scenarios for two managed treatments for the wheat yield and hyperspectral reflectivities dataset. Redundancy filtering was carried out for all multivariate analyses with a threshold $\tau=0.95$.}
	\label{fig:hyper_results}
\end{figure}

The relative performance of the different models clearly depends on the subset of the data and thus the managed treatment (Figure \ref{fig:hyper_results}).
Our proposed pipeline performs best in the HEAT treatment while lsBLUP performs best for B5IR.
All multivariate approaches show CV1 performances that are more or less similar to the performance of the univariate model, with the exception of the deep learner which performs significantly worse in all scenarios.
For the other approaches CV2 performance is clearly higher than the univariate model, especially in the B5IR treatment.
Only including test set hyperspectral measurements from the vegetative phase provided a benefit in B5IR but not in the HEAT treatment, possibly due to the fact that there was only a single measuring point during the vegetative phase for HEAT.

At this point we would like to point out that by using a two-stage modeling approach, i.e., performing the spatial correction first and then using the BLUEs from the first stage in the univariate model, we are likely slightly underestimating the univariate model's accuracy.
It would be trivial to fit a single-stage univariate model in a real world application where a complete trial would be the training set.
This single stage model would perform spatial correction and genomic prediction for test set genotypes not present in the trial simultaneously.
Unfortunately, in a cross-validation setting where training and test set are obtained from the same trial, this is infeasible.

The pattern of wavelengths that were retained after redundancy filtering was not constant across the ten dates for B5IR and four dates for HEAT, although most retained wavelengths represented the blue, red, and near infra-red part of the spectrum.
A graphical overview of what features were retained after filtering can be found in SM Section 7.
\begin{figure}[h]
	\centering
	\includegraphics[width=\textwidth]{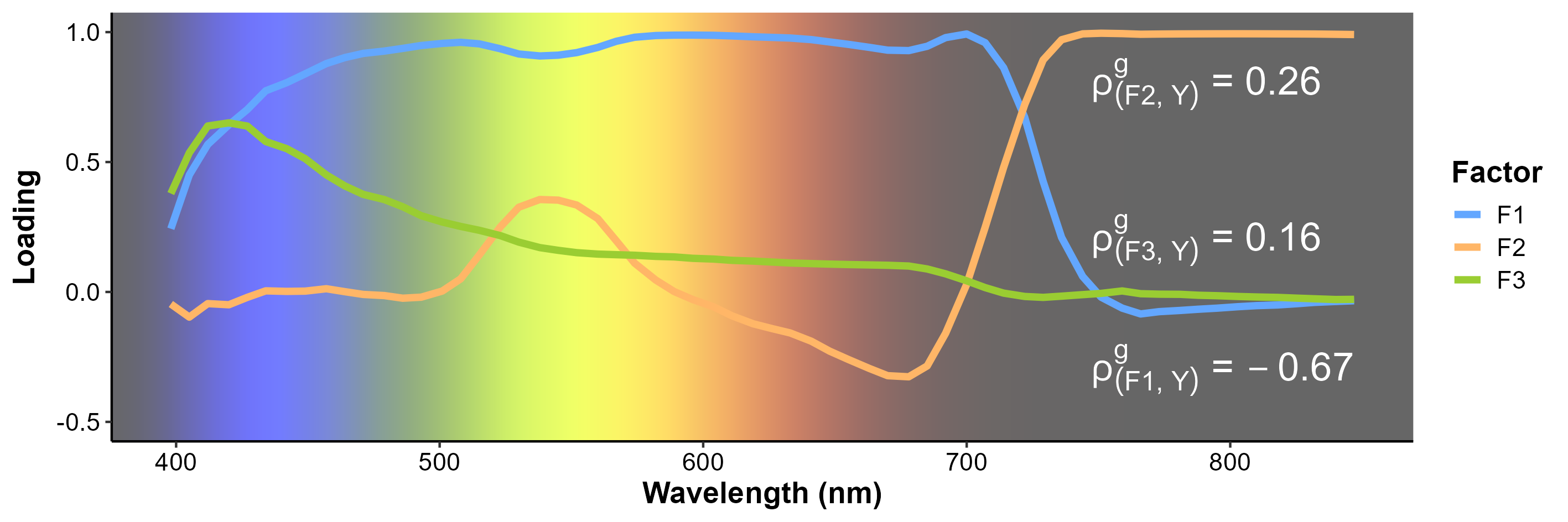}
	\caption{Loadings for the $62$ wavelengths and three factors found using glfBLUP and the hyperspectral data from 10-03-2015. Estimated genetic correlations between the three factors and yield are also shown.}
	\label{fig:hyper_single_date}
\end{figure}

We reanalyzed the data from one of the timepoints (10-03-2015) of the B5IR treatment to assess the interpretability of glfBLUP and the other approaches.
We omitted the redundancy filtering step, resulting in $62$ wavelengths.
Figure \ref{fig:hyper_single_date} shows the glfBLUP factor loadings obtained through this analysis.
All three factors were highly heritable at $H^2=0.98$ for factor $1$, $H^2=0.96$ for factor $2$, and $H^2=0.95$ for factor $3$, where $H^2$ denotes the wide-sense heritability as variances were estimated using sums of squares and are thus not additive only.
Genetic correlations between the factors and yield were $-0.67$ for factor $1$, $0.26$ for factor $2$, and $0.16$ for factor $3$.
Through these correlations and the factor loadings, it is immediately clear that higher yielding genotypes tended to have lower reflectivities in the green-red part of the spectrum or higher reflectivities in the near-infrared (NIR) part.
These results are somewhat reminiscent of the normalized difference vegetation index (NDVI) that is highly correlated with yield and is high itself for crop canopies with high NIR reflectivity and low red light reflectivity \citep{Tealetal2006}.
This showcases the ability of glfBLUP to reveal biologically relevant patterns in high-throughput phenotyping data.
We provide details on the interpretability of the other approaches in SM Section 5.
We did not include deep learning here as it was impossible to extract any biological meaning from the parameters of the neural networks.
We provide some data on the computational complexity of glfBLUP and other methods in SM Section 6.

\section{Discussion}
\label{s:discuss}
With the recent innovations in HTP techniques and the corresponding increase in data available to plant breeders, the development of methodology for integrating these data has become increasingly relevant.
We have developed a pipeline approach that combines classical factor analysis, which is typically applied to phenotypic data, with methods for high-dimensional settings and elements from mixed modeling approaches for genomic prediction.
We have also adapted several of these elements to facilitate the modeling of genetic and non-genetic sources of variation, showing the strength of combining existing methodologies and making critical changes that allow these methods to be applied to different types of data.
We have shown that the proposed methodology glfBLUP performs on par or better than alternatives and is able to integrate secondary HTP data efficiently without loss of interpretability.

We used simulations to evaluate the impact of several genetic parameters on model performance.
Of these parameters, the focal trait heritability is of particular interest.
One of the key motivations for the use of secondary data in multivariate genomic prediction is the belief that these data can improve predictions in cases where the focal trait phenotype has low heritability \citep{Lopez-Cruzetal_2020}.
The proposed glfBLUP performs well in such cases due to its unsupervised nature.
It does not suffer from overfitting at low focal trait heritabilities like some alternative approaches seem to do.

Our approach also does not rely as heavily on hyperparameter tuning as approaches such as deep learning.
The performance of the neural networks on both the simulated data as well as the real data was significantly lower than any of the other models.
This could very well be a result of the chosen architecture and hyperparameters not being optimal.
This result immediately showcases the difficulty with many machine learning approaches.
Proper hyperparameter tuning is either time consuming, computationaly expensive, or both.
We thus view the minimal required tuning in glfBLUP as a major advantage.

Another advantage of glfBLUP over machine learning approaches is the interpretability.
We evaluated interpretability by analyzing a wheat yield and hyperspectral reflectivity dataset.
We found that two of the three obtained factors were responsible for the (co)variation in reflectivities of visible and NIR light.
Combined with the genetic correlations between these factors and yield we found a pattern similar to the NDVI which is known to be a good indicator of yield \citep{Mkhabelaetal2011}.
As large-scale data collection becomes increasingly accessible, the complexity of datasets and the importance of understanding biological implications increases as well \citep{Danileviczetal2022}.
Given the results obtained here, incorporating generative factor analysis in genomic prediction seems an effective way of tackling these challenges.

Regarding factor analysis, we opted to use maximum likelihood estimation and a modified Thomson regression approach to obtain estimates of the loadings, unique variances, and factor scores \citep{Jöreskog1967, Thomson1939}.
Other approaches based on principal component analysis exist as well \citep[e.g.,][]{Zhangetal_2022, Fanetal_2021} and may be an alternative to obtain estimates in cases where the data contains heteroskedastic noise.

There are other implementations of the factor model too \citep{Smithetal_2001, Tolhurstetal_2019}.
These linear mixed models provide an attractive single-step approach to modelling multi-trait data by assuming a factor analytic structure for the genetic covariance matrix of traits and estimating any parameters using restricted maximum likelihood.
Such single-stage approaches have a number of advantages over using a two-stage approach and pseudo-CRD data.
Correcting data for nuisance effects such as spatial trends can be done most effectively by using estimates obtained from a full model including all effects \citep{Piepho_etal2013}.
Furthermore, a single-stage approach will result in valid standard errors for all predictions.
While the single-step nature of linear mixed models is thus attractive, their computational complexity means they are typically not used for data with hundreds or potentially thousands of traits \citep{Runcieetal_2021}.
We demonstrated that using estimates from reduced models to produce pseudo-CRD data for further modeling works well for prediction.
The approach we present here combines the attractive properties of factor analysis with a pipeline for prediction that avoids the computationally complex characteristics of the single-step mixed model approaches needed to obtain quantities such as standard errors.

We recognize several options to extend the proposed methodology.
Our motivation for using an unsupervised dimensionality reduction approach was based on the idea that secondary phenotypes are often collected when the focal trait has a low heritability \citep{Sunetal_2017}.
In these cases, the noisy focal trait data may negatively impact the estimation of latent factors.
While the unsupervised approach of glfBLUP worked well here, likely due to the noisy nature of the focal trait data, there are benefits to using supervised approaches like the other models we evaluated.
We could include the focal trait in the factor analysis we used to reduce the dimensionality, much like MegaLMM.
The factor analysis step could also be replaced by other supervised dimensionality reduction methods such as partial least squares regression to automatically obtain a small set of features that are maximally correlated with the focal trait \citep{Colombani_etal2012}.
Another way of achieving this would be to model secondary features as predictors instead of multivariate response variables by using a factor regression approach \citep{Münchetal_2022}.
Conditioning on the prediction target may lead to higher accuracy when the focal trait training data is of sufficient quality.
Another option would be to consider several ways of integrating multiple data modalities as opposed to using a single (e.g., hyperspectral) modality as we did here.
There are various methods that can generally be categorized into early integration, intermediate integration, or late integration \citep{Pavlidisetal2001}.
These integration approaches each try to balance parsimony and accuracy.
It may be fruitful to develop methodology that finds the right balance between the parsimony and accuracy of different integration approaches within the context of genomic prediction.
Finally, there is the issue of multi-year, multi-environment trials \citep{Cullisetal2014}.
By analyzing a trait of interest measured in several environments using a multivariate model, glfBLUP could reveal latent drivers of environment-specific variety performance.
Combining this approach with dynamic factor analysis \citep{Molenaar1985} would provide a promising framework to model the main source of data in large breeding programs.

In any case, the glfBLUP methodology we propose here is a robust and relatively efficient approach to obtaining multivariate genomic predictions with several promising options regarding extensions and further research.

\vspace*{1pc}

\noindent {\bf{Author Contributions}}

\noindent {\it{Conceptualization: C.F.W.P. and W.K.; methodology and software: K.A.C.M. and J.F.K.; writing--original draft preparation: K.A.C.M.; writing--review and editing: K.A.C.M., J.F.K., J.C., M.R.K., F.A.E., W.K., C.F.W.P. All authors have read and agreed to the published version of the manuscript.}}
\newline
\newline
\noindent {\bf{Conflict of Interest}}

\noindent {\it{The authors have declared no conflict of interest.}}
\newline
\newline
\noindent {\bf{Data Availability Statement}}

\noindent {\it{Code to generate the data and results presented in this paper are available at \url{https://github.com/KillianMelsen/glfBLUP\_2025}.
		The full hyperspectral dataset is available upon reasonable request from J. Crossa. The proposed approach glfBLUP has been implemented in an {\ttfamily R}-package available at \url{https://github.com/KillianMelsen/glfBLUP}.}}

\section*{Appendix: Additional simulation results}
\begin{figure}[h]
	\centering
	\includegraphics[width=\textwidth]{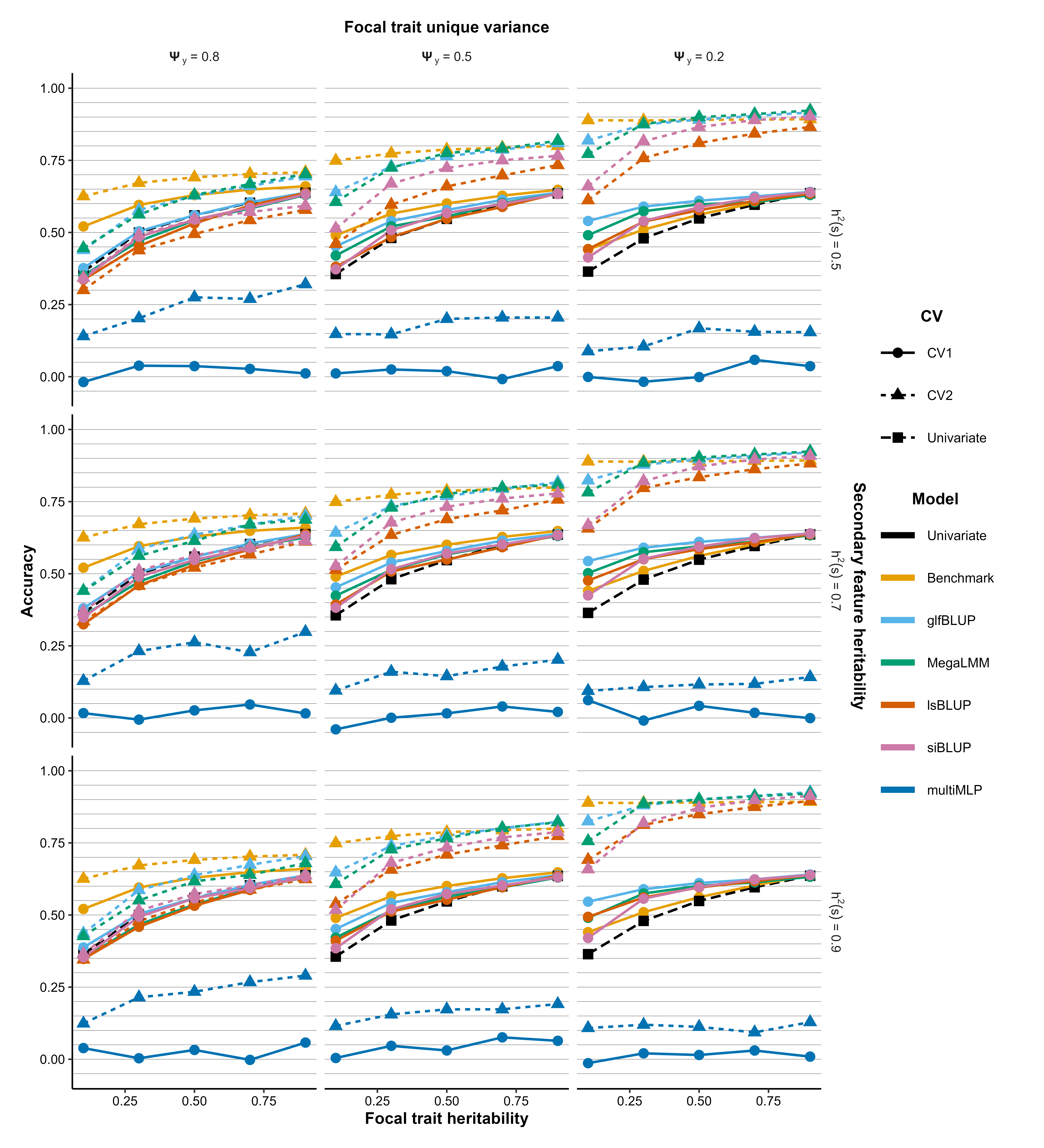}
	\caption{Prediction accuracies for several methods in the CV1 and CV2 scenarios as we vary the focal trait communality, focal trait heritability, and secondary feature heritabilities (random residual correlations). The benchmark predictions use the true simulated factor scores and covariance matrices for multivariate genomic prediction. Further details regarding these simulations can be found in Section \ref{sec:simsetup}.}
	\label{fig:SGP_extra_results}
\end{figure}

Here we provide some additional results using the simulated data (Figure \ref{fig:SGP_extra_results}).
Compared to the $h^2(s)=0.7$ setting the conclusions for $h^2(s)=0.5$ and $h^2(s)=0.9$ are largely similar.
Overall, there is a slight decrease in accuracy when the heritability of the secondary features drops, but the focal trait unique variance and thus the degree to which the focal trait is genetically correlated to the secondary features is far more important.

\begin{figure}[h]
	\centering
	\includegraphics[width=\textwidth]{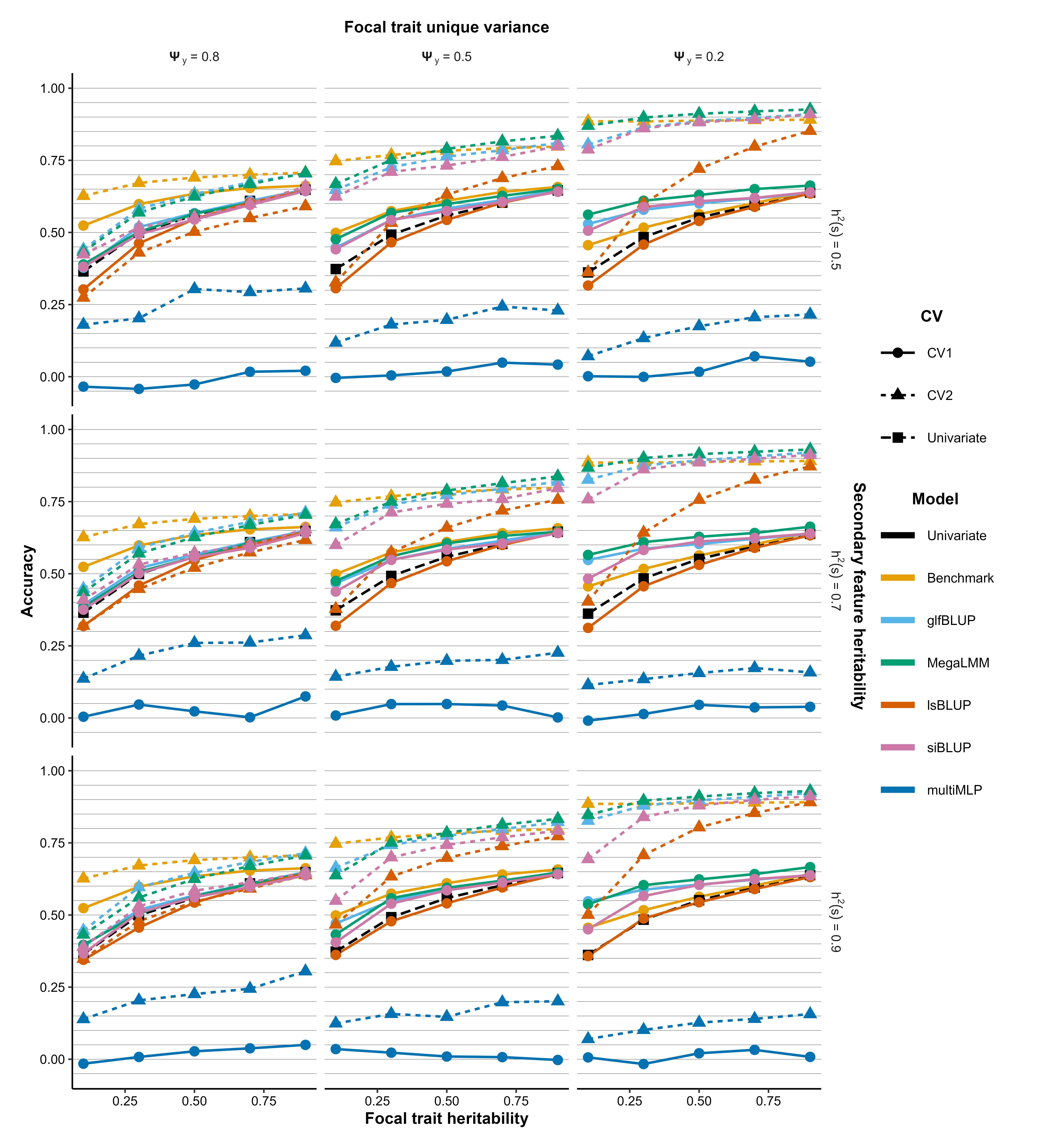}
	\caption{Prediction accuracies for several methods in the CV1 and CV2 scenarios as we vary the focal trait communality, focal trait heritability, and secondary feature heritabilities (low-rank model for the residual correlations). The benchmark predictions use the true simulated factor scores and covariance matrices for multivariate genomic prediction. Further details regarding these simulations can be found in Section \ref{sec:simsetup}.}
	\label{fig:SGP_extra_results_V2}
\end{figure}

Figure \ref{fig:SGP_extra_results_V2} shows results obtained when simulating data using a low-rank structure for the residual correlations instead of small, random correlations (see Figure \ref{fig:SEMS2} and Section \ref{sec:simsetup} for details on the simulation process).

\begin{figure}[h]
	\centering
	\begin{tikzpicture}
		\def\Sx{1.5};
		\def\LFx{5.8};
		\def\Yx{10};
		\def\offset{-1.2};
		
		\node[rectangle, draw=black, very thick, minimum width=70, minimum height=15] (S1) at (\Sx, 0) {$s_1,\dots , s_{100}$};
		\node[rectangle, draw=black, very thick, minimum width=70, minimum height=15] (S2) at (\Sx, -0.8) {$s_{101},\dots , s_{200}$};
		\node[rectangle, draw=black, very thick, minimum width=70, minimum height=15] (S3) at (\Sx, -1.6) {$s_{201},\dots , s_{300}$};
		\node[rectangle, draw=black, very thick, minimum width=70, minimum height=15] (S4) at (\Sx, -2.4) {$s_{301},\dots , s_{400}$};
		
		\node[rectangle, draw=black, very thick, minimum width=70, minimum height=15] (S5) at (\Sx, -4.0) {$s_{401},\dots , s_{500}$};
		\node[rectangle, draw=black, very thick, minimum width=70, minimum height=15] (S6) at (\Sx, -4.8) {$s_{501},\dots , s_{600}$};
		\node[rectangle, draw=black, very thick, minimum width=70, minimum height=15] (S7) at (\Sx, -5.6) {$s_{601},\dots , s_{700}$};
		\node[rectangle, draw=black, very thick, minimum width=70, minimum height=15] (S8) at (\Sx, -6.4) {$s_{701},\dots , s_{800}$};
		
		\node[node, draw=black, minimum size=35] (LsF1) at (\LFx-0.6, 0.0) {$LSF_1$};
		\node[node, draw=black, minimum size=35] (LsF2) at (\LFx+0.6,-0.8) {$LSF_2$};
		\node[node, draw=black, minimum size=35] (LsF3) at (\LFx-0.6,-1.6) {$LSF_3$};
		\node[node, draw=black, minimum size=35] (LsF4) at (\LFx+0.6,-2.4) {$LSF_4$};
		
		\node[node, draw=black, minimum size=35] (LnF1) at (\LFx-0.6,-4.0) {$LNF_1$};
		\node[node, draw=black, minimum size=35] (LnF2) at (\LFx+0.6,-4.8) {$LNF_2$};
		\node[node, draw=black, minimum size=35] (LnF3) at (\LFx-0.6,-5.6) {$LNF_3$};
		\node[node, draw=black, minimum size=35] (LnF4) at (\LFx+0.6,-6.4) {$LNF_4$};
		
		\node[node, draw=black, minimum size=35] (Fe) at (\LFx,-8.0) {$\xi_{\epsilon}$};
		
		\node[square, draw=black, very thick, minimum size=26] (Y) at (\Yx,-1.2) {$y$};
		
		\draw[>=latex,-,thick, dashed] (Fe) to (10,-8);
		\draw[>=latex,->,thick, dashed] (10,-8) .. controls (13,-8) and (13, -1.2) .. (Y);
		
		\draw[>=latex,-,thick, dashed] (Fe) to [out=180,in=0] (\Sx+\offset,-8);
		\draw[>=latex,->,thick, dashed] (\Sx+\offset,-8) .. controls (-0.5,-8) and (-0.5,-6.4) .. (S8);
		\draw[>=latex,->,thick, dashed] (\Sx+\offset,-8) .. controls (-0.7,-8) and (-0.7,-5.6) .. (S7);
		\draw[>=latex,->,thick, dashed] (\Sx+\offset,-8) .. controls (-0.9,-8) and (-0.9,-4.8) .. (S6);
		\draw[>=latex,->,thick, dashed] (\Sx+\offset,-8) .. controls (-1.1,-8) and (-1.1,-4.0) .. (S5);
		\draw[>=latex,->,thick, dashed] (\Sx+\offset,-8) .. controls (-1.3,-8) and (-1.3,-2.4) .. (S4);
		\draw[>=latex,->,thick, dashed] (\Sx+\offset,-8) .. controls (-1.5,-8) and (-1.5,-1.6) .. (S3);
		\draw[>=latex,->,thick, dashed] (\Sx+\offset,-8) .. controls (-1.7,-8) and (-1.7,-0.8) .. (S2);
		\draw[>=latex,->,thick, dashed] (\Sx+\offset,-8) .. controls (-1.9,-8) and (-1.9,0) .. (S1);
		
		\draw[>= latex, ->, thick] (LsF1) to [out=180,in=0] (S1);
		\draw[>= latex, ->, thick] (LsF2) to [out=180,in=0] (S2);
		\draw[>= latex, ->, thick] (LsF3) to [out=180,in=0] (S3);
		\draw[>= latex, ->, thick] (LsF4) to [out=180,in=0] (S4);
		
		\draw[>= latex, ->, thick] (LnF1) to [out=180,in=0] (S5);
		\draw[>= latex, ->, thick] (LnF2) to [out=180,in=0] (S6);
		\draw[>= latex, ->, thick] (LnF3) to [out=180,in=0] (S7);
		\draw[>= latex, ->, thick] (LnF4) to [out=180,in=0] (S8);
		
		\draw[>= latex, ->, thick] (LsF1) to [out=0,in=160] (Y);
		\draw[>= latex, ->, thick] (LsF2) to [out=0,in=172] (Y);
		\draw[>= latex, ->, thick] (LsF3) to [out=0,in=-172] (Y);
		\draw[>= latex, ->, thick] (LsF4) to [out=0,in=-160] (Y);
		
		\node[above=0.5, align=center] at (\Sx, -0.2) {$h^2(s)\in\{0.5,\,0.7,\,0.9\}$\\$\sigma_g^2(s)+\sigma_\epsilon^2(s)=1$};
		\node[above=0.5, align=center] at (\Yx, -1.1) {$h^2(y)\in\{0.1,\,0.3,\,0.5,\,0.7,\,0.9\}$\\$\sigma_g^2(y)+\sigma_\epsilon^2(y)=1$};
		\node[align=center] at (\Yx, -2.5) {$\psi_{p+1}\in\{0.8,\,0.5,\,0.2\}$};
		\node[align=center] at (\Sx+0.8, -3.2) {$\psi_{q}\sim \pm \mathcal{U}\left[1-(0.3)^2,\,1-(0.8)^2\right]$};
	\end{tikzpicture}
	\caption{The basic model underlying the simulated datasets using a low-rank residual correlation structure.
		Groups of observed secondary features are shown as rectangles while latent factors are shown as circles.
		Secondary feature unique variances $\psi_q$ were sampled from a uniform distribution $\psi_{q} \sim \mathcal{U}\left[1-(0.3)^2,\,1-(0.8)^2\right]$.
		The focal trait unique variance $\psi_{p+1}$ was $0.8$, $0.5$, or $0.2$.
		Note that the latent noise factors ($LNF$) and thus $S_{401},\dots,S_{800}$ are not genetically correlated to the focal trait.
		Solid lines denote loadings underlying genetic correlations while dashed lines denote loadings underlying residual correlations.
		While the single factor $\xi_{\epsilon}$ underlying the residual correlations is connected to all secondary feature blocks in the graph, only a random 10\% of the secondary features actually load on $\xi_\epsilon$.
		Loadings and unique variances underlying the residual correlation structure were sampled from the same distributions as those used for the genetic parameters.}
	\label{fig:SEMS2}
\end{figure}


\newpage
\begin{center}
	\textbf{\large Supplementary Material: Improving Genomic Prediction using High-dimensional Secondary Phenotypes: the Genetic Latent Factor Approach}
\end{center}
\setcounter{equation}{0}
\setcounter{figure}{0}
\setcounter{table}{0}
\setcounter{page}{1}
\setcounter{section}{0}
\makeatletter
\renewcommand{\theequation}{S\arabic{equation}}
\renewcommand{\thefigure}{S\arabic{figure}}
\renewcommand{\bibnumfmt}[1]{[S#1]}
\renewcommand{\citenumfont}[1]{S#1}
\renewcommand{\thetable}{S\arabic{table}}
\renewcommand{\thesection}{S.\arabic{section}}
\section{Genomic prediction}
\label{SM:GPS}
We more or less follow the notation of \citet{Runcie&cheng_2019_SM} and \citet{Arouisseetal_2021_SM}.
Let $\Y \in \mathbb{R}^{n \times (p+1)}$ be the data matrix containing unreplicated, mean-centered, and scaled measurements on $p$ secondary features and the focal trait for $n$ genotypes.
We can divide the $n$ genotypes into a test and training set so that $n=n_t+n_o$ ($t$ for test and $o$ for the \emph{observed} training set).
The matrix $\Y$ can then be partitioned as:
\[
\Y=\begin{bmatrix}
	\Y_{st} & \y_{ft} \\
	\Y_{so} & \y_{fo}
\end{bmatrix} =
\begin{bmatrix}
	\Y_{.t} \\
	\Y_{.o}
\end{bmatrix} \text{,}
\]
where $\Y_{st}$ and $\Y_{so}$ ($s$ for secondary) are the sub-matrices containing the HTP features for the test and training set, respectively.
Similarly, $\y_{ft}$ and $\y_{fo}$ ($f$ for focal) are the column vectors containing the focal trait for the test and training set.

The measurements in the data matrix $\Y$ are composed of genetic effects, residual effects, and possibly covariates so that $\Y=\X\boldsymbol{\beta} + \G + \E$.
The random matrices for the genetic and residual effects come from matrix normal distributions:
\begin{equation*}
	\begin{split}
		\G &\sim \mathcal{MN}_{n,p+1}(\mathbf{0},\K,\Covmat^g) \\
		\E &\sim \mathcal{MN}_{n,p+1}(\mathbf{0},\I_{n},\Covmat^\epsilon) \text{,}
	\end{split}
\end{equation*}
where the row-covariances are the kinship matrix $\K$ and the identity matrix $\I_{n}$ for the genetic and residual effects.
The column-covariances are the genetic and residual covariance matrices of the secondary features and focal trait.
Assuming that $\Y$ is only composed of genetic and residual effects, a standard result shows that the column-wise vectorization of $\Y$, denoted as $\vect(\Y)$, follows a multivariate normal distribution:
\begin{equation*}
	\vect(\Y) = \vect(\G) + \vect(\E) \sim \mathcal{N}_{n(p+1)}(\mathbf{0}, \Covmat^g \otimes \K + \Covmat^\epsilon \otimes \I_{n}) \text{,}
\end{equation*}
where $\otimes$ denotes the Kronecker product.
Similarly to the data matrix itself, we can partition the kinship and covariance matrices as well:
\[
\K=\begin{bmatrix}
	\K_{tt} & \K_{to} \\
	\K_{ot} & \K_{oo}
\end{bmatrix} \text{,}
\]
where $\K_{tt}$ and $\K_{oo}$ contain the genetic relatedness coefficients within the test and training sets, respectively.
The genetic relatedness coefficients between test and training genotypes are given by $\K_{to}=(\K_{ot})^\top$.
Furthermore,
\begin{equation*}
	\begin{split}
		\Covmat^g &=\begin{bmatrix}
			\Covmat_{ss}^g & \covmat_{sf}^g \\
			\covmat_{fs}^g & \sigma_{ff}^g \\
		\end{bmatrix} = 
		\begin{bmatrix}
			\covmat_{s.}^g \\
			\covmat_{f.}^g
		\end{bmatrix} \\
		\Covmat^\epsilon &=\begin{bmatrix}
			\Covmat_{ss}^\epsilon & \covmat_{sf}^\epsilon \\
			\covmat_{fs}^\epsilon & \sigma_{ff}^\epsilon \\
		\end{bmatrix} =
		\begin{bmatrix}
			\covmat_{s.}^\epsilon \\
			\covmat_{f.}^\epsilon
		\end{bmatrix} \text{,}
	\end{split}
\end{equation*}
where $\Covmat_{ss}^g$ and $\Covmat_{ss}^\epsilon$ contain the genetic and residual covariances among secondary features, $\sigma_{ff}^g$ and $\sigma_{ff}^\epsilon$ are the scalar focal trait genetic and residual variances, and $\covmat_{sf}^g=(\covmat_{fs}^g)^\top$ and $\covmat_{sf}^\epsilon = (\covmat_{fs}^\epsilon)^\top$ contain the genetic and residual covariances between secondary features and the focal trait.

As mentioned above, there are generally speaking three genomic prediction scenarios depending on whether secondary data is available for just the training set (CV1), both training and test set (CV2), or not at all (univariate genomic prediction).
We now review the expressions for the training ($\G_o$) and test set ($\G_t$) BLUPs in these three scenarios.

In the univariate scenario, we only use information on the focal trait of the training set ($\y_{fo}$) to obtain predictions for the test set:
\begin{equation}\label{univariate_preds}
	\begin{split}
		\Vh&=\hat{\sigma}_{ff}^g\otimes \K_{oo}+\hat{\sigma}_{ff}^\epsilon\otimes \I_{n_o} \\
		\gh^{(uni)}_{fo}&=\left(\hat{\sigma}_{ff}^g\otimes \K_{oo}\right)\Vh^{-1}\left(\y_{fo}-\X_{o}\hat{\boldsymbol{\beta}_f}\right) \\
		\gh^{(uni)}_{ft}&=\mathbb{E}\Bigl(\g_{ft}|\y_{fo}\Bigr) \\
		&=\left(\hat{\sigma}_{ff}^g\otimes \K_{to}\right)\Vh^{-1}\left(\y_{fo}-\X_{o}\hat{\boldsymbol{\beta}_f}\right) \\
		&=\K_{to}\K_{oo}^{-1}\gh^{(uni)}_{fo} \text{,}
	\end{split}
\end{equation}
where $\gh^{(uni)}_{fo}$ and $\gh^{(uni)}_{ft}$ are the focal trait univariate BLUPs for the training and test set.
Note that covariates can be included in the term $\X_{o}\hat{\boldsymbol{\beta}_f}$.
Also note that as $\hat{\sigma}_{ff}^g$ and $\hat{\sigma}_{ff}^\epsilon$ are scalars, $\Vh=\hat{\sigma}_{ff}^g \K_{oo}+\hat{\sigma}_{ff}^\epsilon \I_{n_o}$.

In the multivariate CV1 scenario where secondary features are only available for the training set, predictions can be obtained using:
\begin{equation}\label{CV1_preds}
	\begin{split}
		\Vh&=\Covmath^{g}\otimes \K_{oo}+\Covmath^{\epsilon}\otimes \I_{n_o} \\
		\gh^{(CV1)}_{fo}&=\left(\covmath^{g}_{f.}\otimes \K_{oo}\right)\Vh^{-1}\vect\left(\Y_{.o}-\X_o\hat{\boldsymbol{\beta}}\right) \\
		\gh^{(CV1)}_{ft}&=\mathbb{E}\Bigl(\g_{ft}|\Y_{.o}\Bigr) \\
		&=\left(\covmath^{g}_{f.}\otimes \K_{to}\right)\Vh^{-1}\vect\left(\Yh_{.o}-\X_o\hat{\boldsymbol{\beta}}\right) \\
		&=\K_{to}\K_{oo}^{-1}\gh^{(CV1)}_{fo} \text{,}
	\end{split}
\end{equation}
where $\Y_{.o}-\X_o\hat{\boldsymbol{\beta}}$ is vectorized.
Expression (\ref{CV1_preds}) can be evaluated using a fast algorithm that decreases the time complexity for the calculation of the focal trait training set BLUPs $\gh^{(CV1)}_{fo}$ from $O(n_o^3p^3)$ to $O(n_op^2+p^3)$ by avoiding the naive inversion of $\Vh$ \citep{Dahletal2013_SM}.

The CV2 multivariate predictions can be obtained directly \citep{Arouisseetal_2021_SM} or through a two-step approach \citep{Runcie&cheng_2019_SM}.
We prefer the two-step approach because the fast algorithm can be used to obtain all CV1 training set BLUPs, $\Gh^{(CV1)}_{.o}$:
\begin{equation}\label{CV2_preds_1}
	\begin{split}
		\Vh&=\Covmath^{g}\otimes \K_{oo}+\Covmath^{\epsilon}\otimes \I_{n_o} \\
		\Gh^{(CV1)}_{.o}&=\vect^{-1}\left[\left(\Covmath^{g}\otimes \K_{oo}\right)\Vh^{-1}\vect\left(\Y_{.o}-\X_o\hat{\boldsymbol{\beta}}\right)\right] \text{.}
	\end{split}
\end{equation}
We then convert these training set BLUPs to intermediate CV1 test set BLUPs and combine those with the observed secondary features for the test set to finally obtain CV2 BLUPs for the test set focal trait:
\begin{equation}\label{CV2_preds_2}
	\begin{split}
		\Vh&=\Covmath_{ss}^{g}\otimes \K_{tt}^{-1}+\Covmath_{ss}^{\epsilon}\otimes \I_{n_t} \\
		\gh^{(CV2)}_{ft}&=\mathbb{E}\Bigl(\g_{ft}|\Y_{st},\Y_{.o}\Bigr) \\
		&=\K_{to}\K_{oo}^{-1}\gh^{(CV1)}_{fo}+\left(\covmath^{g}_{fs}\otimes \K_{tt}^{-1}\right)\Vh^{-1}\vect\left(\Y_{st}-\K_{to}\K_{oo}^{-1}\Gh^{(CV1)}_{so}\right) \text{.}
	\end{split}
\end{equation}

In all cases, $\Covmat^g$ and $\Covmat^\epsilon$ must be estimated from the data.

\section{Covariance of BLUEs}
\label{SM:extra:BLUEs}
Consider the secondary feature BLUEs for genotype $j$ which are simply the genotypic means in the case of a completely randomized experimental design:
\begin{equation*}
	\bar{\y}_{s(j)}=\g_{s(j)}+\bar{\e}_{s(j)}=\g_{s(j)}+\dfrac{1}{r}\sum_{i=1}^{r}\e_{s(i|j)} \text{,}
\end{equation*}
where $\g_{s(j)} \sim \mathcal{N}_p(\mathbf{0}, \Covmat_{ss}^g)$ and $\e_{s(i|j)} \sim \mathcal{N}_p(\mathbf{0}, \Covmat_{ss}^\epsilon)$.
All $r$ residual vectors for the replicates of genotype $j$ are assumed \textit{iid}, meaning $\text{var}(\e_{s(i|j)}) = \text{var}(\e_{s(i'|j)}) = \Covmat_{ss}^\epsilon$ and the cross-covariance matrix for any two replicates of genotype $j$, $\text{cov}(\e_{s(i|j)}, \e_{s(i'|j)})$, is a zero matrix.
As a result,
\begin{equation*}
	\text{var}\left(\sum_{i=1}^{r}\e_{s(i|j)}\right) = \sum_{i=1}^{r} \text{var}\left(\e_{s(i|j)}\right) = r\Covmat_{ss}^\epsilon \text{.}
\end{equation*}
Now we divide the sum of the $r$ residual vectors for genotype $j$ by the number of replicates $r$ to obtain the vector of mean residuals for genotype $j$:
\begin{equation*}
	\bar{\e}_{s(j)} = \dfrac{1}{r} \sum_{i=1}^{r}\e_{s(i|j)} \text{.}
\end{equation*}
Using the fact that $\text{var}(a\mathbf{x}) = a^2 \text{var}(\mathbf{x})$ for some scalar $a$ and random vector $\mathbf{x}$, we find that
\begin{equation*}
	\text{var}\left(\bar{\e}_{s(j)}\right) = \left(\dfrac{1}{r}\right)^2 \text{var}\left(\sum_{i=1}^{r}\e_{s(i|j)}\right) = r^{-2}r\Covmat_{ss}^\epsilon = r^{-1}\Covmat_{ss}^\epsilon \text{.}
\end{equation*}
Combined with the assumptions that $\g_{s(j)} \ConIndepNatt \e_{s(i|j)}$, and thus that $\g_{s(j)} \ConIndepNatt \bar{\e}_{s(j)}$, we obtain the covariance of the BLUEs as
\begin{equation*}
	\text{var}\left(\bar{\y}_{s(j)}\right) = \text{var}\left(\g_{s(j)}\right) + \text{var}\left(\bar{\e}_{s(j)}\right) = \Covmat_{ss}^g + r^{-1}\Covmat_{ss}^\epsilon \text{.}
\end{equation*}

\section{Penalty optimization}
\label{SM:extra:pen_optim}
We use Brent's algorithm \citep{Brent1973_SM} implemented in the {\ttfamily optimize()} function in {\ttfamily R} to find the optimal values for $\vartheta_g$ and $\vartheta_\epsilon$ as described in the main text.
In a naive implementation we obtain the mean cross-validated negative log-likelihood at each iteration as:
\begin{equation}\label{SM:pen_opt:min}
	\varphi^K(\vartheta)=\dfrac{1}{K}\sum_{k=1}^{K}w_k \left\{\text{ln}\Big|\Bigl(\Cormat(\vartheta)\Bigr)_{\neg k}\Big|+\tr \left[\Cormat_k\Bigl(\Cormat(\vartheta)\Bigr)_{\neg k}^{-1}\right]\right\} \text{,}
\end{equation}
where we have omitted some sub- and superscripts for clarity.
Note that $(\Cormat(\vartheta))_{\neg k} \in \mathbb{R}^{p \times p}$ is the penalized genetic or residual correlation matrix given the corresponding penalty $\vartheta$ at the current iteration.
Naive minimization thus requires inverting $(\Cormat(\vartheta))_{\neg k} = \left(1 - \vartheta\right)\Cormat_{\neg k} + \vartheta \I$ at each iteration.
Note that
\begin{equation*}
	\Bigl[\Cormat(\vartheta)\Bigr]_{\neg k}^{-1} = \mathbf{U} \Bigl[\mathbf{\Lambda}(\vartheta)\Bigr]^{-1} \mathbf{U}^\top \text{,}
\end{equation*}
where $\mathbf{U}$ contains the eigenvectors of the unregularized correlation matrix $\Cormat_{\neg k}$ and
\begin{equation*}
	\mathbf{\Lambda}(\vartheta) = \left(1 - \vartheta\right)\mathbf{\Lambda} + \vartheta \I_p
\end{equation*}
with $\mathbf{\Lambda}$ a diagonal matrix containing the eigenvalues of $\Cormat_{\neg k}$.
We now rewrite (\ref{SM:pen_opt:min}) as:
\begin{equation}\label{SM:pen_opt:min_2}
	\varphi^K(\vartheta)=\dfrac{1}{K}\sum_{k=1}^{K}w_k \left\{\text{ln}\prod_{i=1}^{p}\mathbf{\Lambda}(\vartheta)_{ii}+\tr \left[\Cormat_k\mathbf{U} \Bigl(\mathbf{\Lambda}(\vartheta)\Bigr)^{-1} \mathbf{U}^\top\right]\right\} \text{.}
\end{equation}
We can then use the invariance of the trace operator to cyclic permutations to rewrite (\ref{SM:pen_opt:min_2}) as:
\begin{equation}\label{SM:pen_opt:min_3}
	\varphi^K(\vartheta)=\dfrac{1}{K}\sum_{k=1}^{K}w_k \left\{\text{ln}\prod_{i=1}^{p}\mathbf{\Lambda}(\vartheta)_{ii}+\tr \left[\Bigl(\mathbf{\Lambda}(\vartheta)\Bigr)^{-1} \mathbf{A}\right]\right\} \text{,}
\end{equation}
where $\mathbf{A} = \mathbf{U}^\top \Cormat_k \mathbf{U}$.
Within the trace operation in (\ref{SM:pen_opt:min_3}) we are now calculating the matrix product of the diagonal matrix $\left(\mathbf{\Lambda}(\vartheta)\right)^{-1}$ and the non-diagonal matrix $\mathbf{A}$, of which the latter is independent of $\vartheta$.
This can be rewritten as a summation of divisions:
\begin{equation}
	\varphi^K(\vartheta)=\dfrac{1}{K}\sum_{k=1}^{K}w_k \left[\text{ln}\prod_{i=1}^{p}\mathbf{\Lambda}(\vartheta)_{ii} + \sum_{j=1}^{p} \dfrac{\mathbf{A}_{jj}}{\mathbf{\Lambda}(\vartheta)_{jj}} \right] \text{.}
\end{equation}
We can calculate the eigenvalues $\mathbf{\Lambda}$ as well as the diagonal entries of $\mathbf{A} = \mathbf{U}^\top \Cormat_k \mathbf{U}$ before starting the optimization.
As a result, each iteration scales linearly with $p$ resulting in faster optimization, especially if the optimal penalty values are near the bounds.

\section{Alternative methods}
\label{SM:competing_methods}
\subsection{Univariate gBLUP}
The baseline model that we compare all multivariate approaches to is the standard univariate gBLUP model:
\begin{equation*}
	\y = \Z \mathbf{u} + \e \text{,}
\end{equation*}
where $\y$ is the $n \times 1$ vector containing the mean-centered and scaled focal trait phenotypes, $\Z$ is the incidence matrix linking individuals to the BLUPs in $\mathbf{u} \sim \mathcal{N}_{n_g}(\mathbf{0}, \sigma_g^2 \K)$, and $\e \sim \mathcal{N}_n(\mathbf{0}, \sigma_\epsilon^2 \I_n)$ contains the residuals.
We used the {\ttfamily R}-package {\ttfamily rrBLUP} version 4.6.1 \citep{Endelman2011_SM} to fit univariate gBLUP models (see \ref{univariate_preds}).
The marker-based kinship matrix was created using the {\ttfamily R}-package {\ttfamily statgenGWAS} version 1.0.9.

\subsection{siBLUP}
The first alternative multivariate approach that we included is a modified version of the bivariate gBLUP model using a regularized selection index as proposed by \citet{Lopez-Cruzetal_2020_SM} and also implemented by \citet{Arouisseetal_2021_SM}.
In short, siBLUP uses a regularized phenotypic covariance matrix of the secondary features and their genetic covariances with the focal trait to obtain a weight for every secondary feature.
We use these weights to calculate a single linear combination of the secondary features for each individual termed the (regularized) selection index.
We then use this index as the single secondary feature in the multivariate gBLUP equations (SM Section \ref{SM:GPS}).

Like glfBLUP, the modified version of siBLUP takes data containing replicates from a completely randomized design.
Next, we obtain $\Covmath_{ss}^g$ and $\Covmath_{ss}^\epsilon$ using sums of squares, after which we redundancy filter the dataset so that no two of the remaining secondary features have an absolute genetic correlation $|\rho^g|$ equal to, or exceeding a threshold $\tau$.

We then convert the redundancy filtered, plot-level \emph{phenotypic} covariance matrix $\Covmath_{ss}^{p*}=\Covmath_{ss}^{g*}+\Covmath_{ss}^{\epsilon*}$ of the remaining secondary features to a correlation matrix which is regularized:
\begin{equation*}\label{eq:identity_reg_phen}
	\Cormath_{ss}^{p*}(\vartheta_p)=\left(1-\vartheta_p\right)\Cormath_{ss}^{p*}+\vartheta_p \I_{p^*} \text{,}
\end{equation*}
where the optimal penalty $\vartheta^\ddagger_p$ is found using 5-fold cross validation as described in the main text using the number of individuals in each fold as the weights $w_k$.
We then reconvert the resulting regularized phenotypic correlation matrix to the covariance matrix $\Covmath_{ss}^{p*}(\vartheta^\ddagger_p)$ and use it to calculate the vector of selection index weights $\hat{\boldsymbol{\gamma}}^{SI}$.

To calculate the selection index weights, we first obtain the $p^*\times 1$ column vector $\covmath_{sf}^{g*}$ containing the genetic covariances between the remaining secondary features and the focal trait using sums of squares.
Next, we calculate the vector of selection index weights $\hat{\boldsymbol{\gamma}}^{SI}$:
\begin{equation*}
	\hat{\boldsymbol{\gamma}}^{SI}=\left[\Covmath_{ss}^{p*}(\vartheta^\ddagger_p)\right]^{-1}\covmath_{sf}^{g*} \text{,}
\end{equation*}
and use it to calculate the plot-level selection indices:
\begin{equation*}
	\hat{\y}_{SI}=\Y_s^*\hat{\boldsymbol{\gamma}}^{SI} \text{,}
\end{equation*}
where $\hat{\y}_{SI}$ is the $n \times 1$ column vector containing the selection index for each individual.
The $n \times p^*$ matrix $\Y_s^*$ contains the original (redundancy filtered) secondary features.

We can now replace the original secondary features in the data matrix $\Y_{sf}$ by the selection indices:
\[
\Y_{SI,f}=\begin{bmatrix}
	\hat{\y}_{SI} & \y_{f}
\end{bmatrix} \text{,}
\]
followed by using sums of squares to obtain $\Covmath^g$ and $\Covmath^\epsilon$ which give us the genetic and residual covariances between the selection index and the focal trait.
We then reduce the data to BLUEs $\bar{\Y}_{SI,y}$ by calculating the genoypic means, divide $\Covmath^\epsilon$ by the number of replicates, and combine the data, covariance matrices, and the kinship $\K$ in the BLUP equations (see SM Section \ref{SM:GPS}) to obtain test set predictions.

\subsection{lsBLUP}
lsBLUP also reduces the secondary features to a single dimension \citep{Arouisseetal_2021_SM}.
For lsBLUP, this secondary feature is a LASSO regression prediction of the focal trait using the redundancy filtered set of secondary features.

We use sums of squares obtained from replicated completely randomized design data to estimate $\Covmat_{ss}^g$.
We then use this genetic covariance matrix of the original secondary features (converted to the correlation scale) to redundancy filter the data.
After filtering, the LASSO regression coefficients $\hat{\boldsymbol{\beta}}^{LS}$ are obtained using the {\ttfamily R}-package {\ttfamily glmnet} version 4.1.6 \citep{Friedmanetal2010_SM}.
To do so, we first reduce the redundancy filtered data to genotypic means.
We then use the {\ttfamily cv.glmnet()} function to determine the optimal (i.e., minimum mean cross-validated error) LASSO penalty $\lambda^\ddagger$ using 5-fold cross-validation, thereby obtaining the vector of coefficients given this optimal penalty $\hat{\boldsymbol{\beta}}^{LS}(\lambda^\ddagger)$.

If all coefficients have been regularized to 0, lsBLUP returns simple univariate gBLUP predictions for the test set using the {\ttfamily rrBLUP} {\ttfamily R}-package \citep{Endelman2011_SM}.
If not, we calculate plot-level LASSO predictions:
\begin{equation*}\label{LSP}
	\hat{\y}_{LS}=\Y_s^*\hat{\boldsymbol{\beta}}^{LS}(\lambda^\ddagger) \text{,}
\end{equation*}
where $\Y_s^*$ is the matrix containing the redundancy filtered observed secondary features.
We then replace the original secondary features in the data matrix $\Y_{sf}$ by the LASSO predictions:
\[
\Y_{LS,y}=\begin{bmatrix}
	\hat{\y}_{LS} & \y_{f}
\end{bmatrix} \text{,}
\]
and compute the CV1 or CV2 test set focal trait BLUPs as described for siBLUP and glfBLUP, using $\hat{\y}_{LSP}$ as the single secondary feature.

\subsection{MegaLMM}
MegaLMM was introduced as a method to make genomic predictions using many secondary features \citep{Runcieetal_2021_SM}.
In essence, it decomposes the $p$ original features into two parts.
One part is projected onto $M$ latent factors while the other part represents the $p$ independent errors for each secondary feature.
Independent linear mixed models are then fit for these $M+p$ decorrelated traits.
For the full details on MegaLMM see \citet{Runcieetal_2021_SM}.

To be able to make consistent comparisons, we redundancy filtered the full set of secondary features as we did for other methods prior to calculating the BLUEs and running MegaLMM.
We initialized and ran MegaLMM with default settings.
The number of modeled latent factors $M$ (argument {\ttfamily K}) was set to $16$ for the simulated data and $20$ for the hyperspectral data.
For the priors we again used the default settings.

After initializing MegaLMM we first ran $1000$ burn-in iterations, reordering the factors after each $100$ iterations, for both simulated and hyperspectral data.
We then ran another 500 iterations for the posterior samples.
Posterior predictions for each genotype were calculated at each iteration, with the final predictions being equal to the posterior means.
We used version 0.1.0 of the {\ttfamily MegaLMM} {\ttfamily R}-package \citep{Runcie2022_SM}.

\subsection{Deep learning}
Deep learning provides a flexible, non-parametric alternative to the models described above.
Here, we limited ourselves to CV1 and CV2 versions of a simple multilayer perceptron (multiMLP).
The only difference between these versions is the inclusion of the scaled and centered, redundancy filtered set of secondary features in the output layer (CV1) or input layer (CV2).

We considered the number of hidden layers, number of neurons per hidden layer, and the dropout regularization rate as the hyperparameters to be tuned separately for each (simulated) dataset.
Other hyperparameters such as the learning rate, early stopping patience, optimizer, etc. were fixed.
After an initial exploration of possible hyperparameter values, we decided to use either $4$ or $8$ hidden layers, with $16$ or $64$ neurons each.
The dropout rate was set to either $0.1$ or $0.3$.
In addition to these values, we opted for the root mean squared propagation optimizer (RMSProp), a learning rate of $5 \times 10^{-4}$, the rectified linear unit (ReLU) activation function in all relevant layers, a batch size of $128$, and mean squared error (mse) for the loss function.

To obtain predictions for each dataset, we first redundancy filtered the secondary features like in the other multivariate approaches, followed by reducing the data to BLUEs.
We then divided the training individuals over five folds to determine the within-training set mean cross validated prediction accuracy for the eight different hyperparameter combinations.
We trained each of the eight networks five times (once for each fold) for at most $25$ epochs.
Training could be stopped early based on the loss for a $10\%$ validation split using a patience of $10$ epochs.
We then determined the combination of hyperparameter values giving the highest mean prediction accuracy to be used for training the final network on the full training set.
This was done for at most $100$ epochs using early stopping with a patience of $20$ epochs and a $10\%$ validation split.
We then obtained test set predictions by feeding the SNP marker matrix (and filtered secondary features for the CV2 multiMLP) for the test set into the trained networks.

We used the {\ttfamily keras} (v2.15.0), {\ttfamily tensorflow} (v2.16.0), and {\ttfamily reticulate} (v1.40.0) {\ttfamily R} packages as an interface to Python (v3.10.16) and TensorFlow (v2.15.1) to implement the deep learning \citep{Cholletetal2015_SM, Abadietal2016_SM}.

\section{Interpretability}
\label{SM:interpretability}
We have stressed the great ease of interpreting glfBLUP's model parameters in terms of biological relevance.
Here we present some additional results with regards to interpretability of the approaches that we compared glfBLUP to.
We ran si-, lsBLUP, and MegaLMM on the non-redundancy filtered B5IR hyperspectral data from 10-03-2015 as we did for glfBLUP.
We then looked at the relevant model parameters that included the selection index weights for siBLUP, LASSO coefficients for lsBLUP, and (mean) posterior loadings for MegaLMM.

\begin{figure}[h]
	\centering
	\includegraphics[width=\textwidth]{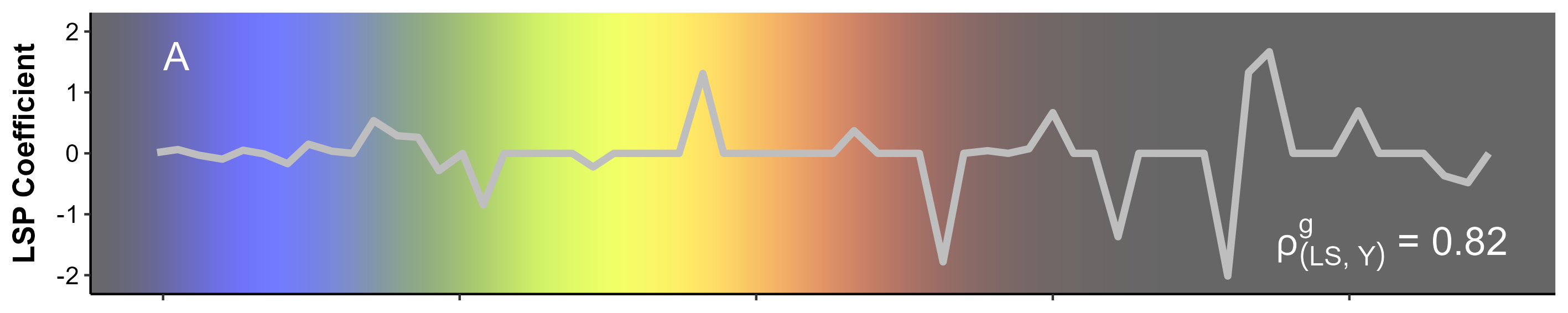}
	\includegraphics[width=\textwidth]{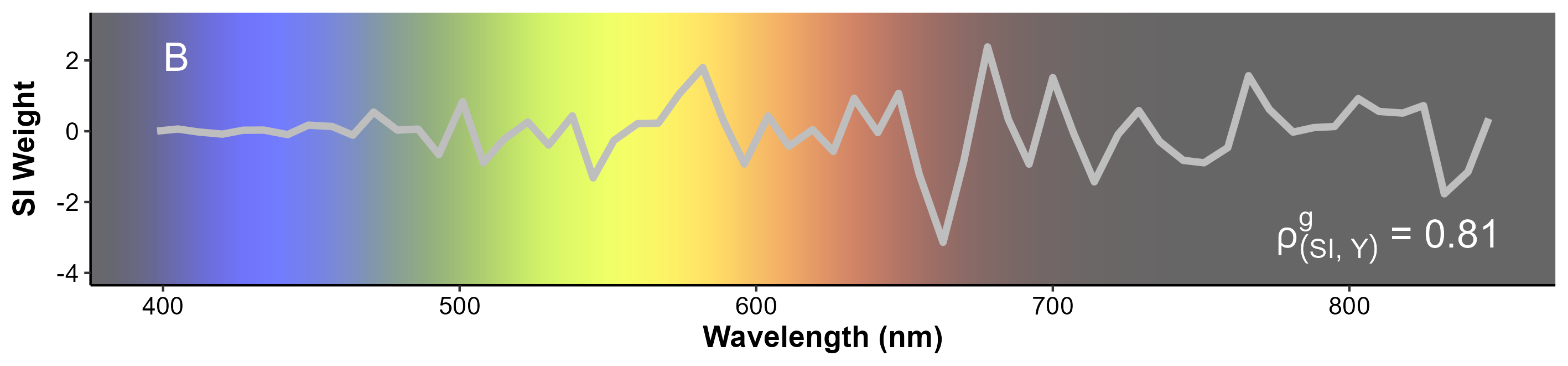}
	\caption{LASSO regression coefficients (A) and selection index weights (B) for the $62$ wavelengths obtained from running lsBLUP and siBLUP on the B5IR hyperspectral data from 10-03-2015. Estimated genetic correlations between the LASSO prediction (LS) or selection index (SI) and yield (Y) are also shown.}
	\label{fig:lssiBLUP_hyper_single_date}
\end{figure}

Starting with lsBLUP and siBLUP (Figure \ref{fig:lssiBLUP_hyper_single_date}) we immediately recognize the difficulties associated with compressing the collinear data to a single dimension.
In the case of lsBLUP we see sign-flipping in the coefficients around the $760$nm wavelength.
Reflectivities around $660$nm, $720$nm, and $760$nm show a negative relationship with yield following from the genetic correlation of $0.81$ between the LASSO prediction and yield.
After $760$nm the relationship rapidly becomes positive, before moving towards $0$ again.

The selection index weights of siBLUP are difficult to interpret as well.
The relationship between hyperspectral reflectivity and yield follows a highly irregular pattern along the spectrum.

While si- and lsBLUP require minimal hyperparameter tuning, their lack of flexibility with regards to the selection of an appropriate number of dimensions clearly limits their interpretability in the case of more complex data.

\begin{figure}[h]
	\centering
	\includegraphics[width=\textwidth]{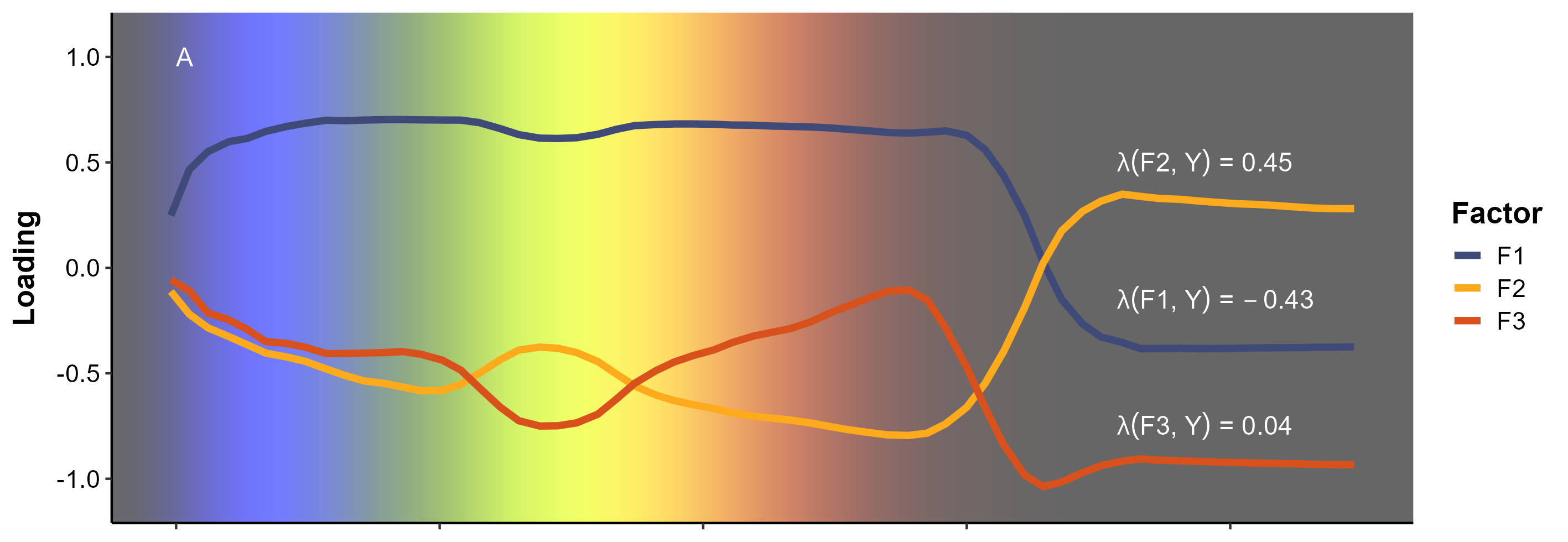}
	\includegraphics[width=\textwidth]{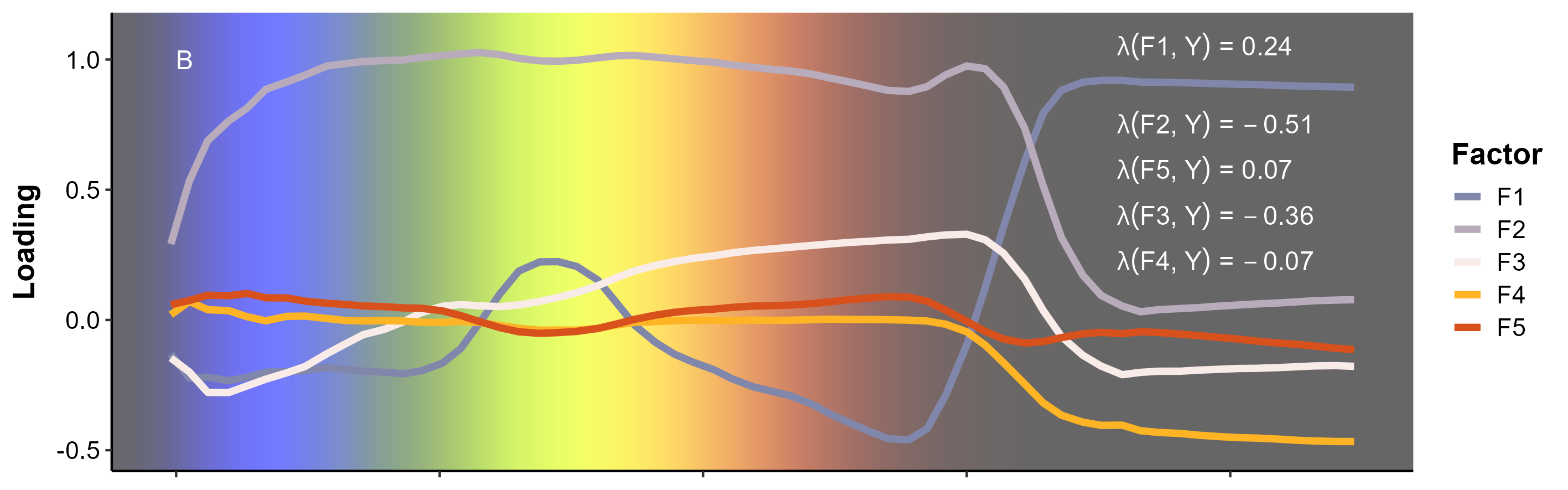}
	\includegraphics[width=\textwidth]{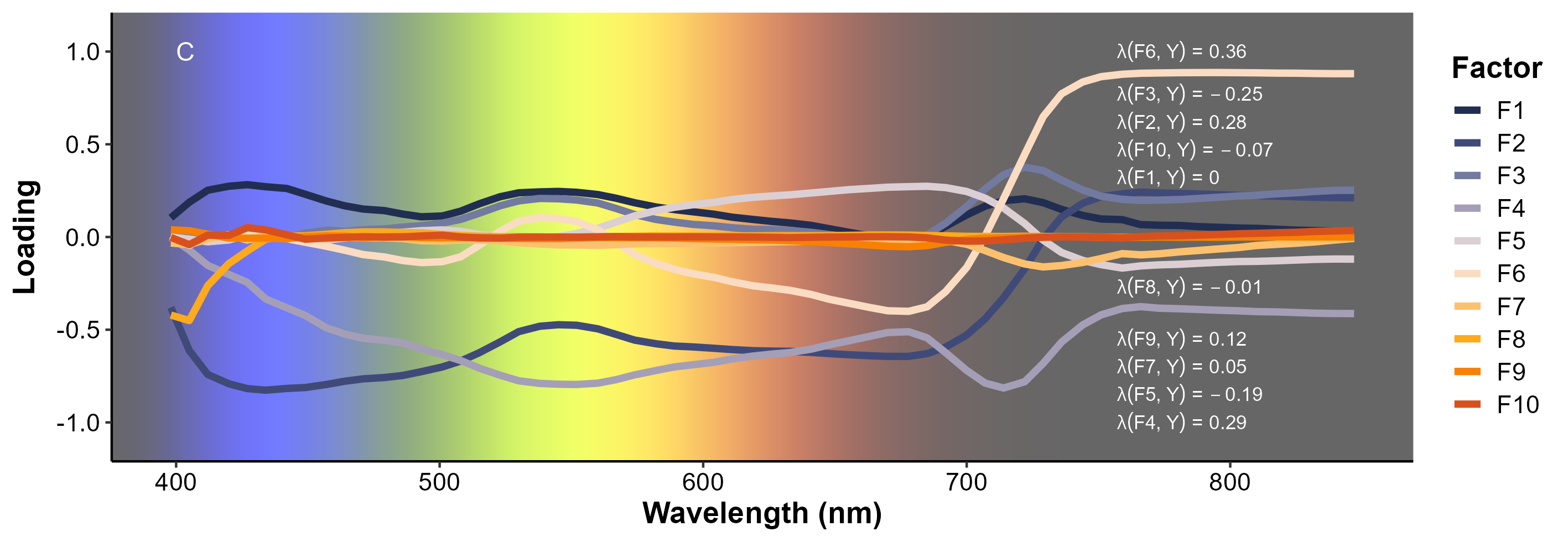}
	\caption{MegaLMM mean posterior loadings for the $62$ wavelengths on the $3$ (A), $5$ (B), and $10$ (C) factors used to run MegaLMM on the B5IR hyperspectral data from 10-03-2015. Mean posterior loadings of yield on the factors are also shown.}
	\label{fig:MegaLMM_hyper_single_date}
\end{figure}

MegaLMM does provide the flexibility of choosing the number of factors ($M$) and thereby the dimensionality, but leaves this choice to the user.
As a result, the ease of interpretation heavily relies on an appropriate choice of $M$, which may not be easy to make without any prior knowledge (Figure \ref{fig:MegaLMM_hyper_single_date}).
We opted for values of $M=3$, $M=5$, and $M=10$, which seemed appropriate given the fact we found $3$ factors using glfBLUP.
Note that while the main factors representing the infrared and visible part of the spectrum are still present (F1 and F2 for $M=3$, F1, F2, and F3 for $M=5$, F6, F3, F2, F5, and F4 for $M=10$), ease of interpretation quickly goes down as the number of factors increases.
This is mainly due to the loadings of yield, which are analogous to the genetic correlations between yield and factors in glfBLUP, becoming divided over a large number of factors.
We also stress that the choice to use a relatively low number of factors here was informed by the results obtained using glfBLUP.
\citet{Runcieetal_2021_SM} use $100$ factors to analyze the hyperspectral data in their paper, which would likely make interpretation even more difficult.

\begin{figure}[h]
	\centering
	\includegraphics[width=\textwidth]{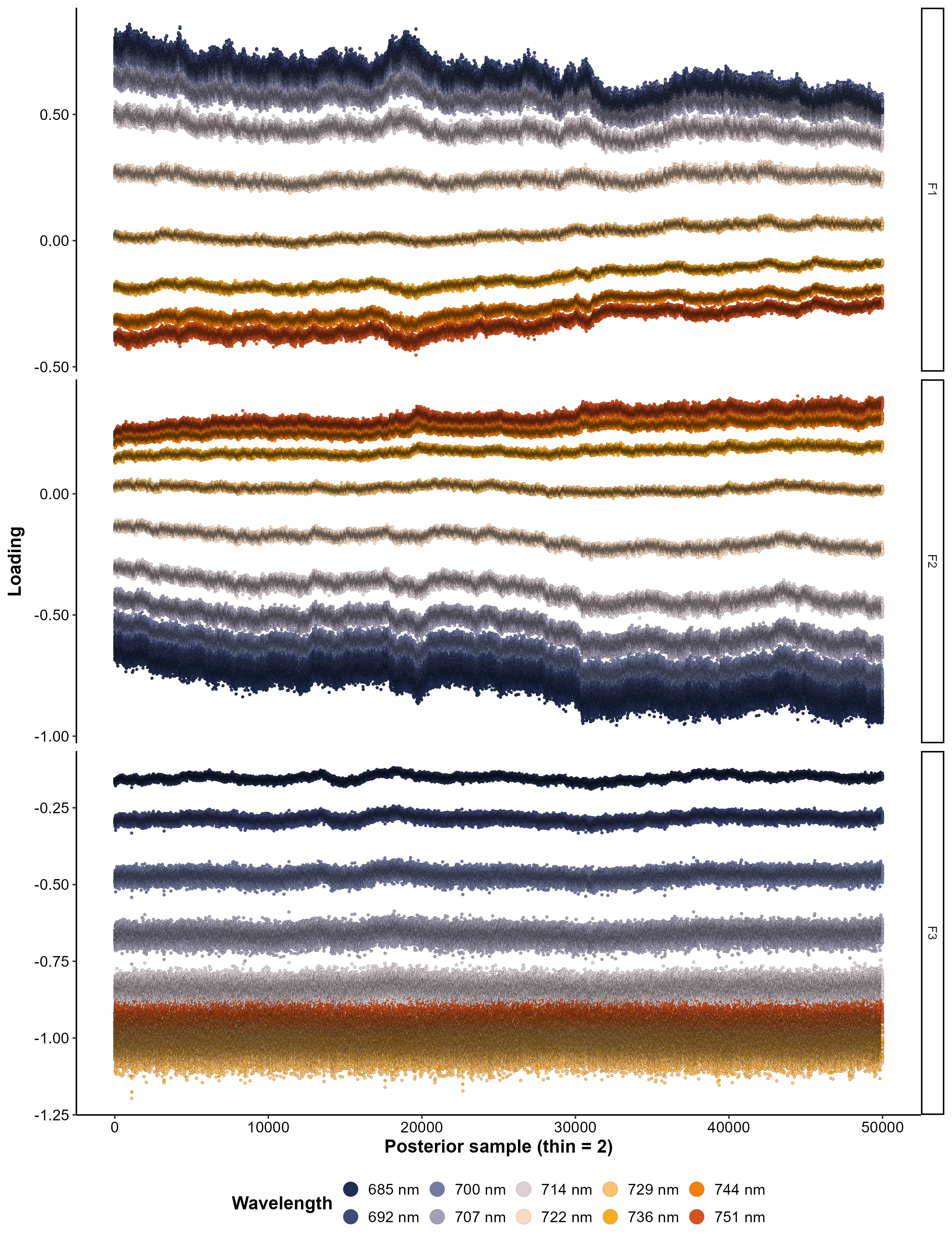}
	\caption{MegaLMM ($3$ factors, $100{,}000$ iterations with thinning rate of $2$ after $10{,}000$ burn-in iterations) traceplots for the posterior samples for loadings of wavelengths around 700 nm on the $3$ factors (top to bottom).}
	\label{fig:MegaLMM_hyper_single_date_traceplots_M3}
\end{figure}

\begin{figure}[h]
	\centering
	\includegraphics[width=\textwidth]{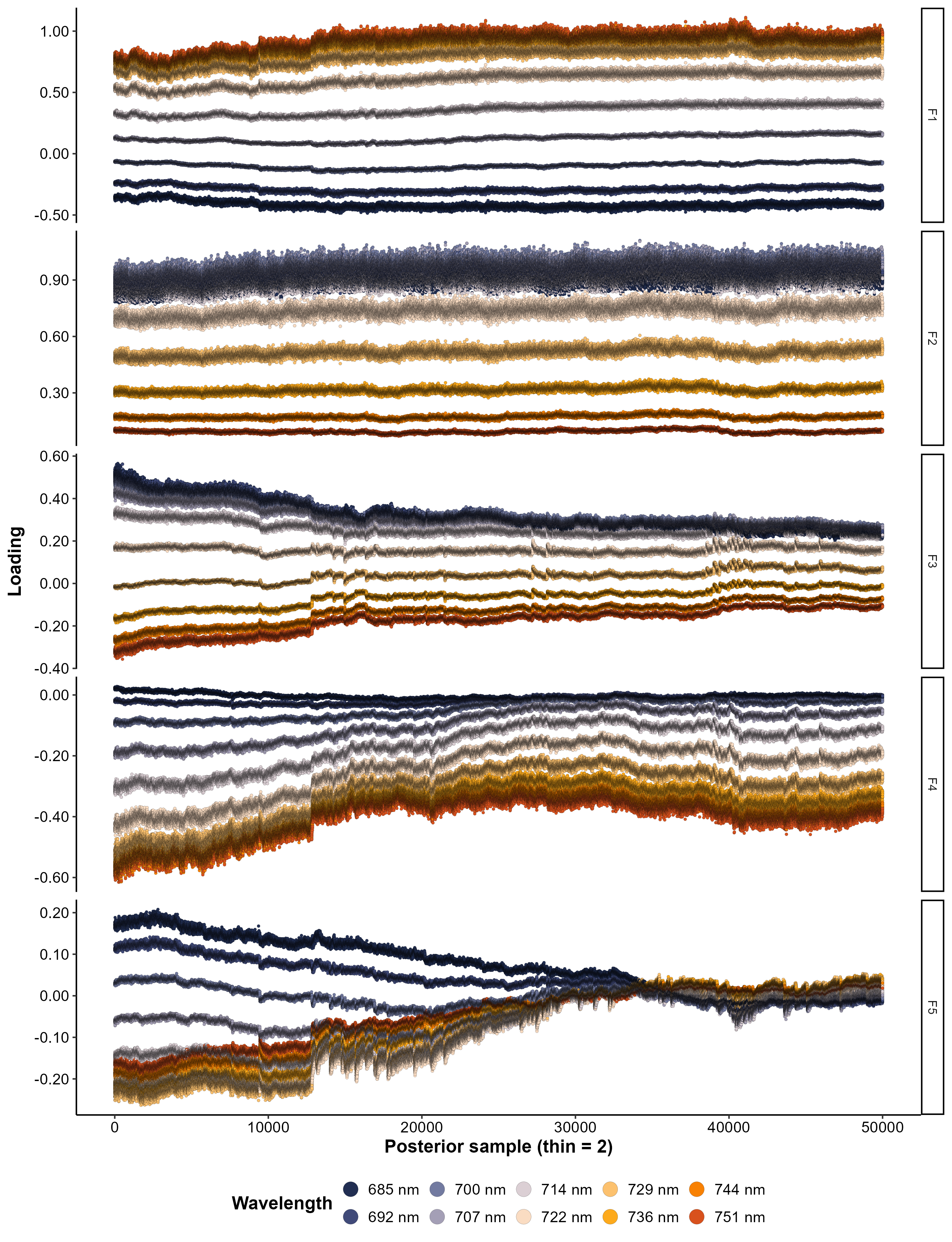}
	\caption{MegaLMM ($5$ factors, $100{,}000$ iterations with thinning rate of $2$ after $10{,}000$ burn-in iterations) traceplots for the posterior samples for loadings of wavelengths around 700 nm on the $5$ factors (top to bottom).}
	\label{fig:MegaLMM_hyper_single_date_traceplots_M5}
\end{figure}

\begin{figure}[h]
	\centering
	\includegraphics[width=\textwidth]{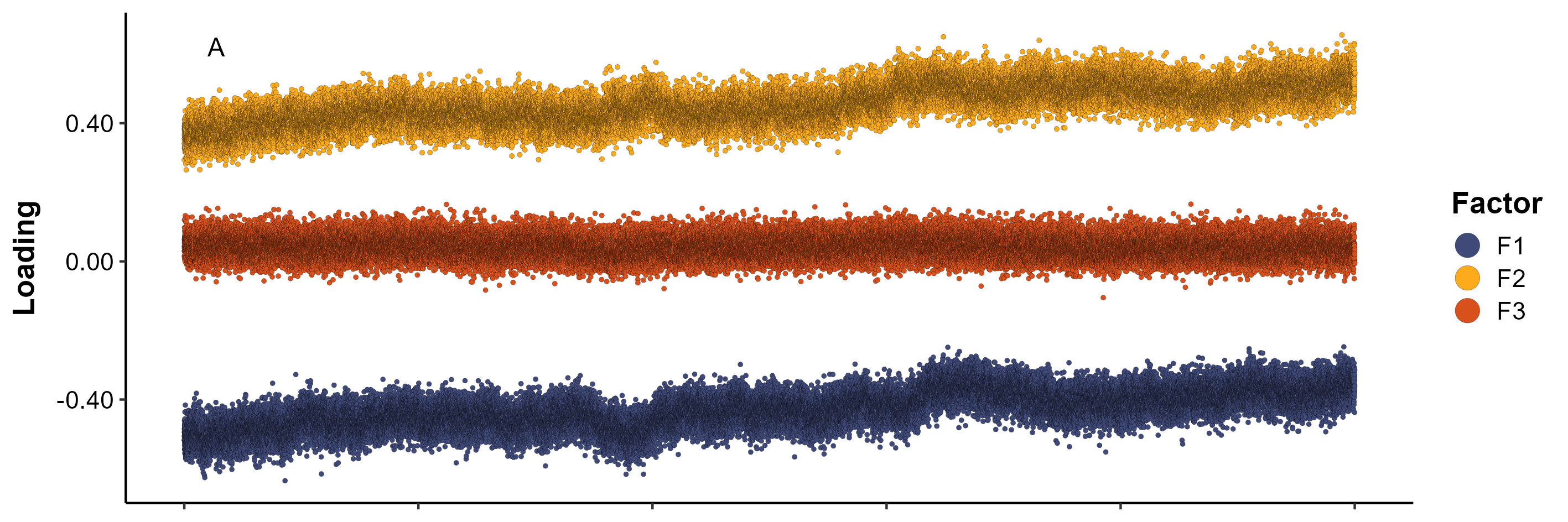}
	\includegraphics[width=\textwidth]{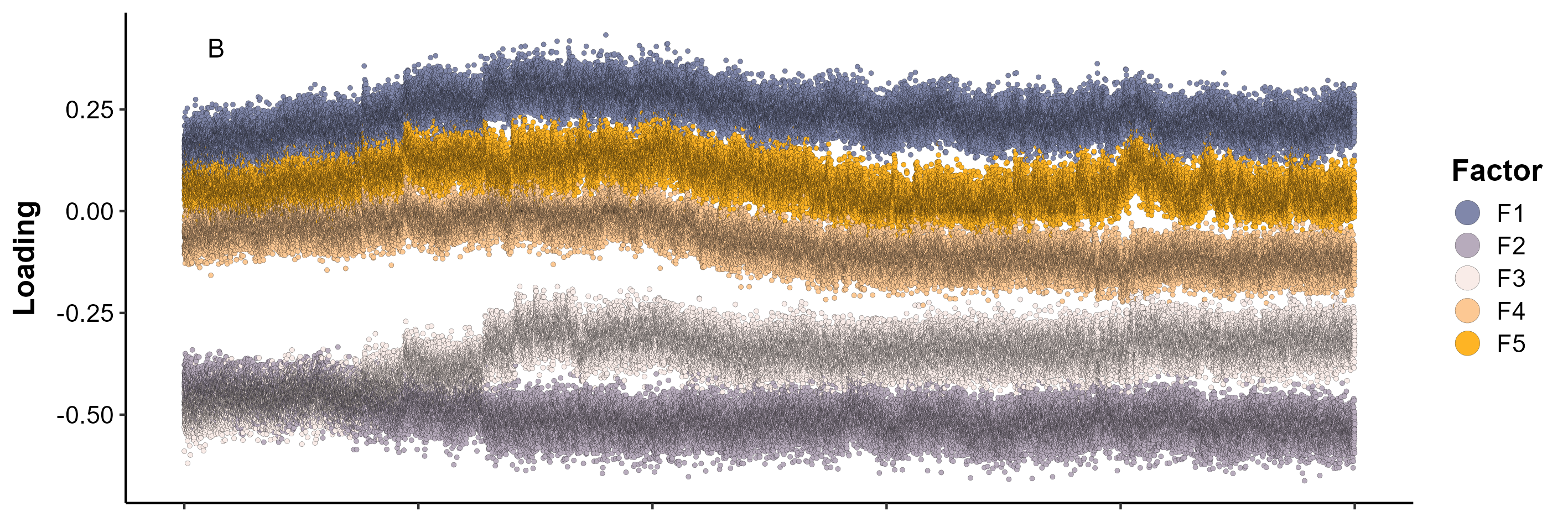}
	\includegraphics[width=\textwidth]{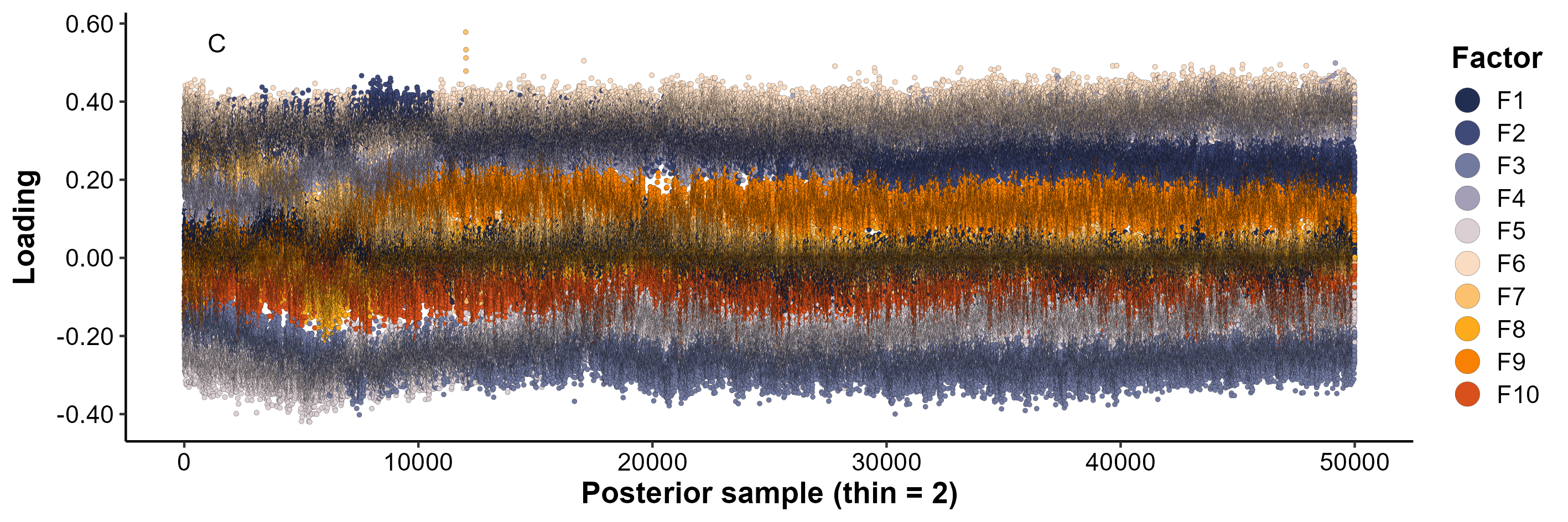}
	\caption{MegaLMM posterior iterations for the loadings of yield on the factors for $M=3$ (A), $M=5$ (B), and $M=10$ (C) for the B5IR hyperspectral data from 10-03-2015.}
	\label{fig:MegaLMM_hyper_single_date_Y_traceplots}
\end{figure}

Furthermore, we observe non-stationary behavior in the loading MCMC chains for $M=3$ (Figure \ref{fig:MegaLMM_hyper_single_date_traceplots_M3}), $M=5$ (Figure \ref{fig:MegaLMM_hyper_single_date_traceplots_M5}), as well as $M=10$ (Figure \ref{fig:MegaLMM_hyper_single_date_traceplots_M10_1} and \ref{fig:MegaLMM_hyper_single_date_traceplots_M10_2}), even after $10{,}000$ burn-in and a further $100{,}000$ posterior iterations for wavelengths around $700$nm (thinning rate of $2$).
Non-stationarity is also visible in the traceplots for the loadings of yield on the different factors (Figure \ref{fig:MegaLMM_hyper_single_date_Y_traceplots}).

\begin{figure}[h]
	\centering
	\includegraphics[width=\textwidth]{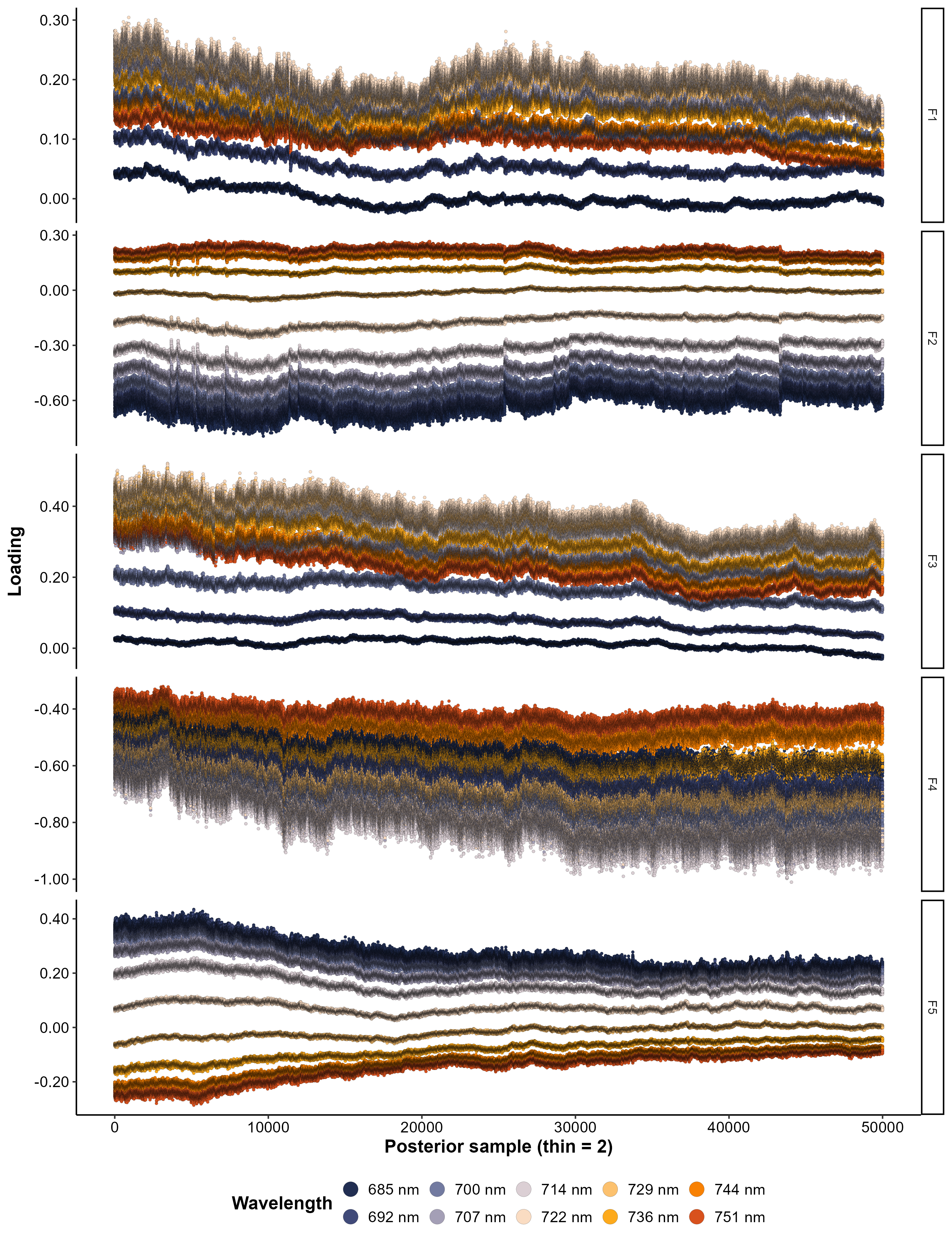}
	\caption{MegaLMM ($10$ factors, $100{,}000$ iterations with thinning rate of $2$ after $10{,}000$ burn-in iterations) traceplots for the posterior samples for loadings of wavelengths around 700 nm on factor $1$ - $5$. \emph{(Continues on next page)}.}
	\label{fig:MegaLMM_hyper_single_date_traceplots_M10_1}
\end{figure}

\begin{figure}[h]
	\centering
	\includegraphics[width=\textwidth]{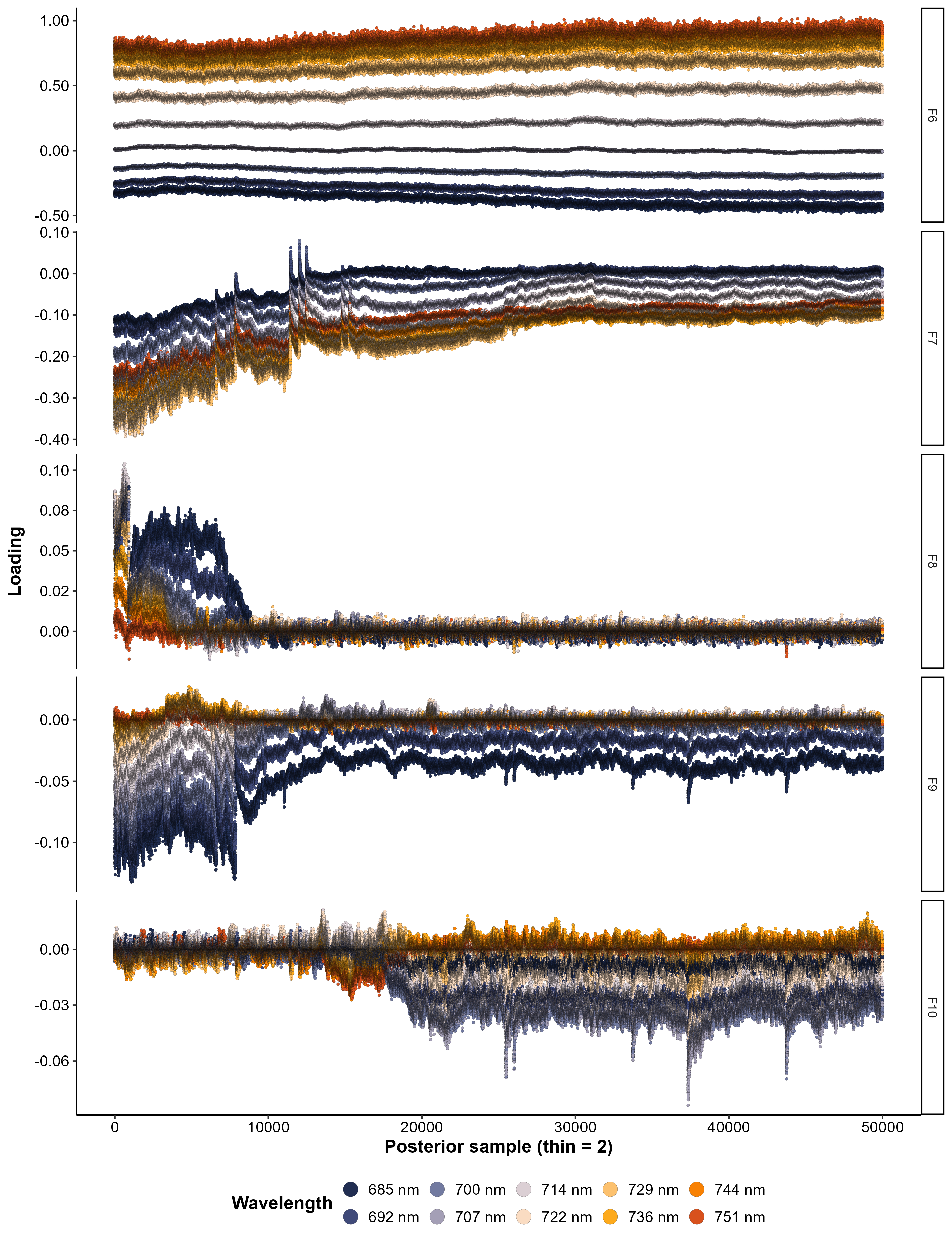}
	\caption{\emph{(Continued)}. MegaLMM ($10$ factors, $100{,}000$ iterations with thinning rate of $2$ after $10{,}000$ burn-in iterations) traceplots for the posterior samples for loadings of wavelengths around 700 nm on factor $6$ - $10$.}
	\label{fig:MegaLMM_hyper_single_date_traceplots_M10_2}
\end{figure}

\section{Computational complexity}
\label{comp}
Obtaining test set predictions for a single CV2 hyperspectral dataset (including all $620$ wavelength-date features) took approximately $1.5$ seconds for the univariate gBLUP model, $7$ seconds for lsBLUP, $200$ seconds for siBLUP, $34$ seconds for glfBLUP, $1090$ seconds for MegaLMM, and $60$ seconds for the CV2 multivariate MLP.

\begin{figure}[h]
	\centering
	\includegraphics[width=\textwidth]{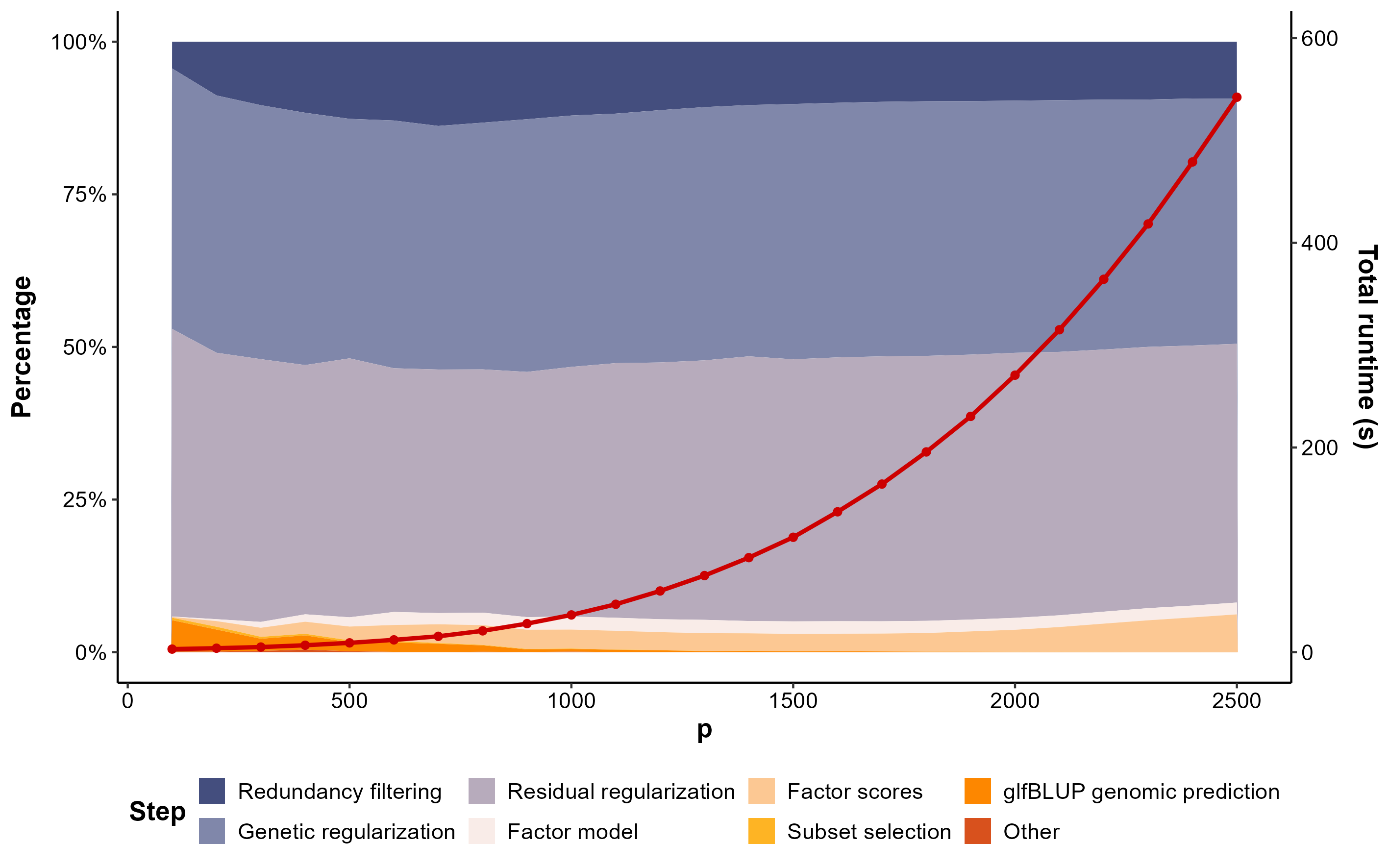}
	\caption{Composition of total runtime of glfBLUP at different numbers of secondary features $p$. The red line shows the total runtime in seconds.}
	\label{fig:timing_percentages}
\end{figure}

We also simulated datasets containing $p = 100$ to $p = 2500$ secondary features representing $10$ underlying factors.
We analyzed these datasets using glfBLUP, obtaining five runtime measurements (after a single warmup run) for each step in the glfBLUP prediction pipeline at each value of $p$, showing that the regularization dominates the pipeline's total runtime throughout the range of $p$ (Figure \ref{fig:timing_percentages}).

All computational times were obtained on Windows 11 Pro 23H2, {\ttfamily R} v4.2.2, using an Intel i5-13600KF CPU.

\section{Redundancy filtering}
\label{SM:filtering}
\begin{figure}[h]
	\centering
	\includegraphics[width=\textwidth]{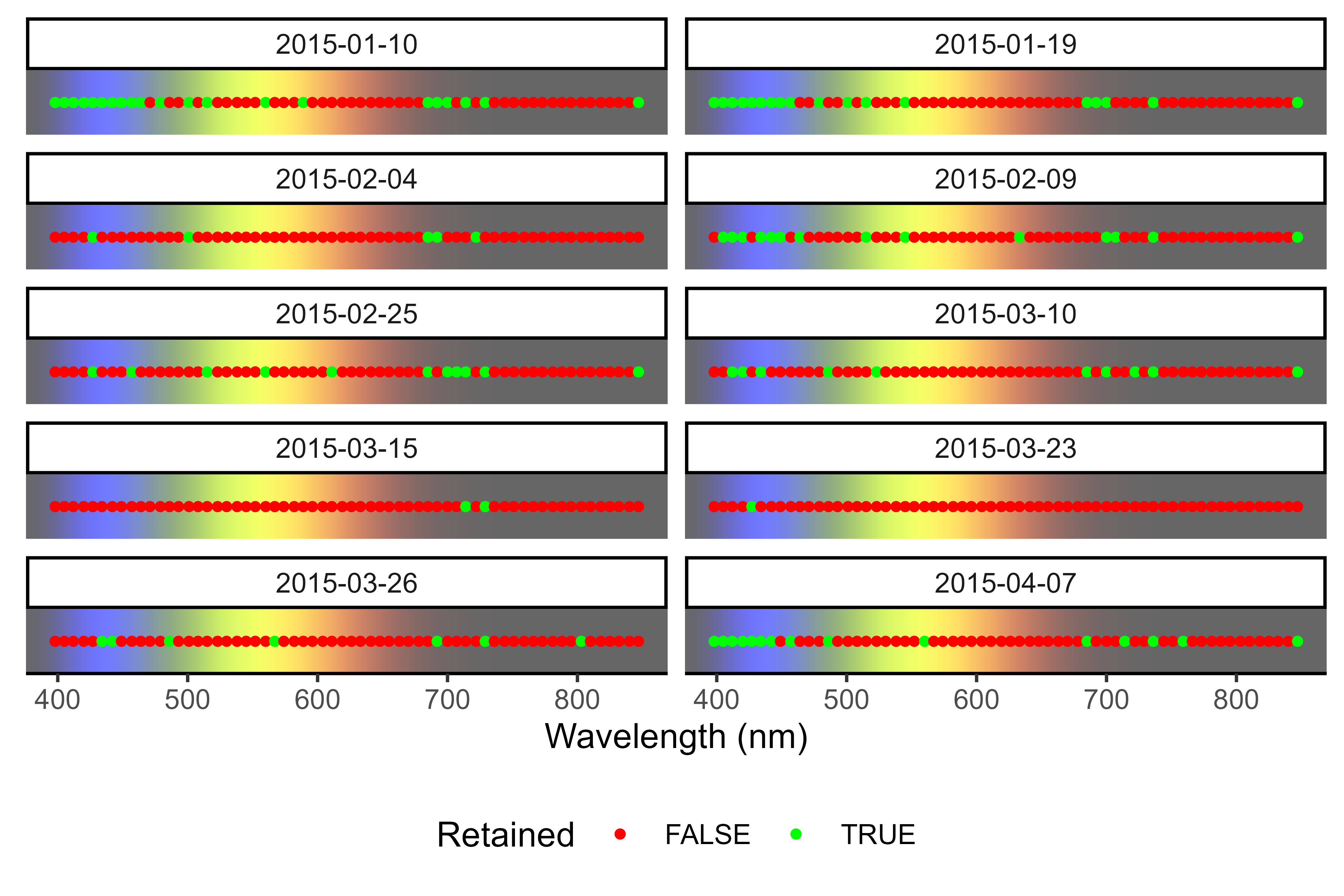}
	\caption{Features (wavelength-date combinations) that remain after redundancy filtering using a threshold of $\tau = 0.95$ for the B5IR treatment.}
	\label{fig:filtering_B5IR}
\end{figure}

\begin{figure}[h]
	\centering
	\includegraphics[width=\textwidth]{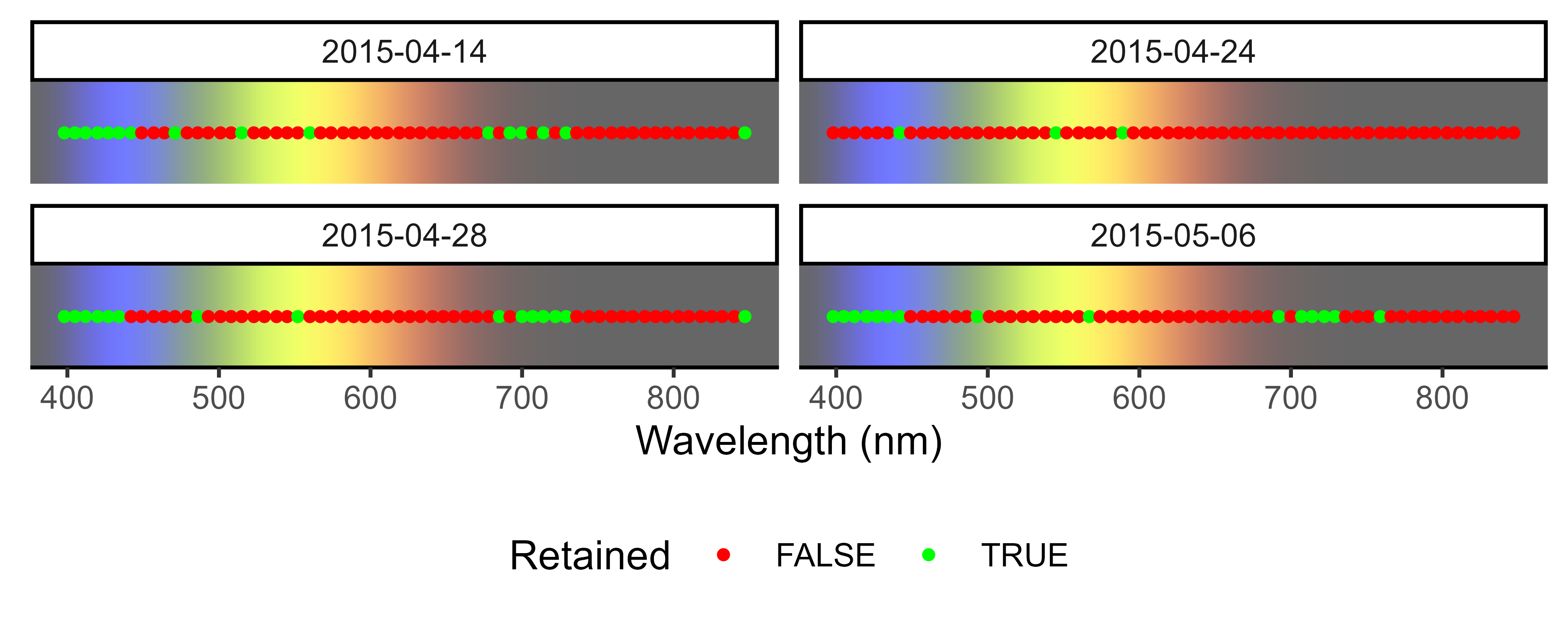}
	\caption{Features (wavelength-date combinations) that remain after redundancy filtering using a threshold of $\tau = 0.95$ for the HEAT treatment.}
	\label{fig:filtering_HEAT}
\end{figure}

Figures \ref{fig:filtering_B5IR} and \ref{fig:filtering_HEAT} provide graphical overviews of the features that remained after redundancy filtering at a threshold of $\tau = 0.95$ for both managed treatments B5IR and HEAT.


\begin{thebibliography}{}
	\bibitem[\protect\citeauthoryear{Adak \it{et~al.}}{2023}]{Adaketal_2023}
	Adak, A., Kang, M., Anderson, S.L., Murray, S.C., Jarquin, D., Wong, R.K.W., and Katzfuß, M. (2023). Phenomic data-driven biological prediction of maize through field-based high-throughput phenotyping integration with genomic data.
	{\it Journal of Experimental Botany} {\bf 74,} 5307--5326.
	\url{https://doi.org/10.1093/jxb/erad216}
	
	\bibitem[\protect\citeauthoryear{Anderson}{2003}]{Anderson2003}
	Anderson, W.T. (2003). An introduction to multivariate statistical analysis. Wiley, 3rd ed., Hoboken, NJ, USA.
	
	\bibitem[\protect\citeauthoryear{Arouisse \it{et~al.}}{2021}]{Arouisseetal_2021}
	Arouisse, B., Theeuwen, T.P.J.M., van Eeuwijk, F.A., and Kruijer, W. (2021). Improving genomic prediction using high-dimensional secondary phenotypes.
	{\it Frontiers in Genetics} {\bf 12,} 667358.
	\url{https://doi.org/10.3389/fgene.2021.667358}
	
	\bibitem[\protect\citeauthoryear{Boer}{2023}]{Boer_2023}
	Boer, M.P. (2023). Tensor product P-splines using a sparse mixed model formulation.
	{\it Statistical Modelling} {\bf 23,} 465--479.
	\url{https://doi.org/10.1177/1471082X231178591}
	
	\bibitem[\protect\citeauthoryear{Brent}{1973}]{Brent1973}
	Brent, R.P. (1973). "Chapter 5: An algorithm with guaranteed convergence for finding a minimum of a function of one variable", Algorithms for minimization without derivatives. Prentice-Hall, Englewood Cliffs, NJ, USA.
	
	\bibitem[\protect\citeauthoryear{Colombani \it{\it{et~al.}}}{2012}]{Colombani_etal2012}
	Colombani, C., Croiseau, P., Fritz, S., Guillaume, F., Legarra, A., Ducrocq, V., and Robert-Granié, C. (2012). A comparison of partial least squares (PLS) and sparse PLS regressions in genomic selection in French dairy cattle.
	{\it Journal of Dairy Science} {\bf 95,} 2120--2131.
	\url{https://doi.org/10.3168/jds.2011-4647}
	
	\bibitem[\protect\citeauthoryear{Crossa \it{\it{et~al.}}}{2013}]{Crossaetal2013}
	Crossa, J., Pérez, P., Hickey, J., Burgue\~{n}o, J., Ornella, L., Cerón-Rojas, J., Zhang, X., Dreisigacker, S., Babu, R., Li, Y., Bonnett, D., and Mathews, K. (2013). Genomic prediction in CIMMYT maize and wheat breeding programs.
	{\it Heredity} {\bf 112,} 48--60.
	\url{https://doi.org/10.1038/hdy.2013.16}
	
	\bibitem[\protect\citeauthoryear{Cullis \it{et~al.}}{2014}]{Cullisetal2014}
	Cullis, B.R., Jefferson, P., Thompson, R., and Smith, A.B. (2014). Factor analytic and reduced animal models for the investigation of additive genotype-by-environment interaction in outcrossing plant species with application to a Pinus radiata breeding programme.
	{\it Theoretical and Applied Genetics} {\bf 127,} 2193--2210.
	\url{https://doi.org/10.1007/s00122-014-2373-0}
	
	\bibitem[\protect\citeauthoryear{Danilevicz \it{et~al.}}{2022}]{Danileviczetal2022}
	Danilevicz, M.F., Gill, M., Anderson, R., Batley, J., Bennamoun, M., Bayer, B.E., and Edwards, D. (2022). Plant genotype to phenotype prediction using machine learning.
	{\it Frontiers in Genetics} {\bf 13,} 822173.
	\url{https://doi.org/10.3389/fgene.2022.822173}
	
	\bibitem[\protect\citeauthoryear{Dahl \it{et~al.}}{2013}]{Dahletal2013}
	Dahl, A., Hore, V., Iotchkova, V., and Marchini, J. (2013). Network inference in matrix-variate Gaussian models with non-independent noise.
	{\it arXiv}, 1312.1622.
	\url{https://doi.org/10.48550/arXiv.1312.1622}
	
	\bibitem[\protect\citeauthoryear{Dai \it{et~al.}}{2020}]{Daietal2020}
	Dai, F., Dutta, S., and Maitra, R. (2020). A matrix-free likelihood method for exploratory factor analysis of high-dimensional gaussian data.
	{\it Journal of Computational and Graphical Statistics} {\bf 29,} 675--680.
	\url{https://doi.org/10.1080/10618600.2019.1704296}
	
	\bibitem[\protect\citeauthoryear{Endelman}{2011}]{Endelman_2011}
	Endelman, J.B. (2011). Ridge regression and other kernels for genomic selection with {\ttfamily R}-package rrBLUP.
	{\it The Plant Genome} {\bf 4,} 250--255.
	\url{https://doi.org/10.3835/plantgenome2011.08.0024}
	
	\bibitem[\protect\citeauthoryear{Fan \it{et~al.}}{2021}]{Fanetal_2021}
	Fan, J., Wang, K., Zhong, Y., and Zhu, Z. (2022). Robust high dimensional factor models with applications to statistical machine learning.
	{\it Statistical Science} {\bf 36,} 303--327.
	\url{https://doi.org/10.1214/20-STS785}
	
	\bibitem[\protect\citeauthoryear{Fotheringham and Oshan}{2016}]{Fotheringham&Oshan_2016}
	Fotheringham, A.S., and Oshan, T.M. (2016). Geographically weighed regression and multicollinearity: dispelling the myth.
	{\it Journal of Geographical Systems} {\bf 18,} 303--329.
	\url{https://doi.org/10.1007/s10109-016-0239-5}
	
	
	\bibitem[\protect\citeauthoryear{Harfouche \it{et~al.}}{2019}]{Harfoucheetal2019}
	Harfouche, A.L., Jacobson, D.A., Kainer, D., Romero, J.C., Harfouche, A.H., Mugnozza, G.S., Moshelion, M., Tuskan, G.A., Keurentjes, J.J.B., and Altman, A. (2019). Accelerating climate resilient plant breeding by applying next-generation artificial intelligence.
	{\it Trends in Biotechnology} {\bf 37,} 1217--1235.
	\url{https://doi.org/10.1016/j.tibtech.2019.05.007}
	
	\bibitem[\protect\citeauthoryear{Higham}{2002}]{Higham2002}
	Higham, N.J. (2002). Computing the nearest correlation matrix - a problem from finance.
	{\it IMA Journal of Numerical Analysis} {\bf 22,} 329--343.
	\url{https://doi.org/10.1093/imanum/22.3.329}
	
	
	\bibitem[\protect\citeauthoryear{Horton \it{et~al.}}{2012}]{Hortonetal2012}
	Horton, M.W., Hancock, A.M., Huang, Y.S., Toomajian, C., Atwell, S., Auton, A., Muliyati, N.W., Platt, A., Sperone, F.G., Vilhjálmsson, B.J., Nordborg, M., Borevitz, J.O., and Bergelson, J. (2012). Genome-wide patterns of genetic variation in worldwide {Arabidopsis thaliana} accessions from the RegMap panel.
	{\it Nature Genetics} {\bf 44,} 212--217.
	\url{https://doi.org/10.1038/ng.1042}
	
	
	\bibitem[\protect\citeauthoryear{Jöreskog}{1967}]{Jöreskog1967}
	Jöreskog, K.G. (1967). Some contributions to maximum likelihood factor analysis.
	{\it Psychometrika} {\bf 32,} 443--482.
	\url{https://doi.org/10.1007/BF02289658}
	
	\bibitem[\protect\citeauthoryear{Kaiser}{1958}]{Kaiser1958}
	Kaiser, H.F. (1958). The varimax criterion for analytic rotation in factor analysis.
	{\it Psychometrika} {\bf 23,} 187--200.
	\url{https://doi.org/10.1007/BF02289233}
	
	\bibitem[\protect\citeauthoryear{Krause \it{et~al.}}{2019}]{Krauseetal_2019}
	Krause, M.R., González-Pérez, L., Crossa, J., Pérez-Rodríguez, P., Montesinos-López, O., Singh, R.P., Dreisigacker, S., Poland, J., Rutkoski, J., Sorrells, M., Gore, M.A., and Mondal, S. (2019). Hyperspectral reflectance-derived relationship matrices for genomic prediction of grain yield in wheat.
	{\it G3} {\bf 9,} 1231--1247.
	\url{https://doi.org/10.1534/g3.118.200856}
	
	\bibitem[\protect\citeauthoryear{Kruijer \it{et~al.}}{2020}]{Kruijeretal2020}
	Kruijer, W., Behrouzi, P., Bustos-Korts, D., Rodríguez-Álvarez, M.X., Mahmoudi, S.M., Yandell, B., Wit, E., and van Eeuwijk, F.A. (2020). Reconstruction of networks with direct and indirect genetic effects.
	{\it Genetics} {\bf 214,} 781--807.
	\url{https://doi.org/10.1534/genetics.119.302949}
	
	\bibitem[\protect\citeauthoryear{Ledermann}{1937}]{Ledermann1937}
	Ledermann, W. (1937). On the rank of the reduced correlational matrix in multiple-factor analysis.
	{\it Psychometrika} {\bf 2,} 85--93.
	\url{https://doi.org/10.1007/BF02288062}
	
	\bibitem[\protect\citeauthoryear{Lopez-Cruz \it{et~al.}}{2020}]{Lopez-Cruzetal_2020}
	Lopez-Cruz, M., Olson, E., Rovere, G., Crossa, J., Dreisigacker, S., Mondal, S., Singh, R., and de los Campos, G. (2020). Regularized selection indices for breeding value prediction using hyper-spectral image data.
	{\it Scientific Reports} {\bf 10,} 8195.
	\url{https://doi.org/10.1038/s41598-020-65011-2}
	
	\bibitem[\protect\citeauthoryear{Lumley}{2020}]{Lumley2020}
	Lumley, T. (2020). leaps: regression subset selection.. {\ttfamily R}-package version 3.1, \url{https://CRAN.R-project.org/package=leaps}
	\url{https://doi.org/10.32614/CRAN.package.leaps}
	
	\bibitem[\protect\citeauthoryear{Marčenko and Pastur}{1967}]{Marcenko&pastur1967}
	Marčenko, V.A., and Pastur, L.A. (1967). Distribution of eigenvalues for some sets of random matrices.
	{\it Mathematics of the USSR-Sbornik} {\bf 1,} 457--483.
	\url{https://doi.org/10.1070/SM1967v001n04ABEH001994}
	
	\bibitem[\protect\citeauthoryear{Meuwissen \it{et~al.}}{2001}]{Meuwissenetal_2001}
	Meuwissen, T.H., Hayes, B.J., and Goddard, M.E. (2001). Prediction of total genetic value using genome-wide dense marker maps.
	{\it Genetics} {\bf 157,} 1819--1829.
	\url{https://doi.org/10.1093/genetics/157.4.1819}
	
	\bibitem[\protect\citeauthoryear{Meyer and Kirkpatrick}{2010}]{Meyer&Kirkpatrick2010}
	Meyer, K., and Kirkpatrick, M. (2010). Better estimates of genetic covariance matrices by "bending" using penalized maximum likelihood.
	{\it Genetics} {\bf 185,} 1097--1110.
	\url{https://doi.org/10.1534/genetics.109.113381}
	
	\bibitem[\protect\citeauthoryear{Mkhabela \it{et~al.}}{2011}]{Mkhabelaetal2011}
	Mkhabela, M.S., Bullock, P., Raj, S., Wang, S., and Yang, Y. Crop yield forecasting on the Canadian prairies using MODIS NDVI data.
	{\it Agricultural and Forest Meteorology} {\bf 151,} 385--393.
	\url{https://doi.org/10.1016/j.agrformet.2010.11.012}
	
	\bibitem[\protect\citeauthoryear{Molenaar}{1985}]{Molenaar1985}
	Molenaar, P.C. (1985). A dynamic factor model for the analysis of multivariate time series.
	{\it Psychometrika} {\bf 50,} 181--202.
	\url{https://doi.org/10.1007/BF02294246}
	
	\bibitem[\protect\citeauthoryear{Münch \it{et~al.}}{2022}]{Münchetal_2022}
	Münch, M.M., van de Wiel, M.A., van der Vaart, A.W., and Peeters, C.F.W. (2022). Semi-supervised empirical Bayes group-regularized factor regression.
	{\it Biometrical Journal} {\bf 64,} 1289--1306.
	\url{https://doi.org/10.1002/bimj.202100105}
	
	
	\bibitem[\protect\citeauthoryear{Ott and Longnecker}{2015}]{Ott&Longnecker2015}
	Ott, R., and Longnecker, M. (2015). An introduction to statistical methods and data analysis. Cengage Learning, 7th ed., Boston, MA, USA.
	
	\bibitem[\protect\citeauthoryear{Pavlidis \it{et~al.}}{2001}]{Pavlidisetal2001}
	Pavlidis, P., Weston, J., Cai, J., and Grundy, W.N. (2001). Gene functional classification from heterogeneous data. In: Proceedings of the fifth annual international conference on computational biology. Association for Computing Machinery, New York, NY, USA.
	\url{https://doi.org/10.1145/369133.369228}
	
	\bibitem[\protect\citeauthoryear{Peeters}{2012}]{Peeters2012}
	Peeters, C.F.W. (2012). Bayesian exploratory and confirmatory factor analysis. PhD thesis, Utrecht University, the Netherlands.
	
	\bibitem[\protect\citeauthoryear{Peeters \it{et~al.}}{2019}]{Peetersetal2019}
	Peeters, C.F.W., Übelhör, C., Mes, S.W., Martens, R., Koopman, T., de Graaf, P., van Velden, F.H.P., Boellaard, R., Castelijns, J.A., te Beest, D.E., Heymans, M.W., and van de Wiel, M.A. (2019). Stable prediction with radiomics data.
	{\it arXiv}, 1903.11696.
	\url{https://doi.org/10.48550/arXiv.1903.11696}
	
	\bibitem[\protect\citeauthoryear{Pérez-Valencia \it{et~al.}}{2022}]{Pérez-Valenciaetal_2022}
	Pérez-Valencia, D.M., Rodríguez-Álvarez, M.X., Boer, M.P., Kronenberg, L., Hund, A., Cabrera-Bosquet, L., Millet, E.J., and van Eeuwijk, F.A. (2022). A two-stage approach for the spatio-temporal analysis of high-throughput phenotyping data.
	{\it Scientific Reports} {\bf 12,} 3177.
	\url{https://doi.org/10.1038/s41598-022-06935-9}
	
	\bibitem[\protect\citeauthoryear{Piepho \it{et~al.}}{2013}]{Piepho_etal2013}
	Piepho, H.P., Williams, E.R., and Ogutu, J. (2013). A two-stage approach to recovery of inter-block information and shrinkage of block effect estimates.
	{\it Communications in Biometry and Crop Science} {\bf 8,} 10--22.
	
	
	\bibitem[\protect\citeauthoryear{Rodríguez-Álvarez \it{et~al.}}{2018}]{Rodríguez-Álvarezetal_2018}
	Rodríguez-Álvarez, M.X., Boer, M.P., van Eeuwijk, F.A., and Eilers, P.H.C. (2018). Correcting for spatial heterogeneity in plant breeding experiments with P-splines.
	{\it Spatial Statistics} {\bf 23,} 52--71.
	\url{https://doi.org/10.1016/j.spasta.2017.10.003}
	
	\bibitem[\protect\citeauthoryear{Runcie and Cheng}{2019}]{Runcie&cheng_2019}
	Runcie, D., and Cheng, H. (2019). Pitfalls and remedies for cross validation with multi-trait genomic prediction methods.
	{\it G3} {\bf 9,} 3727--3741.
	\url{https://doi.org/10.1534/g3.119.400598}
	
	\bibitem[\protect\citeauthoryear{Runcie \it{et~al.}}{2021}]{Runcieetal_2021}
	Runcie, D.E., Qu, J.Q., Cheng, H., Crawford, L. (2021). MegaLMM: Mega-scale linear mixed models for genomic predictions with thousands of traits.
	{\it Genome Biology} {\bf 22,} 213.
	\url{https://doi.org/10.1186/s13059-021-02416-w}
	
	\bibitem[\protect\citeauthoryear{Smith \it{et~al.}}{2001}]{Smithetal_2001}
	Smith, A., Cullis, B., and Thompson, R. (2001). Analyzing variety by environment data using multiplicative mixed models and adjustments for spatial field trend.
	{\it Biometrics} {\bf 57,} 1138--1147.
	\url{https://doi.org/10.1111/j.0006-341X.2001.01138.x}
	
	\bibitem[\protect\citeauthoryear{Sun \it{et~al.}}{2017}]{Sunetal_2017}
	Sun, J., Rutkoski, J.E., Poland, J.A., Crossa, J., Jannink, J.L., and Sorrells, M.E. (2017). Multitrait, random regression, or simple repeatability model in high-throughput phenotyping data improve genomic prediction for wheat grain yield.
	{\it The Plant Genome} {\bf 10,} plantgenome2016.11.0111.
	\url{https://doi.org/10.3835/plantgenome2016.11.0111}
	
	\bibitem[\protect\citeauthoryear{Teal \it{et~al.}}{2006}]{Tealetal2006}
	Teal, R.K., Tubana, B., Girma, K., Freeman, K.W., Arnall, D.B., Walsh, O., and Raun, W.R. (2006). In-season prediction of corn grain yield potential using normalized difference vegetation index.
	{\it Agronomy Journal} {\bf 98,} 1488--1494.
	\url{https://doi.org/10.2134/agronj2006.0103}
	
	\bibitem[\protect\citeauthoryear{Thomson}{1939}]{Thomson1939}
	Thomson, G. (1939). The factorial analysis of human ability. University of London Press, London, UK.
	
	\bibitem[\protect\citeauthoryear{Togninalli \it{et~al.}}{2023}]{Togninallietal_2023}
	Togninalli. M., Wang, X., Kucera, T., Shrestha, S., Juliana, P., Mondal, S., Pinto, F., Govindan, V., Crespo-Herrera, L., Huerta-Espino, J., Singh, R.P., Borgwardt, K., Poland, J. (2023). Multi-modal deep learning improves grain yield prediction in wheat breeding by fusing genomics and phenomics.
	{\it Bioinformatics} {\bf 39,} btad336.
	\url{https://doi.org/10.1093/bioinformatics/btad336}
	
	\bibitem[\protect\citeauthoryear{Tolhurst \it{et~al.}}{2019}]{Tolhurstetal_2019}
	Tolhurst, D.J., Mathews, K.L., Smith, A.B., and Cullis, B.R. (2019). Genomic selection in multi‐environment plant breeding trials using a factor analytic linear mixed model.
	{\it Journal of Animal Breeding and Genetics} {\bf 136,} 279--300.
	\url{https://doi.org/10.1111/jbg.12404}
	
	
	
	\bibitem[\protect\citeauthoryear{van Wieringen and Peeters}{2016}]{vanWieringen&Peeters2016}
	van Wieringen, W.N., and Peeters, C.F.W. (2016). Ridge estimation of inverse covariance matrices from high-dimensional data.
	{\it Computational Statistics \& Data Analysis} {\bf 103,} 284--303.
	\url{https://doi.org/10.1016/j.csda.2016.05.012}
	
	\bibitem[\protect\citeauthoryear{Woodbury}{1950}]{Woodbury1950}
	Woodbury, M.A. (1950). Inverting modified matrices. Memorandum Report 42, Statistical Research Group, Princeton University, Princeton, NJ, USA.
	
	
	
	\bibitem[\protect\citeauthoryear{Zhang \it{et~al.}}{2022}]{Zhangetal_2022}
	Zhang, A.R, Cai, T.T., and Wu, Y. (2022). Heteroskedastic PCA: Algorithm, optimality, and applications.
	{\it Annals of Statistics} {\bf 50,} 53--80.
	\url{https://doi.org/10.1214/21-AOS2074}
	
	\bibitem[\protect\citeauthoryear{Zhou and Stephens}{2014}]{Zhou&Stephens2014}
	Zhou, X, and Stephens, M. (2014). Efficient multivariate linear mixed model algorithms for genome-wide association studies.
	{\it Nature methods} {\bf 11,} 407--409.
	\url{https://doi.org/10.1038/nmeth.2848}
	
\end{thebibliography}

\begin{thebibliography}{}
	
	\bibitem[\protect\citeauthoryear{Abadi \it{et~al.}}{2016}]{Abadietal2016_SM}
	Abadi, M., Barham, P., Chen, J., Chen, Z., Davis, A., Dean, J., Devin, M., Ghemawat, S., Irving, G., Isard, M., Kudlur, M., Levenberg, J., Monga, R., Moore, S., Murray, D.G., Steiner, B., Tucker, P., Vasudevan, V., Warden, P., Wicke, M., Yu, Y., and Zheng, X. (2016). TensorFlow: A system for large-scale machine learning. In
	\emph{12th USENIX Symposium on Operating Systems Design and Implementation (OSDI 16)} (pp. 265-283). USENIX Association.
	\url{https://www.usenix.org/conference/osdi16/technical-sessions/presentation/abadi}
	
	\bibitem[\protect\citeauthoryear{Arouisse \it{et~al.}}{2021}]{Arouisseetal_2021_SM}
	Arouisse, B., Theeuwen, T.P.J.M., van Eeuwijk, F.A., and Kruijer, W. (2021). Improving genomic prediction using high-dimensional secondary phenotypes.
	\emph{Frontiers in Genetics} {\bf 12,} 667358.
	\url{https://doi.org/10.3389/fgene.2021.667358}
	
	\bibitem[\protect\citeauthoryear{Brent}{1973}]{Brent1973_SM}
	Brent, R.P. (1973). "Chapter 5: An algorithm with guaranteed convergence for finding a minimum of a function of one variable", Algorithms for minimization without derivatives. Prentice-Hall, Englewood Cliffs, NJ, USA.
	
	\bibitem[\protect\citeauthoryear{Chollet \it{et~al.}}{2015}]{Cholletetal2015_SM}
	Chollet, F., et al. (2015). \emph{Keras}. \url{https://keras.io/.}
	
	\bibitem[\protect\citeauthoryear{Dahl \it{et~al.}}{2013}]{Dahletal2013_SM}
	Dahl, A., Hore, V., Iotchkova, V., and Marchini, J. (2013). Network inference in matrix-variate Gaussian models with non-independent noise.
	\emph{arXiv}, 1312.1622.
	\url{https://doi.org/10.48550/arXiv.1312.1622}
	
	\bibitem[\protect\citeauthoryear{Endelman}{2011}]{Endelman2011_SM}
	Endelman, J.B. (2011). Ridge regression and other kernels for genomic selection with {\ttfamily R}-package rrBLUP.
	{\it The Plant Genome} {\bf 4,} 250--255.
	\url{https://doi.org/10.3835/plantgenome2011.08.0024}
	
	\bibitem[\protect\citeauthoryear{Friedman \it{et~al.}}{2010}]{Friedmanetal2010_SM}
	Friedman, J., Hastie, T., and Tibshirani, R. (2010). Regularization paths for generalized linear models via coordinate descent.
	{\it Journal of Statistical Software} {\bf 33,} 1--22.
	\url{https://doi.org/10.18637/jss.v033.i01}
	
	\bibitem[\protect\citeauthoryear{Lopez-Cruz \it{et~al.}}{2020}]{Lopez-Cruzetal_2020_SM}
	Lopez-Cruz, M., Olson, E., Rovere, G., Crossa, J., Dreisigacker, S., Mondal, S., Singh, R., and de los Campos, G. (2020). Regularized selection indices for breeding value prediction using hyper-spectral image data.
	\emph{Scientific Reports} {\bf 10,} 8195.
	\url{https://doi.org/10.1038/s41598-020-65011-2}
	
	\bibitem[\protect\citeauthoryear{Runcie and Cheng}{2019}]{Runcie&cheng_2019_SM}
	Runcie, D., and Cheng, H. (2019). Pitfalls and remedies for cross validation with multi-trait genomic prediction methods.
	{\it G3} {\bf 9,} 3727--3741.
	\url{https://doi.org/10.1534/g3.119.400598}
	
	\bibitem[\protect\citeauthoryear{Runcie \it{et~al.}}{2021}]{Runcieetal_2021_SM}
	Runcie, D.E., Qu, J.Q., Cheng, H., Crawford, L. (2021). MegaLMM: Mega-scale linear mixed models for genomic predictions with thousands of traits.
	\emph{Genome Biology} {\bf 22,} 213.
	\url{https://doi.org/10.1186/s13059-021-02416-w}
	
	\bibitem[\protect\citeauthoryear{Runcie}{2022}]{Runcie2022_SM}
	Runcie, D.E. (2022). \_MegaLMM: MegaLMM\_. {\ttfamily R}-package version 0.1.0.
	
\end{thebibliography}
\end{document}